\documentclass[12pt,preprint]{aastex}

\usepackage{apjfonts}
\usepackage{emulateapj5}

\newenvironment{inlinefigure}{%
\def\@captype{figure}%
\noindent\begin{minipage}{0.999\linewidth}\begin{center}\small}
{\end{center}\end{minipage}\smallskip}
\makeatother

\begin{document}
\newcommand{\lya}{Lyman~$\alpha$}
\newcommand{\lyb}{Lyman~$\beta$}
\newcommand{\za}{$z_{\rm abs}$}
\newcommand{\ze}{$z_{\rm em}$}
\newcommand{\cmtwo}{cm$^{-2}$}
\newcommand{\nhi}{$N$(H$^0$)}
\newcommand{\degpoint}{\mbox{$^\circ\mskip-7.0mu.\,$}}
\newcommand{\kms}{\,km~s$^{-1}$}      
\newcommand{\minpoint}{\mbox{$'\mskip-4.7mu.\mskip0.8mu$}}
\newcommand{\peryr}{\mbox{$\>\rm yr^{-1}$}}
\newcommand{\secpoint}{\mbox{$''\mskip-7.6mu.\,$}}
\newcommand{\sqdeg}{\mbox{${\rm deg}^2$}}
\newcommand{\squig}{\sim\!\!}
\newcommand{\subsun}{\mbox{$_{\twelvesy\odot}$}}
\newcommand{\et}{{\rm et al.}~}

\def\ltsima{$\; \buildrel < \over \sim \;$}
\def\simlt{\lower.5ex\hbox{\ltsima}}
\def\gtsima{$\; \buildrel > \over \sim \;$}
\def\simgt{\lower.5ex\hbox{\gtsima}}
\def\arcs{$''~$}
\def\arcm{$'~$}
\def\erf{\mathop{\rm erf}}
\def\erfc{\mathop{\rm erfc}}

\title{REST-FRAME ULTRAVIOLET SPECTRA OF $z\sim 3$ LYMAN BREAK GALAXIES\altaffilmark{1}}
\author{\sc Alice E. Shapley and Charles C. Steidel\altaffilmark{2}} 
\affil{California Institute of Technology, MS 105--24, Pasadena, CA 91125}
\author{\sc Max Pettini}
\affil{Institute of Astronomy, Madingley Road, Cambridge UK}
\author{\sc Kurt L. Adelberger\altaffilmark{3}}
\affil{Harvard-Smithsonian Center for Astrophysics, 60 Garden St., Cambridge, MA 02139}

\altaffiltext{1}{Based, in part, on data obtained at the 
W.M. Keck Observatory, which 
is operated as a scientific partnership among the California Institute of Technology, the
University of California, and NASA, and was made possible by the generous financial
support of the W.M. Keck Foundation.
} 
\altaffiltext{2}{Packard Fellow}
\altaffiltext{3}{Harvard Society Junior Fellow}

\begin{abstract}
We present the results of a systematic study of the rest-frame
UV spectroscopic properties of Lyman Break Galaxies (LBGs).
The database of almost 1000 LBG spectra proves useful for constructing
high S/N composite spectra. The composite spectrum of the
entire sample reveals a wealth of features attributable to hot stars,
H~II regions, dust, and outflowing neutral and ionized gas. 
By grouping the database according
to galaxy parameters such as Ly$\alpha$ equivalent width,
UV spectral slope, and interstellar kinematics, we isolate
some of the major trends in LBG spectra which are
least compromised by selection effects. We find that LBGs with
stronger Ly$\alpha$ emission have bluer UV continua,
weaker low-ionization interstellar absorption lines,
smaller kinematic offsets between Ly$\alpha$ and the interstellar
absorption lines, and lower star-formation rates.
There is a decoupling between the  dependence of low- and
high-ionization outflow features on other spectral
properties.
Additionally, galaxies with rest-frame $W_{{\rm Ly}\alpha} \geq 20$ \AA\
in emission have
weaker than average high-ionization lines, and nebular
emission lines which are significantly stronger than in the
sample as a whole. Most of the above trends can be
explained in terms of the properties of the large-scale outflows
seen in LBGs. According to this scenario, 
the appearance of LBG spectra is determined
by a combination of the covering fraction of outflowing neutral
gas which contains dust, and the range of velocities over which
this gas is absorbing. 
In contrast, the strengths of collisionally excited nebular emission
lines should not be affected by the nature of the outflow,
and variations in these lines may indicate differences
in the temperatures and metallicities in H~II regions of
galaxies with very strong Ly$\alpha$ emission.
Higher sensitivity and
spectral resolution observations are still required
for a full understanding of the 
covering fraction and velocity dispersion
of the outflowing neutral gas in LBGs, and its relationship
to the escape fraction of Lyman continuum radiation in
galaxies at $z\sim 3$.
\end{abstract}

\section{Introduction}
\label{sec:intro}
Until now, the rest-frame UV spectra of Lyman Break Galaxies (LBGs)
have been used primarily to measure redshifts. At first, the
spectra were used to confirm the cosmologically distant nature of
galaxies photometrically selected with the Lyman Break
Technique to be at $z\sim 3$ 
\citep{steidel1996a,steidel1996b,lowenthal1997}.
Spectroscopic redshifts were also
necessary to study the large-scale spatial
distribution and clustering properties of LBGs 
\citep{adelberger1998,steidel1998,giavalisco1998}.
Furthermore, the determination
of the $z\sim3$ rest-frame UV luminosity function
required knowledge of the redshifts of LBGs,
combined with the optical apparent magnitudes
and colors \citep{steidel1999}. Most recently, the redshifts measured from
rest-frame UV spectra have been used to study the
cross-correlation of the large-scale distributions of
galaxies and the inter-galactic medium within the same
cosmic volume \citep{adelbergeret2002a}.
Given the faint nature of LBGs (most have ${\cal R}_{AB} = 24 - 25.5$), 
the desire to observe a large sample results in individual spectra
with low signal-to-noise (S/N) ratios and spectral resolution. In most cases,
the low S/N of LBG spectra
precludes any analysis more detailed than the determination of
redshifts.

One notable exception is the galaxy MS1512-cB58, an $L^*$ LBG at $z=2.73$
with an apparent magnitude of $V=20.6$ due to lensing by a foreground
cluster at $z=0.37$ \citep{yee1996}. Cluster lensing boosts the apparent
luminosity of cB58 by a factor of $\sim 30$, enabling relatively
high-resolution ($R\simeq 5000$) studies of its rest-frame UV spectrum
\citep{pettini2000,pettini2002}.
The velocity profiles of low and high-ionization
interstellar metal absorption features have been characterized in detail;
the weakest interstellar metal transitions have been used together with 
the damped Ly$\alpha$ absorption profile to determine the
abundance pattern in cB58 (an $\alpha$/Fe enhancement
indicative of a young stellar population, and an abundance of 
$\sim 0.4 Z_{\odot}$ for the $\alpha$ elements); C~IV and Si~IV P-Cygni 
stellar wind profiles have been used as independent probes of the stellar
population and metallicity; weak stellar
absorption features have been used to precisely measure the
systemic velocity of the stars, relative to which the
redshifts of Ly$\alpha$ emission and interstellar
absorption indicate offsets of several hundred km s$^{-1}$; 
finally, the strengths of the strongest interstellar absorption features
(which have zero transmission at line center) have been used
to infer a high covering fraction of outflowing neutral material, through
which negligible Lyman continuum emission can escape. 
For the vast majority of unlensed LBGs, it is unfortunately not possible to
obtain individual spectra which
contain the same high-quality information about physical conditions.
Since cB58 is only one object, we need to worry about how ``typical''
its continuum and spectroscopic properties are, relative to the
range seen in the entire sample of LBGs.

Even from the low S/N
spectra used to measure redshifts, we discern
a large variation in the types of spectra associated
with LBGs. Most obviously, there are large ranges of
Ly$\alpha$ profile shapes and UV continuum slopes.
There are also variations among the equivalent widths
of the few strong interstellar absorption lines which we 
detect most of the time in individual spectra, and of the redshift offset
measured between Ly$\alpha$ emission and interstellar absorption
(for spectra in which both types of features are detected).
While there is no hope of collecting data of comparable quality to the
cB58 spectra for individual unlensed LBGs, we have assembled a sample
of almost 1000 spectroscopically confirmed $z\sim 3$ galaxies over
the past six years. By dividing our spectroscopic
database into subsamples according to specific criteria, and creating
high S/N composite spectra of each subsample, we hope to
understand how the LBG spectroscopic properties depend in
a systematic way on other galaxy properties. 

Past uses of composite LBG spectra have proven very illustrative.
For example, a composite of 29 individual LBG spectra at
$\langle z \rangle = 3.4 \pm 0.09$ shows significant residual
flux shortward of the Lyman limit at 912 \AA\ \citep{steidel2001}.
If this composite spectrum is taken to be representative of
LBGs at $z\sim 3$, the LBG contribution to the ionizing background
could exceed that of QSOs at similar redshifts by as much as a factor
of 5. The appearance of this composite spectrum is very different
from that of cB58, with strong Ly$\alpha$ emission,
a continuum slope in the bluest quartile of the total LBG sample,
and interstellar absorption lines roughly half the strength of
those in the cB58 spectrum. Determining the relative numbers of LBGs
that resemble this composite versus those that more closely resemble
cB58 is important for determining the overall contribution of LBGs to
the ionizing background. Composite LBG spectra were also constructed
for galaxy subsamples grouped by the stellar population age inferred
from optical/near-IR photometry \citep{shapley2001}. The ``young''
($t\leq 35$~Myr) and ``old'' ($t\sim 1$~Gyr) composite spectra
exhibited systematic differences, including significantly stronger
Ly$\alpha$ emission in the ``old'' spectrum, and stronger 
interstellar absorption and stellar P-Cygni wind features in
the ``young'' spectrum. Such differences may indicate an
evolutionary sequence for
the appearance of the rest-frame UV spectra.

Based on the promise of these specific composite studies, we are motivated
to undertake a more general study of the spectroscopic properties
of the entire LBG sample. This systematic study is in some ways a
high-redshift analog to
the work of \citet{heckman1998}, which characterized the region
of UV spectroscopic parameter space inhabited by local
starburst galaxies. In particular, we would like to gain
more detailed information about the properties
of the large-scale outflows of interstellar material
which are inferred in LBGs from blueshifted interstellar
absorption and redshifted Ly$\alpha$ emission,
relative to the systemic stellar redshift \citep{pettini2001,pettini2002,franx1997,lowenthal1997}. 
The outflows represent not only an important feedback process which
affects galaxy formation and evolution, but also may have a profound
impact on the metal enrichment, ionization, and physical state 
of the surrounding intergalactic medium
\citep{adelbergeret2002a,adelbergeret2002b,adelberger2002a,steidel2001}.
So far, most of the information about the nature of outflows at high redshift
comes from high-quality observations of a single galaxy -- cB58 -- 
and from studies of the metal content and HI opacity of the 
Ly$\alpha$ forest near LBGs. 
The current survey provides information which is
complementary to both types of observations,
as it explores outflow properties from the galaxy
perspective, but is based on a much larger
(if lower spectral resolution) sample of galaxies.

The LBG spectroscopic sample is described in \S2, while 
\S3 presents the method of generating composite spectra and defining
the ``rest-frame'' in the presence of large-scale velocity fields.
\S4 gives an overview of the types of stellar
and interstellar spectroscopic
features which appear in high S/N composite LBG spectra, {\it some of
which have not previously been observed in UV spectra
of local star-forming regions}. In \S5, we present some
of the important trends observed in LBG spectra, with particular
attention to outflow-related properties. Finally, in \S6
we present a physical picture which is broadly consistent
with the observations, and highlight the need for several
future observations to test this picture.

\section{The LBG Spectroscopic Sample}
\label{sec:sample}
The individual galaxy spectra used to construct the composite spectra in this
paper were drawn from the Lyman Break Galaxy (LBG) survey of $z\sim 3$
galaxies. The details of our survey have been presented elsewhere 
\citep{steidel1996a,steidel1996b,steidel1999}
and will be extensively summarized in a future work
(Steidel \et 2003, in preparation), 
so here we present only a few relevant features.
The full LBG photometric sample consists of 2347 galaxies in 17 
separate fields with optical colors satisfying the following criteria:

\begin{equation}
{\cal R}\le 25.5,\quad G-{\cal R}\le 1.2,\quad U_n-G\ge G-{\cal R}+1
\label{eq:lbg_crit}
\end{equation}

plus an additional 180 galaxies with the same color criteria, but
${\cal R}$ magnitudes which are as faint as ${\cal R}=26$ (which
are located in the LBG survey field containing the bright quasar,
Q1422+239, for which we obtained significantly deeper photometry, 
allowing an extension of the LBG selection technique to fainter magnitudes). 

We have spectroscopically observed 1320 of these photometric 
candidates using the Low Resolution Imaging Spectrometer 
(LRIS) at the W.M.  Keck Observatory \citep{oke1995}. 
Most of the spectra were obtained using a 300 lines mm$^{-1}$ grating blazed
at 5000 \AA, and a multi-object slit mask with 1\secpoint4 slits,
providing a spectral resolution of $8-12$ \AA, depending on the seeing.
Recent spectroscopic data, including some of the observations in
the Q0933+288 and Q1422+2309 fields, and all of the data in the Q0302$-$003
field, were obtained with the new blue arm of LRIS (LRIS-B;
McCarthy \et 1998; Steidel \et 2003),
and dispersed by a
300 lines mm$^{-1}$ grism blazed at 5000 \AA. This setup provided
slightly higher spectral resolution than the 300-line grating setup,
and much higher throughput. 
Typical exposure times for both setups were 
$3 \times 1800 \: {\rm sec}$ with 1\secpoint0
dithers between exposures to provide for adequate sky-subtraction.
The two-dimensional, sky-subtracted, and coadded spectra were then
extracted to one-dimension, and wavelength calibrated using a HgNeAr
arc-lamp spectrum. Spectra were flux-calibrated with observations
of standard stars taken close to the time of the science observations.
Finally, air wavelengths were converted to vacuum wavelengths, in order
to measure redshifts in the vacuum frame. The average S/N
ratio of spectra in the sample is $\sim 4$ per resolution element.

From this sample of spectra, we identify 36 stars, 
2 absorption-line galaxies at $z\sim 0.5$, 2 galaxies with $z\simeq 1.98$,
957 objects at $z>2$, and 292 
objects for which we cannot measure a redshift. 
Of the 957 galaxies at $z>2$, 12 are identified as
broad-lined AGN due to the presence of emission lines with 
FWHM$>2000$ \kms, while 16 are identified as narrow-lined AGN
with strong Ly$\alpha$ emission accompanied by significant
C~IV $\lambda 1549$ emission, but FWHM$<2000$ \kms \citep{steidel2002}. 
We exclude from the spectroscopic sample
the 28 objects identified as AGN on an individual
basis (though we may have included galaxies
with low-level AGN activity which is 
undetected in individual spectra but may become evident in the composites. 
We will treat this point further in 
section~\ref{sec:features_neb_agn}). Furthermore,
we include only galaxies 
whose redshifts were independently and securely
confirmed by at least two members of our group.
This last criterion limits our composite sample to 811 spectra.

\section{Generating the Composite Spectra}
\label{sec:compspec}
There are several steps required to
generate the composite galaxy spectra presented in this work. 
The first step consists
of carefully defining a sample of galaxy spectra to be combined. 
Each extracted, one-dimensional, flux-calibrated spectrum in the sample is
then shifted into the rest frame. The spectra are then
averaged, after being scaled to a common 
mode in the wavelength range 1250-1500 \AA\,
and rebinned to a common dispersion of 1 \AA\ per pixel.
In order to exclude both positive and negative sky subtraction
residuals and cosmic ray events, an equal number of 
positive and negative outliers are rejected at each dispersion pixel, 
totaling less than 10\% of the data.

\subsection{Measuring Redshifts}
\label{sec:z}
One of the non-trivial aspects of generating the composite spectra
consists of defining a systemic rest frame for each galaxy. 
The low S/N ratio
of typical LBG spectra precludes much
more than measuring redshifts from the
very strongest rest-frame UV features. 
These features include H~I Ly$\alpha$, seen either
in emission, absorption, or a combination of both;
low-ionization resonance interstellar metal lines such as 
Si~II $\lambda 1260$, O~I $\lambda 1302$ + Si~II $\lambda 1304$, 
C~II $\lambda 1334$, Si~II $\lambda 1526$, Fe~II $\lambda 1608$, 
and Al~II $\lambda 1670$,
which are associated with the neutral interstellar medium; 
and high-ionization metal lines such as Si~IV $\lambda\lambda 1393,1402$ 
and C~IV $\lambda\lambda 1548,1550$ associated with ionized interstellar 
gas and P-Cygni stellar wind features.
In 28\% of the rest-frame UV LBG spectra in our spectroscopic sample, 
the only spectral feature visible
is Ly$\alpha$ emission. In 32\% of the spectra, 
Ly$\alpha$ appears only as broad absorption, and
multiple low- and/or high- ionization interstellar absorption 
lines are used to measure the redshift 
(broad Ly$\alpha$ absorption is not very precise as a redshift
indicator). Finally, in the remaining 40\% of the spectra, 
both Ly$\alpha$ emission and interstellar absorption 
lines are visible and both can be used to measure redshifts.
In 95\% of the cases where both Ly$\alpha$ emission and
interstellar absorption lines have been used to measure the redshift
of the galaxy, the Ly$\alpha$ emission redshift is higher than the 
interstellar absorption redshift. Figure~\ref{fig:delvhist}
shows the distribution of $\Delta v_{{\rm em-abs}}$, which has
$\langle \Delta v_{{\rm em-abs}}\rangle \sim 650$~\kms.
This velocity difference indicates that at least one 
of the two sets of features is not at rest 
with respect to the stars in the galaxy.
Such kinematics suggest that LBGs are experiencing large-scale outflows
caused by the mechanical
energy input from supernova explosions which are the result of vigorous
massive star-formation rates. It has been shown that, 
in the local universe, any galaxy with a star-formation rate 
surface-density $\Sigma_{*} \geq 0.1 M_{\odot} \: {\rm yr}^{-1} {\rm kpc}^{-2}$
is capable of driving a superwind \citep{heckman2002}. Given their typical
star-formation rates and physical sizes, LBGs easily satisfy and
exceed the criteria for driving a superwind
\citep{shapley2001,giavalisco1996b}.

\begin{inlinefigure}
\centerline{\epsfxsize=9cm\epsffile{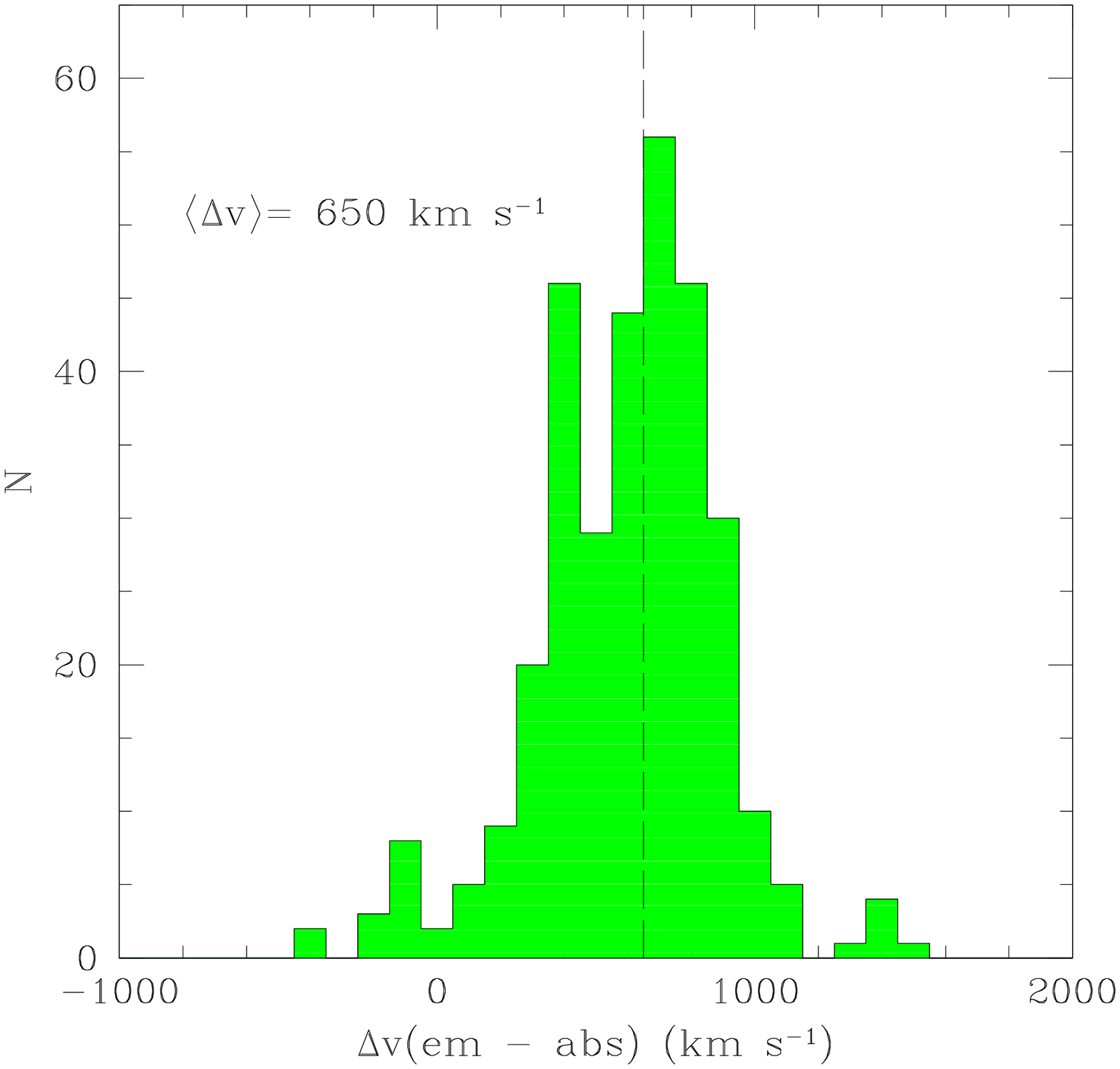}}
\figcaption{The distribution of velocity offsets between Ly$\alpha$ emission
and low-ionization interstellar absorption.
The most straightforward indication that
LBGs are experiencing large-scale outflows of their interstellar material
is the velocity offset measured in individual spectra
between Ly$\alpha$ emission and interstellar absorption lines.
This histogram shows the distribution of velocity offsets for the
323 galaxies with spectra in which both types of features are detected.
The mean velocity offset (redshift difference) is $\Delta v=650$~\kms
($\Delta z=0.008$).
\label{fig:delvhist}
}
\end{inlinefigure}

Since the only features which can be detected in individual LBG spectra 
seem to trace outflow kinematics, the outflows complicate
our effort to assign systemic redshifts. 
In general, stellar systemic redshifts cannot be
measured for individual galaxies because 
UV photospheric features from hot stars are much
too weak to see in typical LBG spectra.
In order to estimate stellar systemic redshifts for individual galaxies,
we applied the formulae presented in \citet{adelbergeret2002a},
which predict a value of the systemic redshift for 
three separate cases: when there is only a Ly$\alpha$ emission
redshift; when there is only an interstellar absorption redshift;
and when there are both Ly$\alpha$ emission and interstellar
absorption redshifts.

While no reference was made to stellar photospheric features
in the estimate of the systemic redshifts of individual galaxies, 
the rest-frame composite spectrum presented in the next section
indicates a mean systemic velocity of $\Delta v = -10 \pm 35$~\kms\ for
the three strongest stellar features, which we have identified as
C~III $\lambda 1176$, O~IV $\lambda 1343$, and S~V $\lambda 1501$.
\footnote{The precise wavelengths are C~III $\lambda 1175.71$,
O~IV $\lambda 1343.35$, which is a blend of lines at 
$\lambda=1342.99$ and $\lambda=1343.51$, and S~V $\lambda 1501.76$}
It is worth noting possible contributions 
to the O~IV $\lambda 1343$ absorption from Si~III $\lambda 1341$, 
and to the S~V $\lambda 1501$ absorption from Si~III $\lambda 1501$,
if there is a significant B-star component in the composite spectrum. 
However, at least the Si~III $\lambda 1501$ contribution will not change 
the inferred negligible systemic velocity of the feature which we have
identified as S~V $\lambda 1501$.
The insignificant velocities of the stellar features demonstrate the
success of the systemic redshift estimates for the LBGs included in
composite spectra, at least on average.
Also, since the stellar features appear at roughly zero velocity, 
the redshifts and blueshifts of other sets
of spectral features measured relative to the rest-frame of
the composite spectrum should
offer a true representation of the average kinematic
properties of the large-scale galactic outflows in LBGs.
The establishment of the velocity zeropoint from the stellar lines
in composite spectra
represents a significant improvement over the kinematic information
contained in individual rest-frame UV spectra, where
only the strongest interstellar outflow-related features are detected.

\section{LBG Rest-frame UV Spectroscopic Features}
\label{sec:features}
\begin{figure*}
\centerline{\epsfxsize=18cm\epsffile{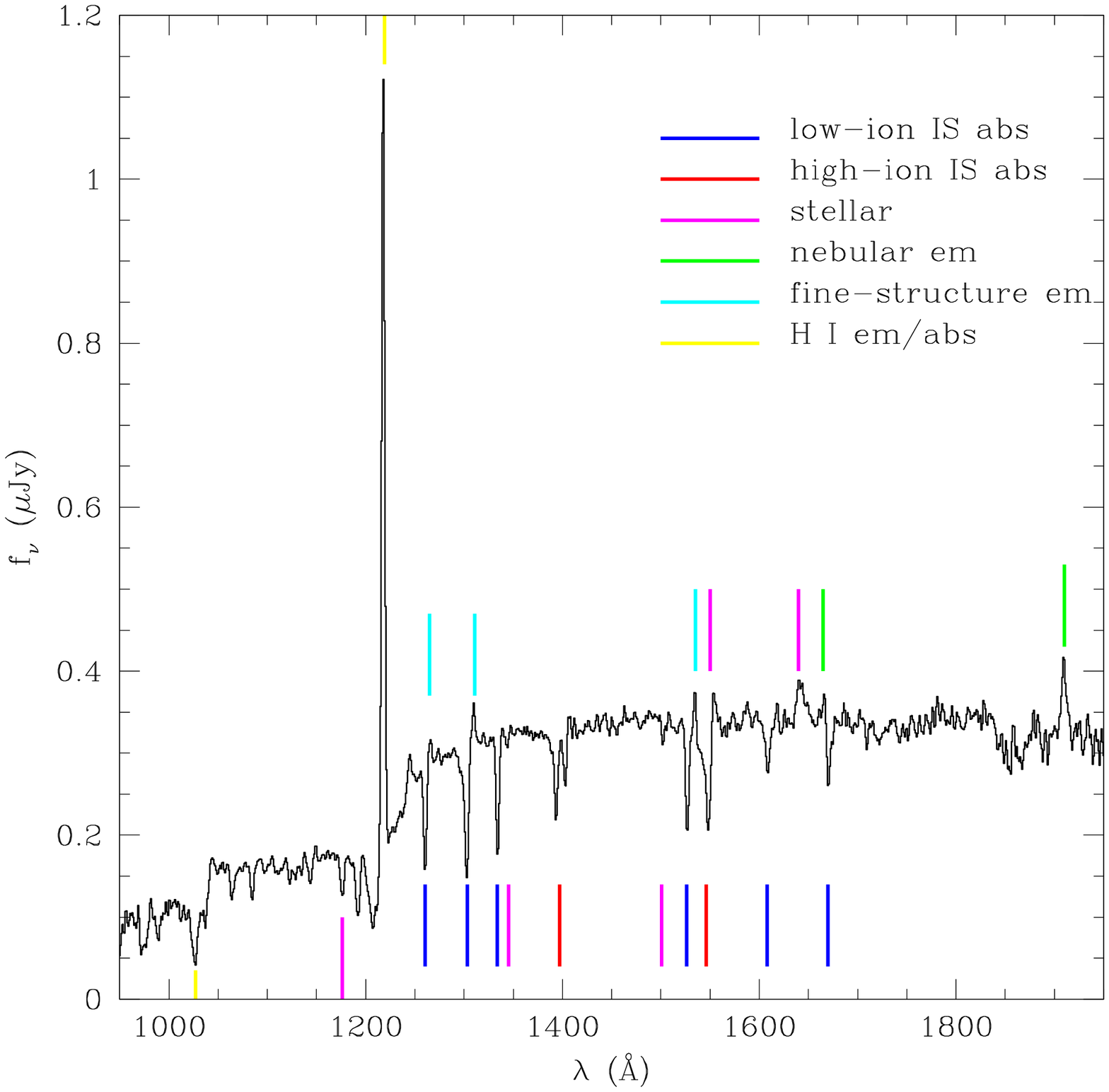}}
\figcaption{A composite rest-frame UV spectrum constructed from 811 individual
LBG spectra. Dominated by the emission from
massive O and B stars, the overall shape of the UV continuum
is modified shortward of Ly$\alpha$
by a decrement due to inter-galactic HI absorption.
Several different sets of UV features
are marked: stellar photospheric and wind,
interstellar low- and high-ionization absorption,
nebular emission from H~II regions, Si~II* fine-structure emission
whose origin is ambiguous, and
emission and absorption due to interstellar HI (Ly$\alpha$ and Ly$\beta$).
There are numerous weak features which are not marked, as well
as several features bluewards of Ly$\alpha$ which only become visible
by averaging over many sightlines through the IGM. The composite LBG spectrum
is available in electronic form from
http://www.astro.caltech.edu/$\sim$aes/lbgspec/.
\label{fig:plotall}
}
\end{figure*}

Figure~\ref{fig:plotall} shows a composite spectrum which is the average 
of our entire spectroscopic sample of 811 LBGs, combined in the
manner described in section~\ref{sec:compspec}.
Rest-frame UV spectra of LBGs are dominated by the emission 
from O and B stars with masses higher than $10 M_{\odot}$ and
$T \geq 25000 K$. The overall shape of the UV spectrum
is modified by dust extinction internal to the galaxy, 
and, at rest wavelengths shorter than $1216$ \AA, 
by inter-galactic HI absorption along the 
line of sight. Composite spectra contain the average of many different
lines of sight through the IGM. Therefore, spectral features 
which are intrinsic to the galaxy at wavelengths shorter than Ly$\alpha$,
and which can be completely wiped out by individual Ly$\alpha$
forest systems along a specific line of sight, become visible in the
composite spectra. While we regain spectroscopic information by averaging
over many different sightlines, we still, however, see the average decrement
of the Ly$\alpha$ forest, $D_{A}$. 
In the following section, we describe the spectroscopic features contained
in the composite spectrum of Figure~\ref{fig:plotall}, which
trace the photospheres and winds of massive stars, 
neutral and ionized gas associated with large-scale outflows, 
and ionized gas in H~II regions where star formation is taking place.

\subsection{Stellar Features}
\label{sec:features_stellar}
The C~III 1176, O~IV 1343, and S~V 1501 stellar photospheric lines
discussed in section~\ref{sec:z}
are marked in Figure~\ref{fig:plotall}.
Also of note (though not marked) is the large number of weak absorption
features between 1400 and 1500 \AA. These include blends of
Fe~V, Si~II, Si~III, and C~III photospheric absorption lines from O and B
stars \citep{bruhweiler1981,demello2000}.
In addition to photospheric absorption features, the spectra
of the most massive hot stars indicate the presence of stellar
winds of 2000-3000 \kms\ due to radiation pressure \citep{groenewegen1989}. 
These wind features appear as broad blue-shifted absorption for weaker winds, 
or as a P-Cygni type profile if the wind density
is high enough \citep{leitherer1995}. 
The most prominent stellar wind features are 
N~V~$\lambda \lambda 1238,1242$, 
Si~IV~$\lambda \lambda 1393,1402$, C~IV~$\lambda \lambda 1548, 1550$, and
He~II~$\lambda 1640$. 
The shape of the N~V wind profile, especially the absorption component,
is affected by its close proximity to the Ly$\alpha$ region of the 
spectrum, and is therefore difficult to characterize in detail, though
we do see both emission and absorption qualitatively consistent with
a P-Cygni type profile.
While clear of the large-scale continuum effects of Ly$\alpha$, 
the Si~IV and C~IV transitions contain the combination of stellar 
wind and photospheric absorption, plus a strong interstellar 
absorption component, which are difficult to disentangle. 
The stellar wind feature only becomes
apparent in Si~IV for blue giant and supergiant stars, while,
in contrast, the C~IV wind feature is strong in main sequence,
giant, and supergiant O stars \citep{walborn1984}. 
Consequently, the interstellar, rather than the wind component,
seems to dominate the Si~IV doublet in the LBG composite spectrum,
while the C~IV feature exhibits both the 
blue-shifted broad absorption and redshifted
emission associated with stellar winds, in addition to a strong, narrower,
interstellar absorption component.
The redshifted emission indicates the presence of
stars with $M\geq 30M_{\odot}$ \citep{leitherer1995,pettini2000}.
In Figure~\ref{fig:plotallzoomnorm}, a 
Starburst99 model spectrum \citep{leitherer1999}
is plotted over the zoomed-in C~IV $\lambda\lambda 1548,1550$ region of
the LBG composite spectrum for comparison.
The model spectrum is for a 300 Myr old episode of continuous
star formation (the median stellar population age
inferred from the optical/near-IR colors of LBGs, Shapley \et 2001),
and is constructed from a library of HST FOS and STIS observations
of massive hot stars in the Magellanic Clouds with a mean metallicity 
$Z=0.25 Z_{\odot}$ (consistent with the limited information
on LBG metallicities). The model and data agree quite well
in the emission component of the P-Cygni profile. However, the model
overpredicts the strength of the broad wind absorption. This discrepancy 
may be due to a combination of age and metallicity effects.

The composite LBG spectrum also shows He~II $\lambda 1640$
emission which is quite strong compared with observations
in local starburst galaxies \citep{heckman1998}. 
Narrow He~II $\lambda 1640$ emission
is seen in galaxy spectra with strong nebular emission. If massive
stars are forming in He~III regions, He~II $\lambda 1640$ can appear
as a nebular recombination emission line (from supernovae remnants
or superbubbles) \citep{leitherer1995}. The He~II $\lambda 1640$
emission in the composite LBG spectrum, though, is quite
broad, with ${\rm FWHM} \sim 1500$ \kms. This is
visibly broader than the most of the other weak emission lines 
in the spectrum originating in H~II nebular regions, so
we favor a stellar wind origin for
the He~II feature. Broad stellar He~II
emission is predominantly produced in fast, dense
winds from Wolf-Rayet (W-R) stars, which 
are the evolved descendants
of O stars more massive than $M>20-30M_{\odot}$. 
Therefore, the strength of the
He~II emission should provide information about the ratio of
W-R to O stars \citep{schaerer1998}.

In order to interpret the He~II line, we use
Starburst99 population synthesis codes \citep{leitherer1999}.
Model UV spectra of both solar and  $0.25\: \times$~solar 
metallicity are produced by Starburst99. The solar metallicity
spectrum extends from $1200 - 1800$~\AA, which includes
the He~II line, but the sub-solar metallicity model only
extends from $1200-1600$~\AA.
While the current best estimates of LBG metallicities 
\citep{pettini2002,pettini2001} are closer to
$Z=0.25 Z_{\odot}$, we can still use the solar metallicity
model spectra to make some interesting inferences.
The observed strength of the LBG He~II emission line can be matched
using a 3~Myr old instantaneous burst model with a standard Salpeter IMF.
This is the brief period, following a burst of star
formation, when the fractional contribution of
Wolf-Rayet stars to the integrated UV luminosity
reaches a maximum.
Such young burst ages have indeed been derived for Wolf-Rayet galaxies,
whose strong optical and UV He~II emission lines indicate the
presence of numerous high-mass stars and a high W-R/O star ratio
\citep{conti1996,leitherer1996}. It is not clear which
average stellar population is represented by
the composite spectrum in Figure~\ref{fig:plotall}, 
but since LBG rest-frame UV/optical SEDs indicate a wide range
of properties and ages, and are generally
not well-represented by instantaneous burst models
\citep{shapley2001,papovich2001},
it seems extremely unlikely that the 3~Myr old
burst model can be representative of the whole
population and accordingly we rule out this
interpretation.

If we then consider 300 Myr continuous star-formation models
(based on the median age derived from SED-fitting), 
the only way to produce a high enough W-R/O star ratio is to invoke
an IMF slope much flatter than the standard Salpeter form
of $N(m)\propto m^{-\alpha}$ with $\alpha=2.35$. Slopes of 
$\alpha \leq 1$ match the He~II strength, but then the C~IV
P-Cygni emission is overproduced by a factor of $\geq 4$.
The discrepancy will be even worse for the sub-solar metallicity
models. The ratio of W-R/O stars decreases as a strong
function of decreasing metallicity,
since the lower-limit on the masses of O stars which evolve into
W-R stars moves towards higher masses as the metallicity
increases \citep{maeder1991,meynet1995}. 
Theoretical predictions of W-R line luminosities
\citep{schaerer1998,leitherer1999} show that,
for a given IMF, 300 Myr
continuous star-formation models with
$Z=0.25 Z_{\odot}$ have He~II $\lambda 1640$ line
strengths only half as strong as in the case of
solar-metallicity. A more drastic adjustment to the
slope of the IMF would be required for the sub-solar metallicity
model to match the He~II emission
strength, which would then lead to an even larger
discrepancy for the C~IV P-Cygni emission (since, empirically, 
we measure that the C~IV P-Cygni emission does not depend
very significantly on metallicity). Clearly, current
population synthesis models do not 
simultaneously reproduce the C~IV and He~II stellar
wind features in LBGs for reasonable choices of star-formation history,
age, metallicity, and IMF slope. 
We will be able to test this discrepancy better
with high-quality spectra for individual
LBGs whose stellar population ages and star-formation histories
we know more accurately than our rough estimate of
the stellar population represented
by the composite of all the LBGs.

\subsection{Outflow-related Features}
\label{sec:features_outflow}
The large-scale outflow of interstellar material
appears to be a generic feature of LBGs, one implied
both by the typical LBG star-formation rate per unit area, and also
by the fairly ubiquitous observed offset in velocity between 
Ly$\alpha$ emission and interstellar absorption lines when both 
sets of lines are seen. Spectral features probing neutral outflowing
gas are H~I Ly$\alpha$ and Ly$\beta$, and neutral and 
singly ionized metal absorption lines. More highly ionized
metal lines probe the ionized phase of the 
outflow \citep{heckman2001b,pettini2002}.
While the interstellar absorption lines
in $\sim 70 \%$ of individual LBG spectra are 
strong enough that an absorption redshift can be assigned to 
at least one low- or high-ionization feature,
it is not possible to obtain robust equivalent width
measurements for multiple features due to the typical
low S/N (and prevalent sky-subtraction residuals). 
However, in composite spectra containing hundreds of LBGs, these
features are detected with high significance.
In this section, we describe the average properties of the 
spectral features related to the large-scale outflows in LBGs.

\vskip 0.1in
\begin{inlinefigure}
\centerline{\epsfxsize=9cm\epsffile{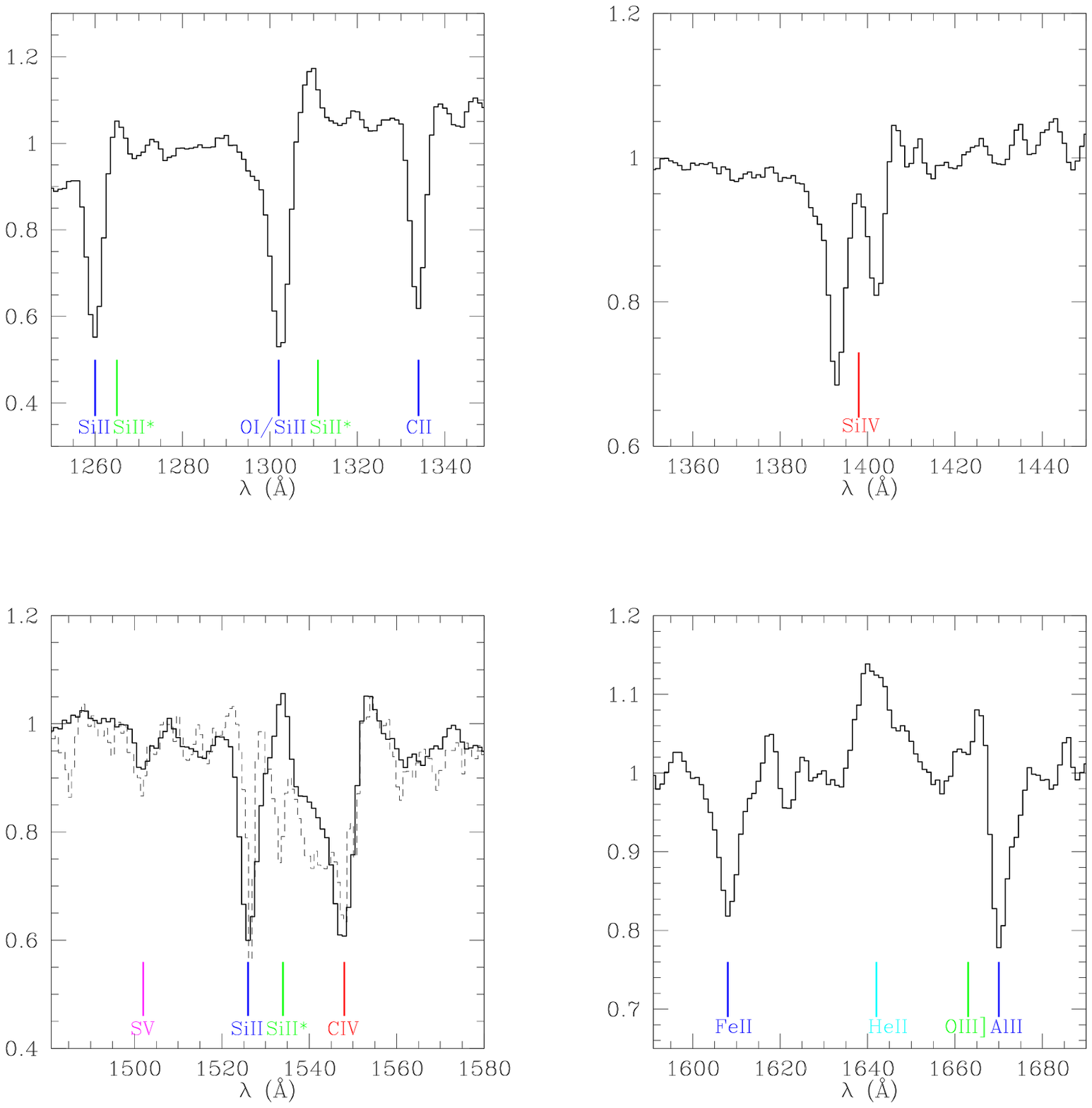}}
\figcaption{Four zoomed-in regions from the composite spectrum
of Figure 2. The zoomed-in vertical scale allows a more detailed
look at weak nebular emission features such as Si~II*
$\lambda\lambda\lambda 1265, 1309, 1533$, and
O~III] $\lambda\lambda 1661, 1666$,
which are produced in H~II regions. The lower left-hand box shows the zoomed-in
region near C~IV~$\lambda\lambda 1548,1550$. The observed composite spectrum
is plotted as a solid line, while the dashed line is
a Starburst99 model spectrum for 300 Myr of continuous star-formation with
$Z=\frac{1}{4} Z_{\odot}$ \citep{leitherer1999,leitherer2001}.
The model and data agree well
for the emission component of the C~IV P-Cygni stellar wind line,
but the model overpredicts the amount of broad, blue-shifted
absorption. This may be due to the lower average metallicity or
older age of the LBGs included in the composite spectrum, relative
to the model. The lower right-hand box shows the zoomed-in region near
the He~II $\lambda 1640$ stellar wind line,
whose large strength we have trouble reproducing
using Starburst99 models with reasonable parameters.
\label{fig:plotallzoomnorm}
}
\end{inlinefigure}

\subsubsection{Low-ionization Lines Associated with Neutral Gas}
\label{sec:features_outflow_lis}
The strongest low-ionization interstellar lines probing outflowing
interstellar material are marked in Figure~\ref{fig:plotall}, and include
Si~II $\lambda 1260$, O~I+Si~II $\lambda 1303$ (a blend at the spectral
resolution of our sample), C~II $\lambda 1334$,
Si~II $\lambda 1526$, Fe~II $\lambda 1608$, and Al~II $\lambda 1670$.
While there are additional interstellar
absorption lines at wavelengths shorter than Ly$\alpha$, as
well as a host of other, weaker, low-ionization interstellar absorption lines,
we confine the analysis in section~\ref{sec:trends} to these six strongest
features redward of Ly$\alpha$, which we detect at very high significance.
The features marked have been well-studied in local starburst
galaxies \citep{gonzalez1998,heckman1998},
as well as in the lensed LBG, MS1512-cB58 
\citep{pettini2000,pettini2002}. Relative to the stellar, systemic
redshift, we measure an average blueshift for the strong
low-ionization interstellar features 
of $\Delta v =-150 \pm 60$\kms.
Using the C~II $\lambda 1334$ and Si~II $\lambda 1526$
features, which we assume to be the least affected by blends,
and assuming an effective spectral resolution of 2.6 \AA\ for the
LBG composite spectrum, we compute the average deconvolved
velocity full-width \footnote{By deconvolved velocity full-width,
${\rm FWHM}_{int}$, we mean the
square-root of the difference in quadrature between the observed
full-width, ${\rm FWHM}_{obs}$ and the instrumental resolution, 
${\rm FWHM}_{inst}$:
${\rm FWHM}_{int}=({\rm FWHM}_{obs}^2-{\rm FWHM}_{inst}^2)^{1/2}$}
for the low-ionization interstellar (LIS) lines,
${\rm FWHM(LIS)} = 560 \pm 150$ \kms. The largest uncertainty
affecting this measurement is the uncertainty of the effective spectral
resolution of the composite spectrum, which we conservatively
estimate to range between $2-3.25$ \AA\ (roughly $400-700$\kms). 
The upper bound on the
spectral resolution is set by the minimum FWHM value measured
for any of the strong interstellar absorption lines. The lower bound
of 2 \AA\ is set by our estimate of the spectral resolution provided by 
the optimum observing conditions.\footnote{ The spectra were obtained
through 1\secpoint4 slits, which is 
much larger than the LBG size in typical seeing conditions 
(${\rm 0\secpoint5}-{\rm 1\secpoint0}$). Under such conditions,
the spectral resolution was dictated by the angular sizes of
objects falling within slits (i.e. the seeing), rather than by the slit-width.}

We list rest-frame equivalent widths and relative systemic redshifts 
for the six strongest low-ionization interstellar absorption lines in 
Table~\ref{tab:is}.  The strength of these features
makes them ideal for measuring interstellar absorption 
redshifts in noisy individual spectra. However, 
they are not useful for measuring chemical abundances,
due to the fact that all of the strong lines are saturated. 
The saturation of the strong lines is most easily
demonstrated by comparing the equivalent widths for two different
Si~II transitions: Si~II $\lambda 1260$ and $\lambda 1526$. 
On the linear part of the curve of growth, $W\propto Nf\lambda^{2}$,
where $N$ is the column density of the ionic species and $\lambda$ is
the rest-frame wavelength of the transition. According to the
relative oscillator strengths and wavelengths of the two Si~II transitions, 
the ratio $W_0(1260)/W_0(1526) > 5 $ on the linear part of
the curve of growth. We measure $W_0(1260)/W_0(1526)=0.95$, 
consistent with a ratio of unity, given the uncertainties,
thus demonstrating that the Si~II transitions are optically thick.
There are weaker features detected in the composite LBG spectrum 
which probe the linear part of the curve of growth, 
and which have been used to derive metal abundances in the outflow 
of MS1512-cB58 \citep{pettini2000,pettini2002}. These include  
S~II $\lambda\lambda\lambda 1250,1253,1259$, Si~II $\lambda 1808$, 
Fe~II $\lambda 1144$, Ni~II $\lambda 1317, \lambda 1370,
\lambda 1703, \lambda 1709, \lambda 1741, \; {\rm and} \; \lambda 1751$.
Most of these features are detected with only marginal significance in the
composite LBG spectrum due to low spectral resolution.
An abundance determination also requires an estimate of the H~I column
density, which is not easily measured from the composite spectrum, due
to the way in which it was combined.
In this paper, the analysis of the low ions in the
outflowing gas is confined to the properties of the strong transitions.

\subsubsection{High-ionization Lines Associated with Ionized Gas}
\label{sec:features_outflow_his}
In addition to the low-ionization features associated with neutral
outflowing gas, we detect high-ionization interstellar features
such as Si~IV $\lambda \lambda 1393, 1402$, C~IV 
$\lambda\lambda 1548, 1550$, and N~V $\lambda \lambda 1238,1242$. 
These features predominantly trace gas at $T\simgt 10^4 K$, which has
been ionized by a combination of radiation from massive stars
and collisional processes associated with the outflow. 
We also detect O~VI $\lambda\lambda 1032, 1038$ in absorption.
If the radiation field is dominated by the
spectrum of hot stars, rather than an AGN, 
O~VI is likely to arise in collisionally ionized gas,
indicating the presence of an even hotter
phase with $T \simgt 10^5 K$ \citep{heckman2001b}.

Section~\ref{sec:features_stellar} included a discussion of
the stellar winds indicated by the Si~IV, C~IV, and N~V profiles.
In this section we consider the properties of the
interstellar contributions to the same high-ionization transitions.
Again, the proximity of N~V to Ly$\alpha$ prevents us
from studying this transition in detail, though we do see evidence 
for an interstellar absorption component.
Also, it is difficult to characterize the properties of the hot phase
traced by O~VI, due to the fact that the O~VI absorption
is fairly weak, blended with C~II $\lambda 1036$ absorption, and resides
in the red wing of the strong Ly$\beta$ profile.
However, we measure $\Delta v=-180$~\kms\ for both members of the Si~IV transition, 
which has a doublet ratio of roughly 2:1, indicating that the lines
are on the linear part of the curve of growth.
We measure $\Delta v= -390$~\kms, for 
what we isolate as the interstellar component of the C~IV absorption,
assuming a saturated doublet ratio of 1:1 and therefore a
rest-frame centroid of $\lambda=1549.479$~\AA.
The measurement of the C~IV interstellar velocity is fairly uncertain, 
due to the combination of P-Cygni emission and broad absorption,
possible nebular emission, and interstellar absorption all superposed
on one another.  The properties of the Si~IV and C~IV interstellar
absorption lines are listed in Table~\ref{tab:is}.

In light of the complexities associated with the C~IV doublet,
we emphasize the comparison between the Si~IV doublet and
the low-ionization lines.  The blueshift
of the Si~IV doublet ($\Delta v=-180$~\kms) agrees quite well with the average
blueshift of the strong low-ionization interstellar lines
($\Delta v=-150$~\kms). Furthermore, 
the average deconvolved velocity full-width for the two members 
of the Si~IV doublet is
${\rm FWHM(Si~IV)}=590 \pm 140$~\kms, very similar to the average
FWHM for the low-ionization lines.
The deconvolved velocities associated with either the Si~IV 
or low-ionization lines are very uncertain due to the uncertainties
in spectral resolution. Independent of 
the resolution of the composite spectrum, however,
both the blue-shifts and the un-deconvolved velocity widths of the low and high
ionization lines are consistent with each other. 

The properties of low and high ionization absorption 
profiles were compared in the spectrum of the gravitationally 
lensed LBG, MS1512-cB58, using $\sim 10$ times higher
spectral resolution \citep{pettini2002}. In the case of cB58, which has much
stronger than average low-ionization interstellar absorption
lines, saturated Si~IV and C~IV transitions, and 
Ly$\alpha$ dominated by a damped absorption
profile, absorptions from low-ions and high-ions span the same
overall velocity range. Also, the material with the highest optical
depth (i.e. where the lines are black) is blueshifted
by roughly the same amount for the low ions and the high ions, to
within 20 \kms. One distinction highlighted by \citet{pettini2002}
is that the high-ionization lines show smoother absorption profiles,
while the low-ionization profiles break up into a number of discrete
components. Composite LBG spectra do not have sufficient
spectral resolution to discern qualities such as profile smoothness
or clumpiness. However, the overall agreement between low and high
ion stages in mean blueshift and velocity
FWHM is consistent with the high-resolution results from cB58.
In contrast, \citet{wolfe2000a} find distinct kinematic
properties for low and high-ionization transitions associated
with damped Ly$\alpha$ absorbers. 
Specifically, the mean velocities of low
and intermediate (Al~III) ionization stages
differ from those of the high ions, although
within each of these two sets of lines there
is normally very good internal velocity agreement.
Also, in 29 out of 32 cases, the velocity width of the 
high-ionization absorption exceeds
that of the low-ionization absorption.
The kinematic differences between LBGs and DLAs
are another manifestation of the differing characteristics
of these two populations of high redshift galaxies,
which also exhibit distinct clustering properties
\citep{adelbergeret2002a} and
metallicities \citep{mpettini2002}.
Such differences will eventually help us
clarify the true nature of damped Ly$\alpha$
absorption systems.

\subsubsection{Ly$\alpha$}
\label{sec:features_outflow_lya}
By far the most prominent feature in individual LBG spectra
is H~I Ly$\alpha$. The original sources of most Ly$\alpha$ photons
are recombinations in H~II regions. While the Ly$\alpha$ equivalent
width is sensitive to conditions in the H~II regions such as 
temperature, metallicity, star-formation rate, and star-formation history,
resonant scattering of Ly$\alpha$ in LBGs by interstellar H~I
makes the emergent Ly$\alpha$ profile at least as sensitive 
to the geometry, kinematics, and dust content of the 
large-scale outflows. When seen in emission, 
Ly$\alpha$ can be used to measure
a redshift.  In the composite LBG spectrum shown in Figure~\ref{fig:plotall}, 
we measure a Ly$\alpha$ emission redshift of $\Delta v = +360$~\kms.
This relative redshift reflects the fact that a Ly$\alpha$ photon
has a much better chance of escaping a galaxy if its last scattering occurs
off of an atom which is redshifted with respect to the bulk of the
neutral material in the galaxy, imparting a Doppler shift 
which takes the Ly$\alpha$ photon off resonance. We measure 
a deconvolved emission full-width of 
${\rm FWHM(Ly}\alpha)=450 \pm 150$~\kms.
When seen in absorption, the Ly$\alpha$ feature can be quite broad, 
with blueshifted absorption extending from zero velocity
down to $\Delta v \leq -5000$~\kms. Broad Ly$\alpha$ absorption
is therefore not a precise redshift indicator.
In spectra with Ly$\alpha$ seen only in absorption, interstellar
metal lines can be used to measure the redshift more precisely.

A wide distribution of Ly$\alpha$ profiles
is seen in the LBG spectroscopic sample, ranging from
damped absorption to emission an order of 
magnitude stronger than the feature shown in 
Figure~\ref{fig:plotall} \citep{steidel2000}.
In section~\ref{sec:trends_lya}, we will discuss how several LBG
spectroscopic properties depend on Ly$\alpha$ equivalent width,
and what inferences can be drawn about the physical conditions which determine
the emergent Ly$\alpha$ profile.
The composite spectrum shown in Figure~\ref{fig:plotall}  has
a Ly$\alpha$ feature dominated by emission. The total rest-frame
equivalent width is $W_0=14.3$~\AA\, which includes both redshifted
emission, and much weaker blueshifted absorption.
This spectrum is the average of all the LBG
spectra in the spectroscopic sample, yet there are selection
effects which depend on both apparent ${\cal R}$ magnitude,
color, and spectroscopic type,
which bias the spectrum relative to a true ``average''
of the total LBG photometric sample. For example, the spectroscopic sample
over-represents bright objects relative to faint objects, 
and the number of objects with Ly$\alpha$
emission relative to those with only Ly$\alpha$ absorption. As discussed
in section~\ref{sec:trends_seleffects}, it becomes much more 
difficult at fainter magnitudes to identify spectroscopically a galaxy with no 
Ly$\alpha$ emission. Therefore, as the number of spectroscopically
unidentified objects increases at fainter magnitudes,
so does the ratio of emission to
absorption line galaxies with measured redshifts. 
Accordingly, the Ly$\alpha$ feature in Figure~\ref{fig:plotall}
may be biased towards stronger emission than the true average for
the total photometric sample. In order to quantify this bias,
a more detailed treatment of selection effects is required.

\subsection{Emission Lines}
\label{sec:features_neb}
One of the benefits of producing a high S/N ratio composite
spectrum is that it can reveal weak spectral features which would
have remained undetected in individual spectra.
We detect several weak emission
lines, some of which we attribute to nebular regions photoionized by
radiation from massive stars, and others whose origin is still
ambiguous. The weak emission lines in 
the LBG composite spectra are: Si~II*~$\lambda 1265$, Si~II*~$\lambda 1309$,
Si~II*~$\lambda 1533$, O~III]~$\lambda \lambda 1661, 1666$, 
and C~III]~$\lambda \lambda 1907, 1909$. We measure a mean velocity of
$\Delta v = 100\pm 35$ \kms\ for the Si~II* transitions. 
The centroids of the fine-structure emission lines may be
biased to the red by weak fine-structure absorption lines
or neighboring saturated resonance absorption features
(Si~II $\lambda 1260$, O~I+Si~II $\lambda 1303$, and Si~II $\lambda 1526$)
associated with the outflow, which attenuate the blue 
edges of the fine-structure emission profiles. 
We measure a velocity of $\Delta v = 0$~\kms\ for the 
O~III]~$\lambda 1663$ doublet, and
a velocity of $\Delta v = 40$~\kms\ for the C~III] $\lambda 1909$ transition, 
both of which agree very well with the stellar systemic velocities
(section~\ref{sec:z}). 
O~III] $\lambda 1663$ and C~III] $\lambda 1909$ are both collisionally excited,
semi-forbidden transitions, so there is no absorption from these ions
in the large-scale outflow of gas. 
While the Al~II $\lambda 1670$ resonance
absorption feature is fairly close to O~III], the C~III] transition
should be clear of any absorption line. 
There may be a nebular emission component in 
C~IV $\lambda\lambda 1548, 1550$, but it is difficult to isolate 
nebular C~IV emission from the stellar P-Cygni emission. 
The properties of the emission lines are summarized in 
Table~\ref{tab:nebem}.

\begin{inlinefigure}
\centerline{\epsfxsize=9cm\epsffile{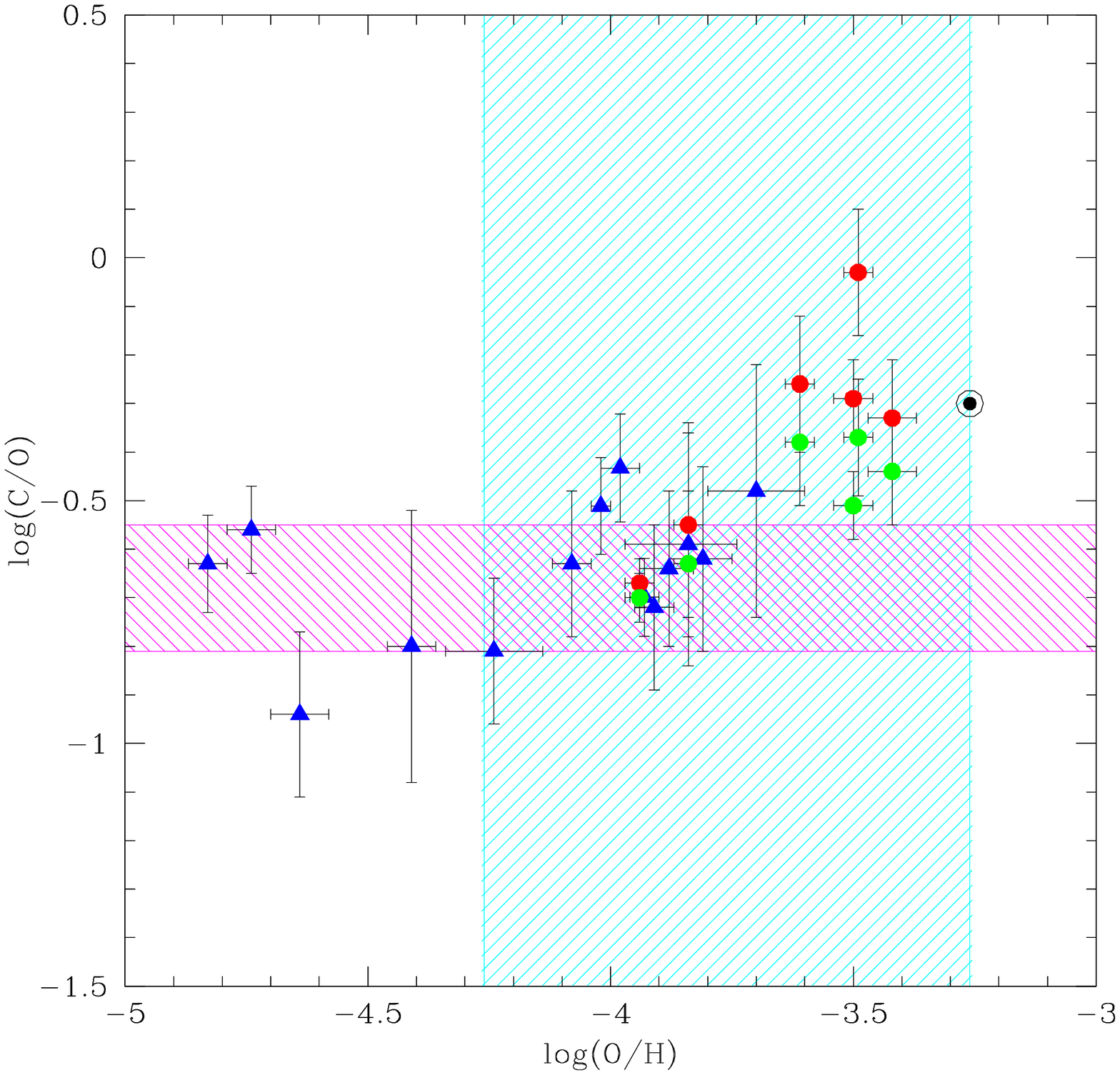}}
\figcaption{$\log ({\rm C/O})$ vs. $\log ({\rm O/H})$ for local H~II
regions. Blue triangles are data from dwarf irregular galaxies
and the Magellanic Clouds \citep{garnett1995,garnett1997,kobulnicky1998}.
Spiral galaxy data are shown with circles \cite{garnett1999}. Red symbols
assume a shallow ($R_V=3.1$) Milky Way extinction curve, whereas
green symbols indicate a steeper $(R_V=5.0)$ one. The solar
abundances \citep{holweger2001,allende2002}
are indicated by the large bulls-eye. The horizontal
(magenta) shaded area indicates the $\log({\rm C/O})$ confidence region
derived from the total composite LBG spectrum. The vertical
(cyan) shaded area indicates the range of $\log({\rm O/H})$ implied
by the ratio of [O~III], [O~II], and H$\beta$ line-strengths
in a small sample of bright LBGs \citep{pettini2001}.
\label{fig:co}
}
\end{inlinefigure}

\subsubsection{AGN Contribution?}
\label{sec:features_neb_agn}
Even though all galaxy spectra flagged as narrow- or broad-lined
AGN on an individual basis were removed from the LBG composite spectra
sample, the presence of weak emission lines in the LBG composite spectrum
may reveal some average low-level
of nuclear activity in LBGs. To address this issue, we examined
the emission line ratios in a composite spectrum of 198 LBGs with 
rest-frame $W_{{\rm Ly}\alpha}\geq 20$~\AA\ in emission
(see section~\ref{sec:trends_lya}). 
The nebular emission lines in this composite spectrum are
also stronger than those marked in the total LBG spectrum of
Figure~\ref{fig:plotall}.  We compared the emission line ratios 
in this strong-emission spectrum with those in a composite
spectrum of 16 LBGs flagged as narrow-lined AGN on an individual basis
\citep{steidel2002}. The average Ly$\alpha$
emission equivalent width in this strong-emission (yet
not AGN-flagged) subsample is only half as large the emission
strength in the narrow-line AGN spectrum. 
The narrow-line AGN spectrum has line
intensity ratios of C~IV/Ly$\alpha$~$\sim 0.25$, 
and C~III]/Ly$\alpha$~$\sim 0.125$, which are very similar to the mean 
ratios of C~IV/Ly$\alpha=0.21\pm0.09$ and C~III]/Ly$\alpha=0.10\pm 0.05$, 
measured for a sample of four local Seyfert 2 galaxies by 
\citet{ferland1986}. In contrast, the non-AGN spectrum
has intensity ratios of 
C~IV/Ly$\alpha \leq 0.02$ and C~III]/Ly$\alpha=0.05$.
The $2 \%$ represents a strict upper limit on the ratio of 
nebular C~IV/Ly$\alpha$ (which is probably much smaller), 
since the total C~IV emission 
represents the sum of nebular emission plus stellar P-Cygni emission.
Clearly, the ratios of both C~III] and C~IV to Ly$\alpha$
are much smaller in the non-AGN LBG spectrum than in the 
narrow-line AGN spectrum. Additionally, the ratio of C~III]/C~IV is
significantly higher in the non-AGN spectrum than in the AGN spectrum,
indicating a softer photoionizing radiation field, 
more likely dominated by the emission from hot stars 
rather than non-thermal processes.

\subsubsection{C/O Abundance}
\label{sec:features_neb_co}
Ultraviolet spectroscopic observations 
of H~II regions in nearby irregular and spiral galaxies
with the {\it Hubble Space Telescope} ({\it HST}) have been used to study
how the relative abundances of carbon and oxygen (C/O) depend
on oxygen abundance (O/H) \citep{garnett1995,garnett1997,garnett1999,kobulnicky1998}.
Ranging in
O/H from less than 0.1 to 1.0 times $({\rm O/H})_{\odot}$, a compilation 
of these data is shown in figure~\ref{fig:co} and demonstrates
a clear trend of increasing C/O with increasing O/H \citep{garnett1999}. 
For example, in spiral galaxy H~II regions, $\log ({\rm C/O}) \simeq -0.7$ at
$\log ({\rm O/H}) = -4.0$, and increases to $\log ({\rm C/O}) \simeq -0.2$
at $\log ({\rm O/H}) = -3.4$\footnote{The solar oxygen abundance
is $\log({\rm O/H})_{\odot}=-3.26$ \citep{holweger2001}}. While the behavior at the
smallest metallicities is less clear, due the unexpectedly
high C/O ratio in the extremely metal-poor galaxy, I Zwicky 18, 
the H~II regions in other sub-solar metallicity irregular galaxies show the same
overall trend as the spiral galaxies \citep{garnett1995}. 
Stellar evolution models including stellar winds
from massive stars predict that the carbon yield from massive stars
increases relative to the oxygen yield, with increasing
metallicity \citep{maeder1992}. These models successfully
reproduce both the observed local trend in C/O with O/H, and also
C/O abundance gradients in Galactic stars and H~II regions
(e.g. Carigi 2000). While the rise of C/O with O/H in
local star-forming regions
is mainly due to yields from massive stars, there 
is also the fact that oxygen is primarily
synthesized in stars with $M>10 M_{\odot}$, whereas
carbon is produced in both high and intermediate 
mass $(2-8M_{\odot})$, delaying in time
the ejection of some fraction of carbon into the ISM,
relative to oxygen. In relatively
young galaxies at high redshift, the observed C/O
ratio may thus also reflect the average time since the onset of
star-formation. In such circumstances, 
younger stellar populations may be
dominated by the chemical yields from the most massive stars, 
whereas after a few 100 Myr since the onset of star formation,
an increased C/O ratio reflects the fact
that intermediate-mass stars
have had a chance to release their carbon into the ISM.

Since we detect both C~III] $\lambda 1909$ and O~III] $\lambda 1663$ in
the LBG composite spectrum shown in Figure~\ref{fig:plotall},
we follow the analysis of \citet{garnett1995},
assuming that the electron densities in these H~II regions 
are well below the critical limit for the C~III] and O~III] 
transitions. In the low-density limit, we can express the relative
abundances of ${\rm C}^{+2}$ and ${\rm O}^{+2}$ as:

\begin{equation}
 \frac{{\rm C}^{+2}}{{\rm O}^{+2}} = \frac{1}{9} \times \frac{\Omega_{{\rm OIII]}}(^3{\rm P},^5{\rm S}_2) }{\Omega_{{\rm CIII]}}(^1{\rm S},^3{\rm P})} \times \frac{\lambda_{{\rm CIII]}1909}}{\lambda_{{\rm OIII]}1663}} \times e^{-11054/T} \times \frac{I({\rm CIII]} \lambda 1909) }{I({\rm OIII]} \lambda 1663)}
\label{eq:co}
\end{equation}

In this expression $\Omega_{{\rm OIII]}}(^3{\rm P},^5{\rm S}_2)$ and 
$\Omega_{{\rm CIII]}}(^1{\rm S},^3{\rm P})$ are the multiplet collision
strengths for the O~III] $\lambda 1663$ and C~III] $\lambda 1909$ doublets,
which have a weak temperature dependence (see Garnett \et 1995 for
specific collision-strength values);
the prefactor of $\frac{1}{9}$ represents a combination of statistical
weights; $\lambda_{{\rm CIII]}1909}$ and $\lambda_{{\rm OIII]}1663}$
are the effective wavelengths of the C~III] and O~III] doublets;
$T$ is the H~II region electron temperature; and $I({\rm CIII]} \lambda 1909)$
and $I({\rm OIII]} \lambda 1663)$ are the line intensities. To compute
$I({\rm OIII]} \lambda 1663)$ we integrate the flux from both members
of the resolved O~III] $\lambda 1663$ doublet (the C~III]
doublet is unresolved).
Strictly speaking, 
${\rm C/O} = {\rm C}^{+2}/ {\rm O}^{+2} \times {\rm ICF}$, where ICF is an
ionization correction factor which takes into account the fact that
while C~III] and O~III] are similar ionization states,
${\rm C}^{+2}$ has a slightly lower ionization potential
than ${\rm O}^{+2}$. In practice, \citet{garnett1995} find
ICF only ranges between $1.06 - 1.33$, and so, for lack of any detailed
information, we make the approximation ${\rm ICF}=1$ and 
${\rm C/O} \simeq {\rm C}^{+2}/ {\rm O}^{+2}$. Based on the measured
ratio of C~III] and O~III] line strengths in the total composite
LBG spectrum, we measure $\log ({\rm C/O}) = -0.68 \pm 0.13$. 
This confidence interval is marked by the horizontal magenta shaded region
in Figure~\ref{fig:co}.
The quoted uncertainty corresponds to a temperature range
$T = 10000-20000$~K, but does not include the
uncertainty in the determination of the continuum level,
which may amount to an additional error of about a factor
of two in the determination of C/O.
We measure a very similar value of $\log ({\rm C/O}) = -0.74 \pm 0.14$ in the 
LBG composite spectrum constructed from the sub-sample which includes only
strong emission-line galaxies (see sections~\ref{sec:features_neb_agn},~\ref{sec:trends_lya}). Similar
C/O ratios are found by \citet{garnett1999} for galaxies with 
${\rm O/H} \sim 0.2 \times ({\rm O/H})_{\odot}$. There are large uncertainties
associated with the LBG C/O measurements. However, 
we note that the corresponding O/H values are consistent with --
if towards the low-end of the confidence interval of --
the O/H metallicity determinations for LBGs
based on the ratio of rest-frame optical nebular [O~II], [O~III], 
and H$\beta$ emission lines ($\sim 0.1 - 1 \times ({\rm O/H})_{\odot}$)
\citep{pettini2001}.\footnote{The galaxies
with rest-frame optical spectroscopic observations were drawn from the
bright end of the LBG UV luminosity function. It is not clear
how this selection criterion limits the range of
metallicities probed, relative to the abundance range in the 
LBG sample as a whole.}
The range in $\log({\rm O/H})$ deduced by \citet{pettini2001} is marked by 
the vertical cyan shaded region in Figure~\ref{fig:co}. In view of the large
uncertainties associated with both C/O and O/H determinations
in LBGs, we do not interpret these results any further, except
to point out that the comparison of C and O emission strengths
will be an interesting diagnostic of chemical evolution and
stellar populations at high-redshift in future, higher-quality data.

\subsubsection{Si~II* Lines}
\label{sec:features_neb_siII}
In addition to C~III] and O~III], which have been studied in local
star-forming regions, we also detect lines at which we identify as
Si~II* fine-structure emission lines. Unfortunately,
our spectral resolution is too coarse to determine if we detect
C~II* $\lambda 1335$ \AA\ emission as well (the C~II $\lambda 1334$
resonance absorption line from outflowing neutral gas swamps 
any signal at that wavelength).
A literature search of local ultraviolet observations
reveals very few references to Si~II* fine-structure emission lines.
The {\it International Ultraviolet Explorer} ({\it IUE}) atlas
of star-forming galaxies compiled by \citet{kinney1993} contains
a census of ultraviolet emission lines associated with nebular
objects such as H~II regions, planetary nebulae, and supernova remnants, 
but does not include the Si~II* features. 
The spectral resolution of IUE is $\sim 1000$ \kms, coarser than
the resolution of the LBG composite spectra. When the LBG
composite spectra are smoothed to the resolution of the IUE data,
the Si~II* lines are still visible at 5\% to 10\% above the continuum
level. Presumably, the individual IUE spectra contained in the 
\citet{kinney1993} atlas are of insufficient S/N to see these features. 
However, the Si~II* features are also not detected in composite
starburst spectra containing 20 individual galaxy spectra 
drawn from the IUE atlas, each of which has continuum S/N of at least 10
\citep{heckman1998}.

The Einstein A coefficients associated with the Si~II* transitions
range from $10^8-10^9$~s$^{-1}$, more than six orders of magnitude
larger than the A-values for the semi-forbidden 
O~III] and C~III] transitions. 
Only in very high-density environments ($N_e = 10^9 - 10^{13}$~cm$^{-3}$)
are the Si~II excited level populations determined by
collisional excitations and de-excitations \citep{keenan1992}. 
In H~II regions, where the electron densities are
typically $N_e = 10^2 - 10^{3}$~cm$^{-3}$, collisions
are therefore not the dominant mechanism governing the Si~II
level populations. Also, when $T\sim 10^4$~K and Si~II and Si~III
have comparable abundances (as appropriate here), it can be shown that
the recombination rate of Si~III into the excited
Si~II levels is of the same order as the 
Si~II collisional excitation rates \citep{shull1982}.
In order to understand the origin of the
Si~II* emission lines in a more systematic way,
we modeled all observed LBG nebular emission lines
using the CLOUDY96 software package \citep{ferland1998}.
The observational constraints are the average rest-frame
optical strengths of [O~III], [O~II], and H$\beta$
\citep{pettini2001}, as well as the rest-frame UV O~III] and C~III]
strengths (section~\ref{sec:features_neb_co}). 
We found that {\it any}
model which provides a satisfactory fit to the O, C, and H$\beta$
line ratios simultaneously predicts 
Si~II* emission line strengths which are more than an order of
magnitude weaker than observed. This result
seems to exclude an origin in photoionized H~II regions
for the Si~II* emission lines.

An alternative explanation for these lines is
that they are produced in the large-scale outflows in LBGs. 
UV photons with $\lambda=1260,1304, 1526$ are absorbed by ground state Si~II
in the outflowing neutral gas, as indicated by the 
saturated Si~II resonance absorption transitions.
Each photon absorbed in
a resonance or fine-structure transition is re-emitted in a transition
either to the ground state or the excited ground state, with 
the relative probabilities determined by the Einstein A-coefficients
for the different transitions. 
In the absence of dust extinction, and if the slit used to 
observe the galaxy/outflow system is as large as the region emitting
in Si~II and Si~II*,
the sum of the resonant and fine-structure emission 
equivalent widths should be equal in strength to the
absorption from the same transitions. This is clearly not the
case, in that the Si~II~*$\lambda 1265, 1309, 1533$ 
fine-structure emission lines are an order of magnitude
weaker than the Si~II~$\lambda 1260, 1304, 1526$ resonant absorption lines. 
We don't even detect any resonant Si~II emission,
as saturated blue-shifted Si~II absorption dominates over emission
at $\Delta v=+100$\kms, the mean velocity of the Si~II*
emission lines.
The dominance of resonant
absorption over fine-structure emission
may be due to dust attenuation of Si~II photons
during resonant scattering. It also may indicate that 
the slit used in LBG spectroscopic observations only
subtends a small fraction of the Si~II* emitting region.
One problem with interpreting the Si~II*
emission lines as being produced in the outflowing gas is their
kinematic properties. The outflow
is optically thin to Si~II* (negligible absorption is detected
in these transitions), so we expect 
Si~II* emission lines produced in outflowing gas
to probe the full range of approaching and receding velocities
($\geq 1000$~\kms). In fact, the Si~II* lines are 
barely resolved in the composite spectrum, with ${\rm FWHM} \leq 500$~\kms.
While Si~II* emission may be biased towards positive velocities
by blue-shifted fine-structure, or even broad resonance, absorption, 
we expect that it should be at least as broad 
as the Ly$\alpha$ emission, which is attenuated on the blue 
edge by much stronger absorption over a wider range of velocities.
However, the Si~II* lines are narrower than the Ly$\alpha$ emission
line, which extends to much more redshifted velocities. 
In summary, both the outflow and H~II region models of the Si~II* features
fail to explain all of the observed properties of the emission lines,
whose true nature remains to be determined.

\section{LBG Spectroscopic Trends}
\label{sec:trends}
We have characterized the basic features of the 
composite spectrum formed from the $z\sim 3$ LBG spectroscopic sample.
We are now ready to examine how
these spectroscopic features vary across the
sample as functions of different galaxy parameters. 
Some of the parameters which can be measured for individual galaxies
are: redshift, $z$; rest-frame UV apparent magnitude, ${\cal R}$;
rest-frame UV color corrected for IGM absorption, $(G-R)_0$,
which can be parameterized in terms of a reddening, $E(B-V)$,
given an assumed form for the intrinsic spectrum; 
Ly$\alpha$ rest-frame equivalent width, $W_{{\rm Ly}\alpha}$;
and interstellar kinematics, $\Delta v_{{\rm em-abs}}$. 
Our spectroscopic sample is large enough that 
we can divide the total sample into several subgroups based on
each of the above parameters, and
still create a high S/N composite spectrum for each subgroup.
Individual spectra are not of sufficient
S/N to be able to reliably measure low- and high-ionization  
interstellar absorption equivalent
widths, $W_{{\rm LIS}}$ and $W_{{\rm HIS}}$. 
Therefore, we do not bin the
sample according to interstellar absorption line strength,
but we can measure the interstellar absorption strengths with
high significance in all of the composite spectra.

\subsection{Selection Effects}
\label{sec:trends_seleffects}

Before considering LBG spectroscopic trends, we must
isolate which parameters are sensitive to the variance
in the underlying galaxy population and which are more
sensitive to our photometric and spectroscopic selection
criteria. To illustrate the importance of selection effects,
we consider how our photometric and spectroscopic biases
limit the range of galaxy parameter space which can be 
sampled as a function of $z$ and ${\cal R}$ magnitude.

First, we examine biases resulting from the LBG photometric 
selection criteria.  One of the LBG color criteria 
is $G-{\cal R} \leq 1.2$, which
affects the range of intrinsic UV colors selected with the LBG
technique as a function of redshift. The reason
for this effect is that absorption by HI in
the IGM attenuates the flux in the $G$-band for galaxies
with $z\geq 2.4$. The average amount of attenuation is an increasing
function of redshift, ranging from $\Delta G= 0$ magnitudes at $z=2.4$ to
$\Delta G = 0.2$ magnitudes at $z=3.0$, to as much as $\Delta G= 0.5$ 
magnitudes at $z=3.4$. Since LBGs are photometrically pre-selected on the 
basis of observed
$G-R$ color, which includes the effect of IGM absorption,
the range of intrinsic colors which can be included in the LBG
sample is a strong function of redshift. 
As discussed in section~\ref{sec:trends_dust},
the rest-frame UV colors of continuously star-forming galaxies are
largely determined by the amount of reddening affecting the 
stellar continuum, so the intrinsic UV color, $(G-R)_0$ can also be 
parameterized in terms of $E(B-V)$, once a form of the attenuation curve is 
assumed \citep{calzetti1997,calzetti2000}.  
In subsequent discussion, we use $E(B-V)$ to 
represent both intrinsic UV color and the amount of dust extinction, since
there is almost a one-to-one correspondence between the two parameters.
Figure~\ref{fig:ebv_vs_z}  shows 
$E(B-V)$ vs. $z$ for the LBG spectroscopic sample,
demonstrating the strong apparent correlation of UV color with redshift,
induced by our photometric selection effects. 

\begin{inlinefigure}
\centerline{\epsfxsize=9cm\epsffile{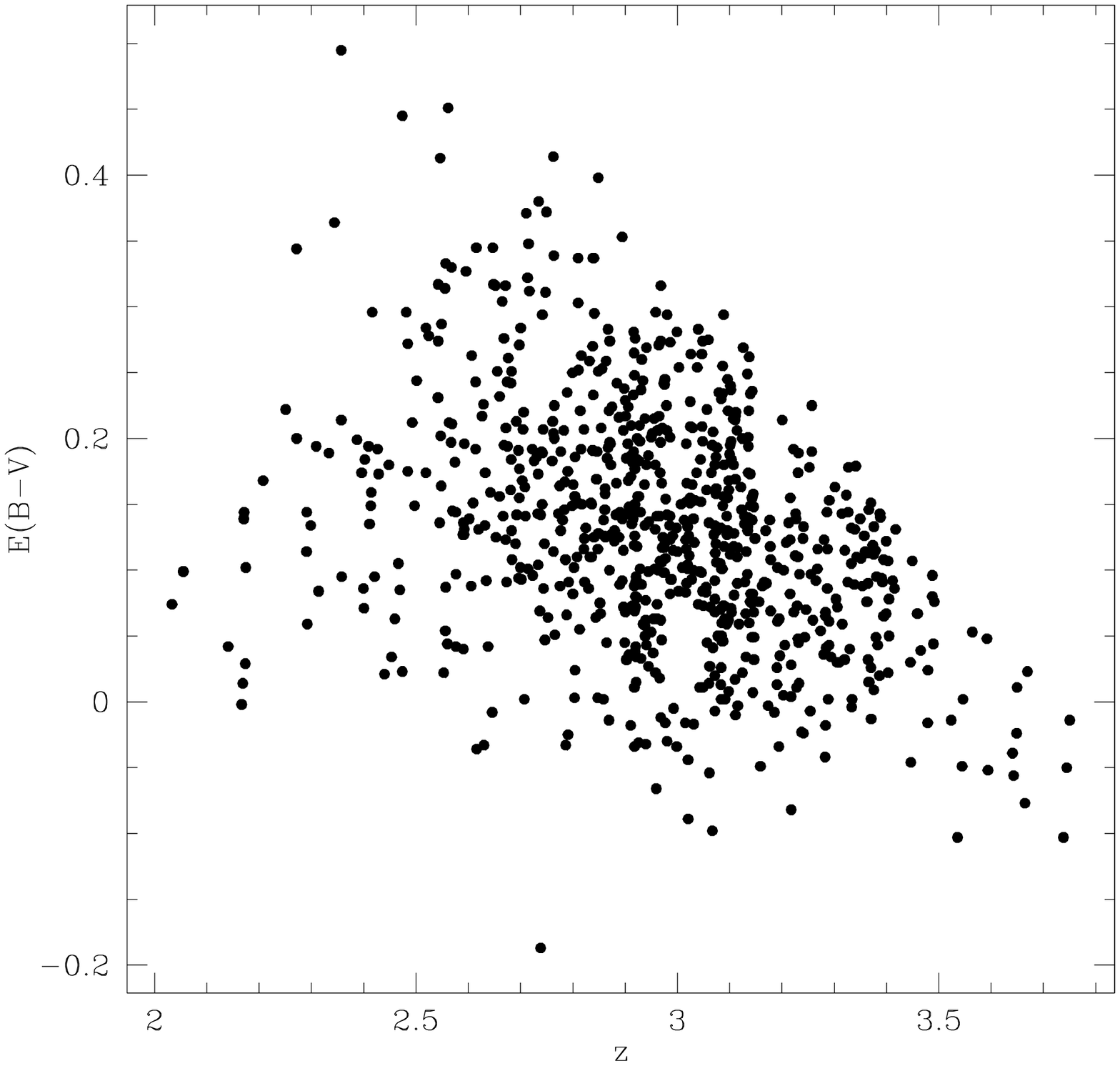}}
\figcaption{The distribution of $E(B-V)$ values as a function of $z$.
Due to the way in which LBGs are color-selected to
have {\it observed} $G-{\cal R}$ colors which lie within a specific
range, the increased IGM absorption affecting the G-band flux
as a function of $z$ limits the range of {\it intrinsic} colors of
LBGs (parameterized by $E(B-V)$) at higher $z$.
\label{fig:ebv_vs_z} 
}
\end{inlinefigure}

There are also important photometric selection effects
associated with apparent ${\cal R}$ magnitude. 
Due to the nature of LBG color criteria, and the fact the $U_n$ images
have finite depth, the dynamic range in 
$G-{\cal R}$ color is limited on average to bluer colors
at fainter ${\cal R}$ magnitudes. Accordingly,
there is a weak apparent correlation between ${\cal R}$ and
$E(B-V)$, in the sense that fainter LBGs
are bluer on average. However, \citet{adelberger2002b}
demonstrates that, once selection effects are accounted for, 
${\cal R}$ and  $E(B-V)$
are consistent with being independently distributed.
The biases against UV color as a function of
$z$ and ${\cal R}$ magnitude result from the LBG
photometric selection criteria and uncertainties, and
have been discussed extensively and quantified 
for the purpose of constructing the LBG rest-frame UV luminosity function
\citep{steidel1999,adelberger2000,adelberger2002b}.

There are two additional
spectroscopic sources of incompleteness which affect the
sample under consideration. First, not all galaxies in the photometric sample
were assigned to slitmasks and observed spectroscopically.
This incompleteness is primarily a function of ${\cal R}$ magnitude,
i.e. a larger fraction of bright galaxies in the photometric
sample were observed spectroscopically, relative to the fraction of 
faint galaxies. 
The distribution of $G-{\cal R}$ colors of 
spectroscopically observed galaxies
is relatively unbiased with respect to the photometric
sample. Second, not all galaxies observed
spectroscopically had successfully measured redshifts.
This type of incompleteness also depends on ${\cal R}$ magnitude,
but the subtlety lies in the fact that that galaxies with
different spectroscopic types (i.e. those with and without 
Ly$\alpha$ emission) have different spectroscopic
success rates as a function of ${\cal R}$ magnitude. Figure~\ref{fig:rhistq} 
shows the spectroscopic incompleteness as a function of magnitude,
and Figure~\ref{fig:rtypehistq} 
shows the incompleteness as a function of both magnitude and spectral type. 

\begin{inlinefigure}
\centerline{\epsfxsize=9cm\epsffile{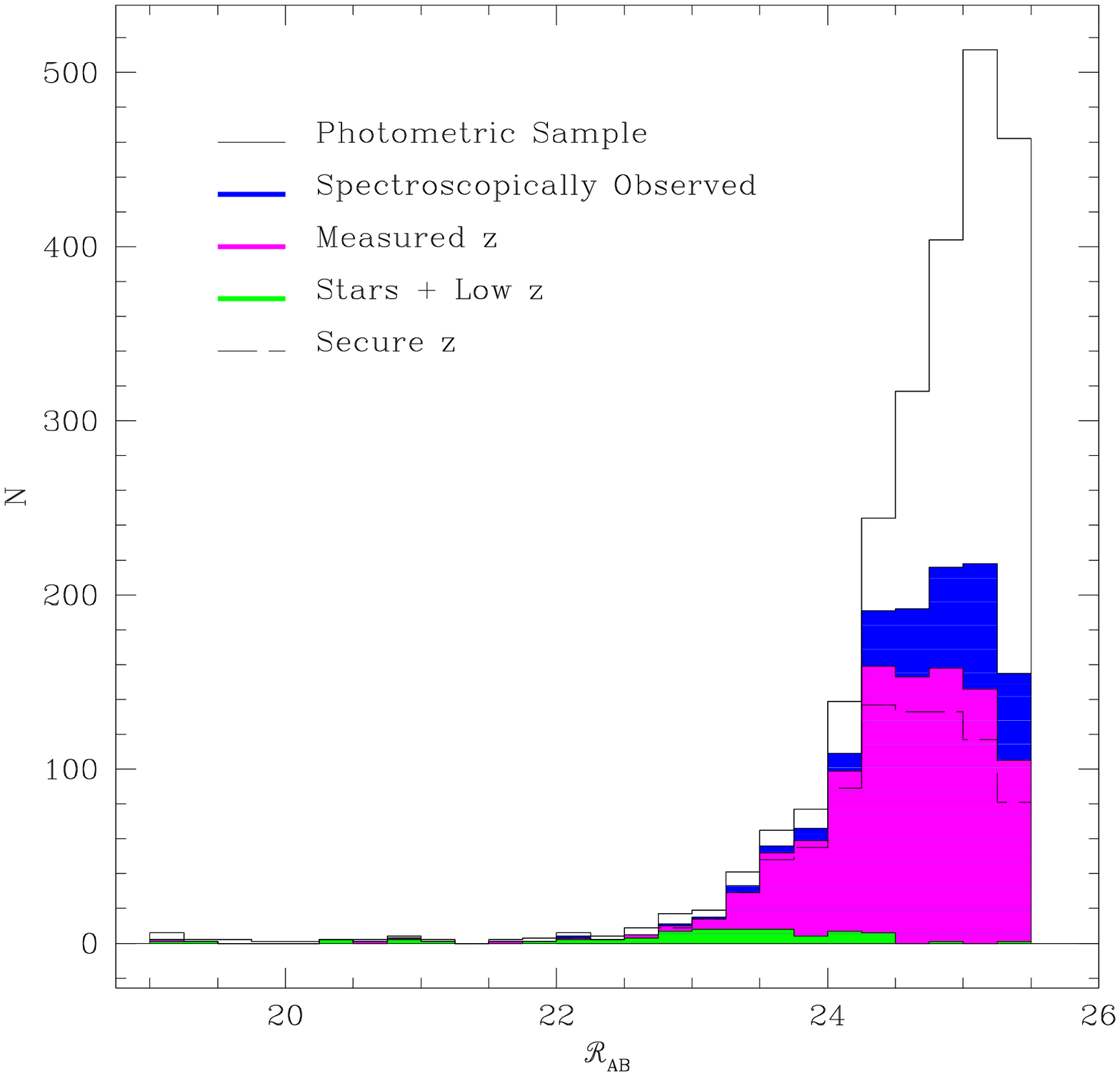}}
\figcaption{Photometric and Spectroscopic incompleteness of the LBG sample
as a function of ${\cal R}$ magnitude. The empty histogram represents
galaxies photometrically selected by their colors to be at $z\sim 3$.
The blue histogram
consists of galaxies from the photometric sample which were
observed spectroscopically. The pink histogram indicates spectroscopically
observed objects for which a redshift was successfully measured.
The green histogram shows the small fraction of spectroscopically
confirmed objects $(\sim 4\%)$ which turn out to be stars or low-z
galaxies. The dashed histogram shows the subset of spectroscopically
confirmed objects which have secure redshifts confirmed by
at least two independent members of our group. Non-AGN objects in this
secure-z sample with $z>2$ were included in composite spectra.
\label{fig:rhistq}
}
\end{inlinefigure}

There are three subgroups identified in Figure~\ref{fig:rtypehistq}: 
spectra for which the redshift was measured with Ly$\alpha$ emission
(some of which also have interstellar absorption redshifts, but which
have at least one identifiable emission line); spectra for which
the redshift was only measured with interstellar absorption lines;
spectra for which no redshift was measured. Not surprisingly, the fraction
of galaxies with spectroscopic failures increases from $\sim 10\%$ at
${\cal R}=24$ to $> 30\%$ at ${\cal R}=25-25.5$. 
The fraction of spectroscopic successes for which the redshift
was measured from features including 
Ly$\alpha$ emission increases from $\sim 60 \%$
at ${\cal R}=24$ to $\sim 75 \%$ at ${\cal R}=25-25.5$. Simultaneously,
the fraction of successes for which the redshift was measured only from
absorption lines decreases from $\sim 40 \%$ at ${\cal R}=24$ down
to $\sim 25 \%$ at ${\cal R}=25-25.5$. These percentages reflect the
somewhat obvious fact that it is easier to measure a redshift
from an emission line than from an absorption line at fainter magnitudes.
As a result, the observed fractions of spectroscopically confirmed
objects with and without emission lines in the faintest 
magnitude bin of Figure~\ref{fig:rtypehistq} 
do not represent the true underlying fractions,
in the sense that emission line objects are overrepresented.
We can try to draw inferences about the underlying proportion of emission
and absorption objects from the fraction of objects without 
measured spectroscopic redshifts. The simplest possible assumption
is that these galaxies must have Ly$\alpha$ emission fluxes
below some limiting value as a function of magnitude. This limiting
Ly$\alpha$ flux depends on other factors in addition to ${\cal R}$ magnitude, 
most notably
time-dependent observing conditions such as sky transparency and seeing. 
It is difficult to quantify these effects exactly, so we use the
conservative assumption that the galaxies without redshifts have
$W_{{\rm Ly}\alpha} < 0$, i.e. no Ly$\alpha$ emission,
and that if redshifts had been measured for these galaxies they would
have only been measured from interstellar absorption lines. If
the fraction of galaxies with no redshifts is added to the 
fraction of galaxies with only interstellar absorption redshifts,
we see that the implied true proportion of emission and absorption objects
remains roughly constant and equal as a function of magnitude (pink dots in
Figure~\ref{fig:rtypehistq}), 
in contrast to the observed proportion (black dots in 
Figure~\ref{fig:rtypehistq}).
This exercise gives a rough indication of the degree to which
objects with Ly$\alpha$ emission lines are overrepresented 
relative to absorption-line-only objects within the 
spectroscopic sample as a function of magnitude.

The photometric and spectroscopic biases presented
above affect determinations of both the total
LBG spectrum and of the ways in which spectroscopic properties
depend on galaxy parameters. Luminosity and redshift
are especially prone to these selection effects. Therefore,
the discussion of LBG spectroscopic trends is limited
to the parameters $W_{{\rm Ly}\alpha}$, $E(B-V)$, and $\Delta v_{{\rm em-abs}}$.
There are highly significant patterns in the spectroscopic
properties of LBGs as functions of these parameters.
With simple arguments, we show why the strong dependences in
the spectroscopic parameters reflect the physical conditions
in LBGs, and not the nature of our selection effects.

\subsection{Uncertainties}
\label{sec:trends_uncertainties}

In order to assess the significance of the spectroscopic
trends in the sample, it is necessary to assign error bars
to the spectroscopic measurements from 
each composite spectrum. The uncertainty on an equivalent
width measurement from a composite spectrum
includes not only the finite S/N
of the composite spectrum, but also the range of equivalent
widths in the sample of galaxies used to construct the composite.
While Ly$\alpha$ equivalent widths were measured for
almost all of the individual galaxies in the spectroscopic sample,
only a subset of the strong interstellar absorption
lines was detectable in typical individual spectra. 
In fact, in 231 cases Ly$\alpha$ emission was the only
spectroscopic feature identified. Additionally,
the Si~II*, C~III], and O~III] nebular emission lines are far too weak
to be measured in individual spectra.

\begin{inlinefigure}
\centerline{\epsfxsize=9cm\epsffile{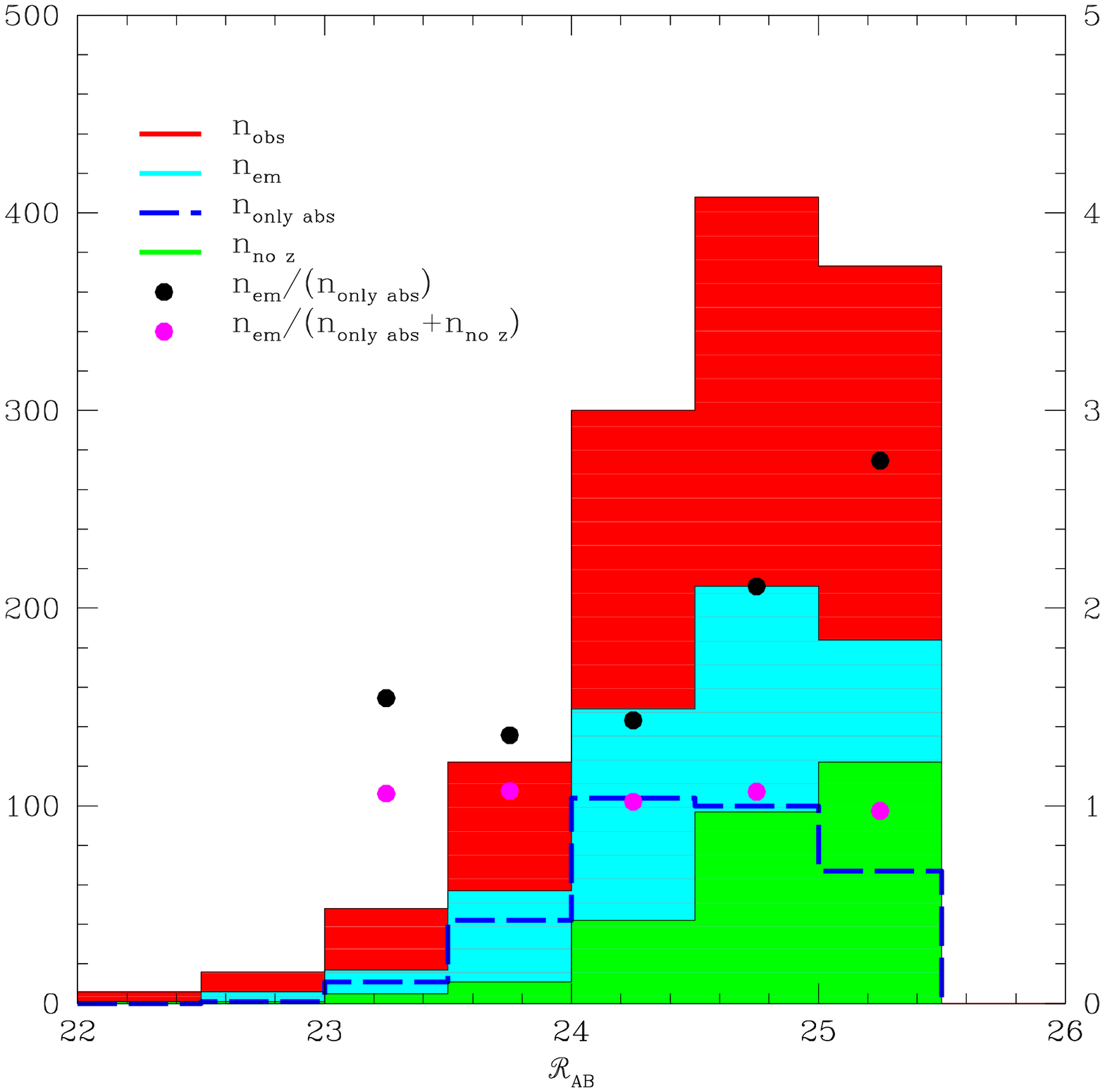}}
\figcaption{Photometric and Spectroscopic incompleteness of the LBG sample
as a function of ${\cal R}$ magnitude and spectroscopic type.
The red histogram shows the number of galaxies observed
spectroscopically. The cyan histogram shows the number of
galaxies with redshifts measured from features which include Ly$\alpha$
emission.  The blue dashed histogram shows the number of objects with
no detectable Ly$\alpha$ emission and redshifts measured only from
interstellar absorption lines. The green histogram shows the number
of objects for which no redshift was successfully measured.
The black dots indicate the ratio of galaxies with
Ly$\alpha$ emission to those with only absorption line
redshifts, which increases steeply as a function of ${\cal R}$
magnitude. Based on the assumption that the unidentified objects
in the green histogram are $z\sim3$ galaxies that have $W_{{\rm Ly}\alpha} < 0$
and spectra with insufficient S/N to identify absorption features,
we see that the ratio of objects with Ly$\alpha$ emission to those
with only absorption features remains roughly constant as a function
of magnitude (pink dots).
\label{fig:rtypehistq}
}
\end{inlinefigure}

Given that most of the spectroscopic features analyzed
in the composites could not be measured reliably
in individual spectra, we estimated
the sample variance for these features using bootstrap techniques,
as follows.
For each composite spectrum, we generated 500 fake
composite spectra, each one constructed from a sample of galaxies
drawn with replacement from the sample used for the real composite spectrum
($\sim 37\%$ of the sample is replaced with duplicates). 
Using a measurement technique identical to the analysis applied
to the real composite spectrum,  we continuum
normalized each fake composite spectrum, and measured the equivalent
widths of Ly$\alpha$, the six strongest low-ionization interstellar
absorption lines, the
Si~IV and C~IV high-ionization interstellar absorption lines,
and the weak nebular emission lines. To take into account 
the noise in each fake composite spectrum, each fake
equivalent width measurement was perturbed by an amount drawn
from a Gaussian distribution with standard deviation
$\sigma=\frac{\sqrt n}{S/N}$,
where $n$ is the number of pixels over which the equivalent width was measured,
and $S/N$ is the signal-to-noise ratio of the real composite spectrum,
measured in a relatively featureless portion of the continuum. For the
strong interstellar absorption lines and weak emission lines,
the contributions to the uncertainty from sample variance and finite S/N
were roughly comparable, while in the case of Ly$\alpha$,
sample variance dominated the uncertainty estimate.
The total uncertainty for each real equivalent width measurement
was then equal to the standard deviation of the distribution of 500
perturbed fake equivalent width measurements (which were
distributed around the actual measured value with no systematic
offset). The uncertainties for all quantities measured from composite
spectra were derived with the above technique.
Bootstrap techniques can also be used to estimate the
uncertainty in the mean continuum properties such as 
$E(B-V)$ and ${\cal R}$ magnitude, of each composite sample. 
\footnote{Of course,
it is also possible to estimate the uncertainty in the
mean, $\langle x \rangle $,  as 
$\sigma_{\langle x \rangle}=\frac{\sigma_x}{\sqrt N_x}$,
where $\sigma_x$ is the standard deviation of the distribution
of $x$, and $N_x$ is the number galaxies in the sample
from which $x$ was measured. This technique yields roughly
the same uncertainty in $E(B-V)$ and ${\cal R}$ magnitude
for each composite sample as the value estimated from
the bootstrap method.} 

\begin{inlinefigure}
\centerline{\epsfxsize=9cm\epsffile{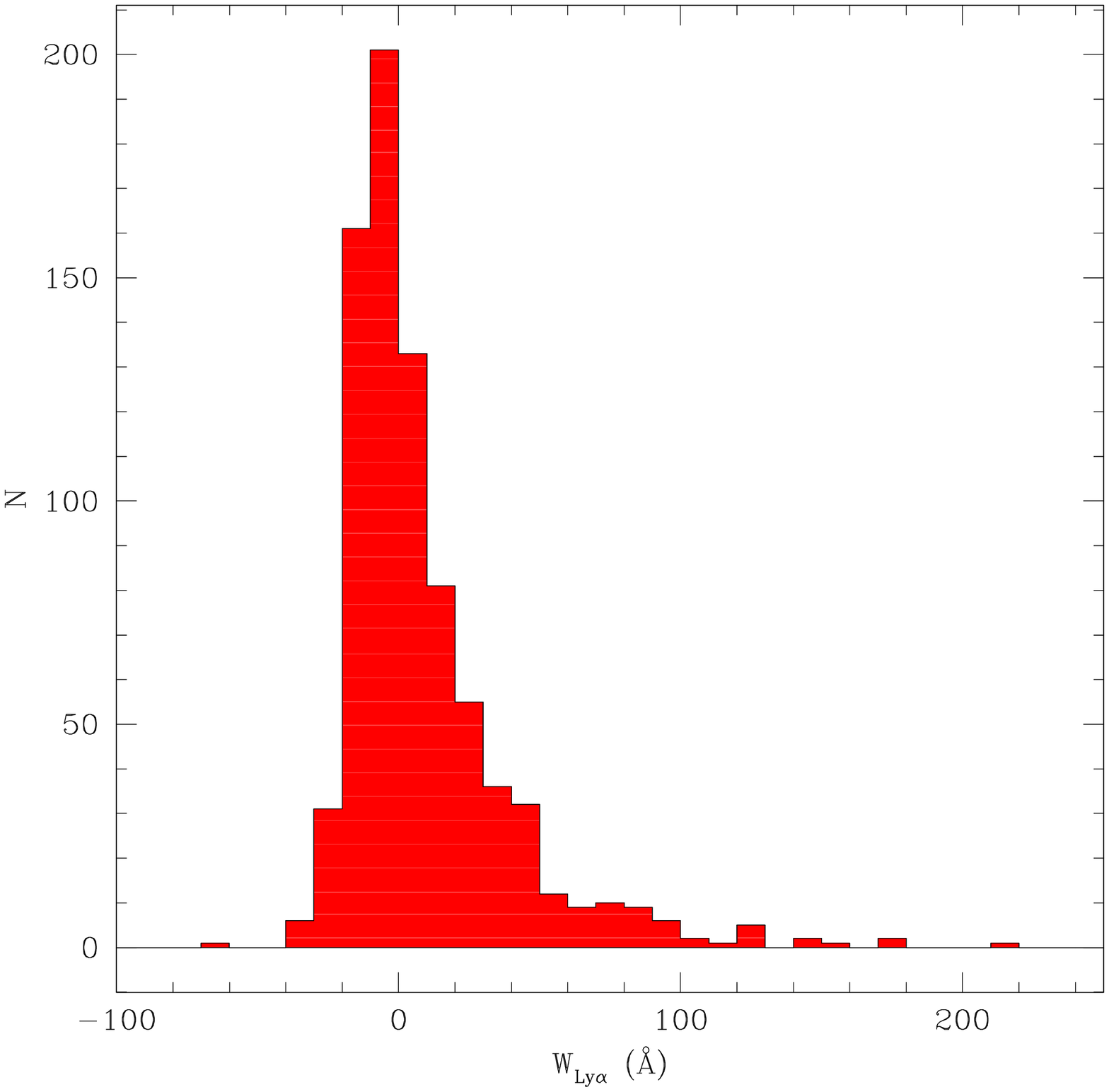}}
\figcaption{The distribution of Ly$\alpha$ equivalent widths for the
LBG spectroscopic sample. This sample contains a broad range of
equivalent widths with a median of $\sim 0$~\AA. Only $25\%$ of the
sample has rest-frame $W_{{\rm Ly}\alpha} \geq 20$~\AA, large enough
to be selected by narrow band excess techniques, given
the depth of current surveys. There is a correlation
between Ly$\alpha$ emission equivalent width and ${\cal R}$
magnitude among sources with $W_{{\rm Ly}\alpha} \geq 20$~\AA, such that
fainter galaxies have larger $W_{{\rm Ly}\alpha}$.
\label{fig:lyatothist}
}
\end{inlinefigure}

\subsection{Ly$\alpha$ Dependences}
\label{sec:trends_lya}
One of the most striking characteristics of the LBG
spectroscopic sample is the
broad distribution of Ly$\alpha$ strengths 
and profile-types, ranging from pure damped absorption, to emission plus
absorption, to pure strong emission. There is a large body of theoretical and 
observational work concerning the physical processes 
which determine the emergent Ly$\alpha$ profile in 
local and high-redshift star-forming galaxies. 
In local star-forming galaxies, early observations indicated
Ly$\alpha$ emission equivalent widths 
much smaller than expected from recombination theory,
given the star-formation rates inferred from optical Balmer emission
lines \citep{meier1981,hartmann1984,hartmann1988}.  Additionally,
these observations offered evidence for a 
correlation between Ly$\alpha$/H$\beta$ 
and metallicity (measured from the nebular oxygen abundance, O/H), 
in the sense that galaxies with lower 
Ly$\alpha$/H$\beta$ also had higher O/H \citep{hartmann1988,charlot1993}.
The observations were first explained
with the presence of interstellar dust in a uniform medium, 
which preferentially destroys
Ly$\alpha$ photons relative to non-resonant UV continuum photons,
due to the increased path-length traversed by the 
resonantly scattered Ly$\alpha$ photons.
In addition to affecting Ly$\alpha$ photons
produced by recombinations in H~II regions, dust extinction
can also attenuate stellar continuum photons in the immediate vicinity
of Ly$\alpha$, which are also resonantly scattered
\citep{charlot1993,chen1994}.
The combination of resonant scattering and dust attenuation was thought
to significantly reduce the emergent Ly$\alpha$ emission.

The models of a uniform scattering medium 
clearly oversimplify the structure of the ISM in star-forming galaxies.
There are other factors which introduce complexity into the 
description of Ly$\alpha$ radiative transfer, including the
geometry and kinematics of the ISM. 
As described by \citet{neufeld1991} and \citet{charlot1993}, the
relative geometries of interstellar H~I and H~II regions significantly
affect the transfer of resonantly scattered photons, but in ways
which can either suppress or enhance the Ly$\alpha$ line relative to the 
continuum.
The importance of the geometry of the neutral phase of the ISM is
emphasized by the observational results of \citet{giavalisco1996a}.
A lack of correlation between
the equivalent width of Ly$\alpha$ and the UV continuum slope,
$\beta$ (a measure of continuum extinction),
is interpreted as evidence for the decoupling of the reddening of 
line and continuum photons. This decoupling can occur if the 
neutral ISM (where the dust resides) is inhomogeneous. 
Due to resonant scattering, for example, Ly$\alpha$ photons can propagate
into paths with less than average gas and dust and escape, whereas the 
continuum photons reflect the average dust obscuration along the line of 
sight. The kinematics of the the neutral ISM can also affect
the emergent Ly$\alpha$ profile. 
Inspired by such examples as the metal-poor local
starburst I~Zwicky~18, which has Ly$\alpha$ only in absorption 
\citep{kunth1994}, and the dustier and more metal-rich Haro~2 which shows a 
redshifted P-Cygni 
Ly$\alpha$ emission feature \citep{lequeux1995}, \citet{kunth1998} survey a 
small sample of local starburst galaxies in the UV, and find that
the objects with Ly$\alpha$ only in absorption also
have interstellar absorption lines 
(O~I~$\lambda 1302$ and Si~II~$\lambda 1304$) which are at rest with respect
to the H~II regions, whereas galaxies with
Ly$\alpha$ emission exhibit asymmetric, P-Cygni Ly$\alpha$ profiles,
with redshifted Ly$\alpha$ emission, and blueshifted interstellar 
absorption lines. These observations support a picture in which
Ly$\alpha$ photons mainly escape when they are produced by --
or scatter off of --  material which is offset in velocity from
the bulk of the scattering neutral medium. If the neutral
ISM is static with respect to the sources of Ly$\alpha$ photons,
then the covering factor of the neutral gas becomes important
\citep{kunth1998}. Using such observational evidence,
\citet{tenorio1999} have developed a detailed
model for the way in which outflow kinematics determine the emergent
Ly$\alpha$ profile in starburst galaxies.

\begin{figure*}
\centerline{\epsfxsize=14cm\epsffile{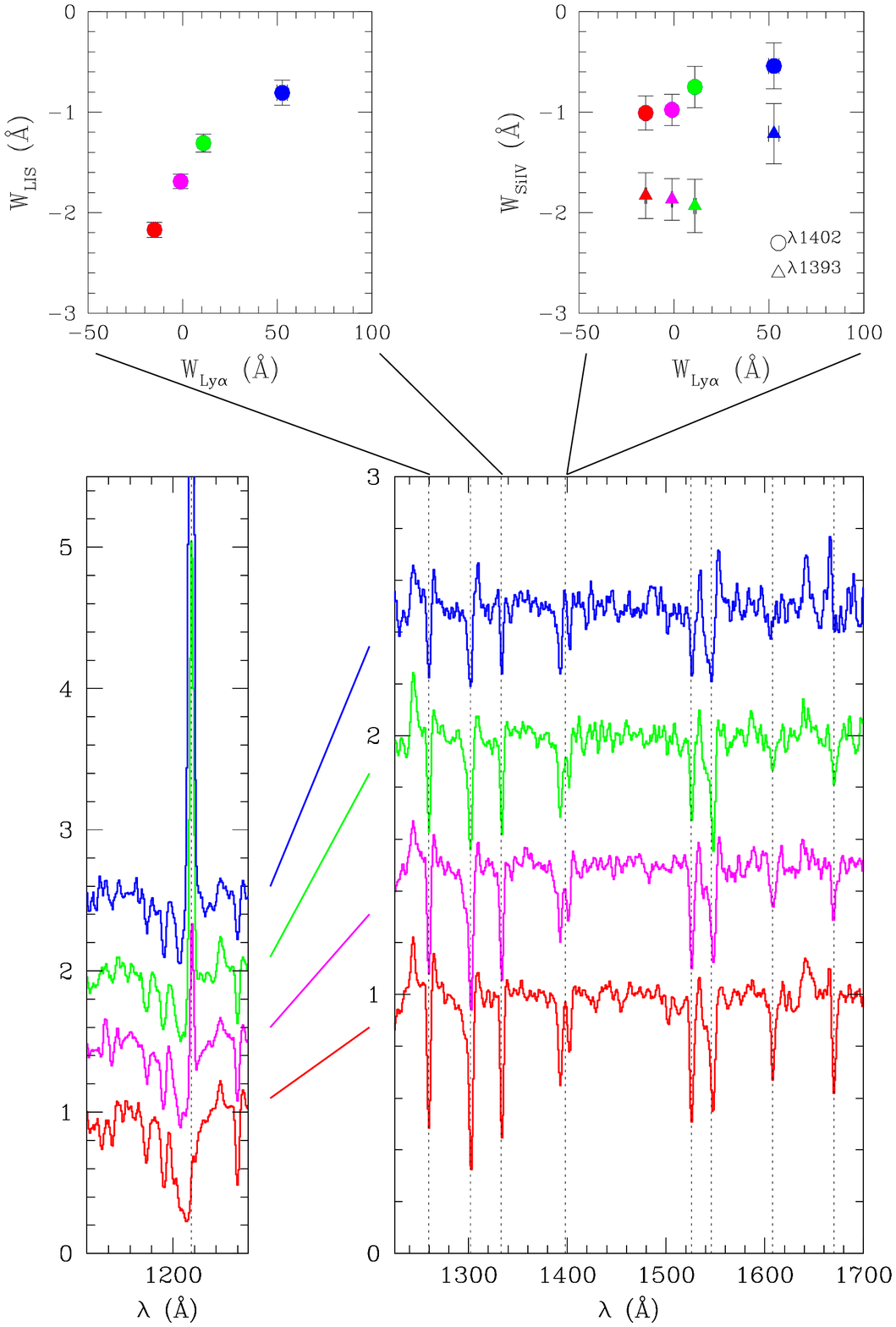}}
\figcaption{Bottom: A sequence of 4 continuum-normalized
composite spectra, constructed from the
4 quartiles of LBG spectroscopic sample grouped according to Ly$\alpha$
equivalent width. The spectra have been offset
by regular vertical intervals for easier viewing,
in order of increasing $W_{{\rm Ly}\alpha}$.
The lower left-hand panel zooms in on the region near Ly$\alpha$, while
the right-hand panel focuses on the region redwards of Ly$\alpha$,
where the strongest features are blue-shifted low-ionization and
high-ionization interstellar absorption features associated
with large-scale outflows of interstellar material. Top: The
behavior of low- and high-ionization interstellar absorption
lines as a function of Ly$\alpha$ equivalent width. These plots
confirm quantitatively what the bottom panels indicate visually:
the average low-ionization absorption equivalent width,
$W_{{\rm LIS}}$, decreases
dramatically as $W_{{\rm Ly}\alpha}$ varies from strong absorption to strong
emission,
while the high-ionization Si~IV absorption equivalent width, $W_{{\rm SiIV}}$,
remains roughly constant (except in the quartile of the
sample with strong Ly$\alpha$ emission, in which $W_{{\rm HIS}}$ is
slightly weaker).
\label{fig:lya}
}
\end{figure*}

\begin{figure*}
\centerline{\epsfxsize=18cm\epsffile{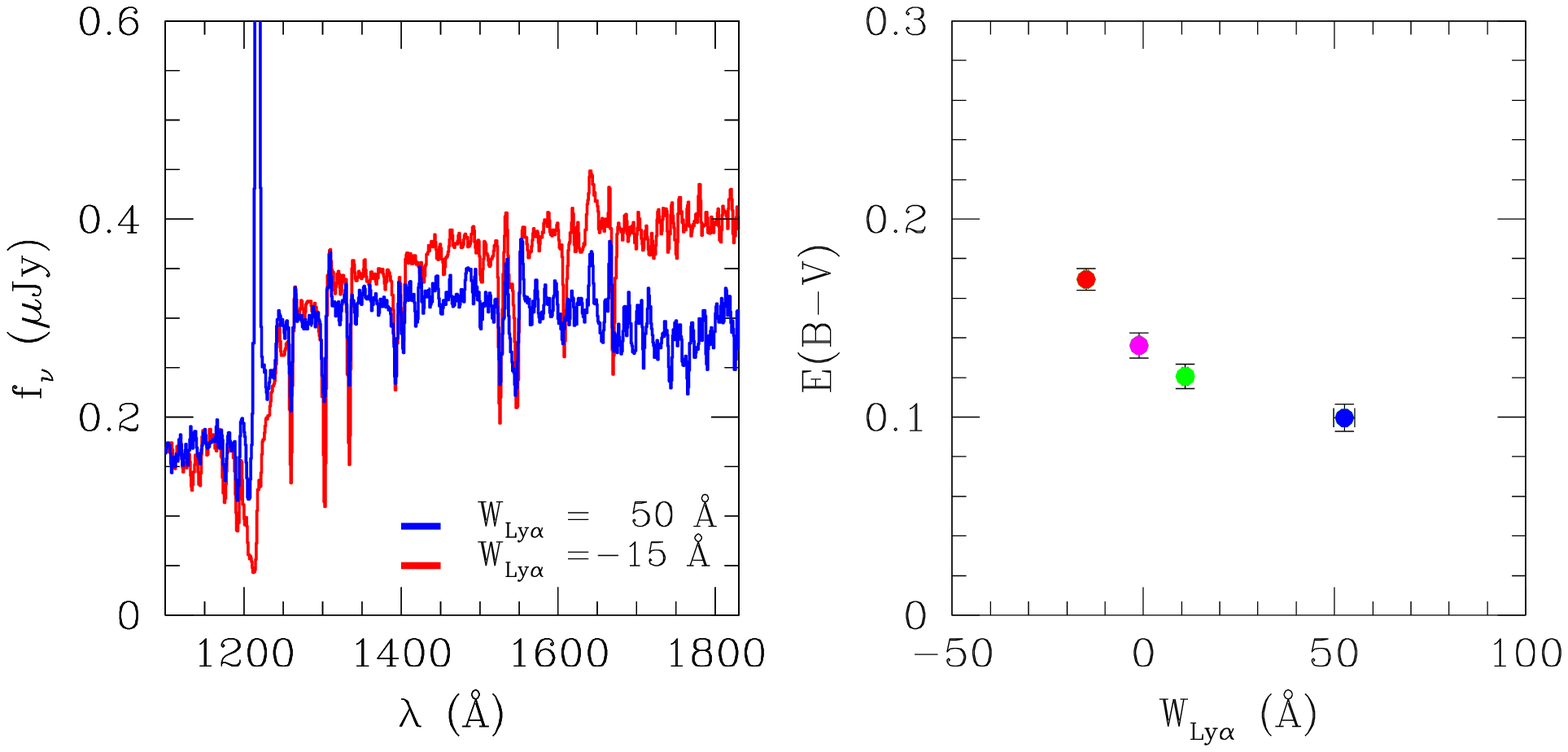}}
\figcaption{The dependence of UV continuum shape on Ly$\alpha$ equivalent
width. (left) Plotted in blue, the composite spectrum of the quartile
of galaxies with the strongest Ly$\alpha$ emission and the weakest
low-ionization interstellar absorption lines is also significantly
bluer in spectral slope than the composite spectrum of the quartile of
galaxies with the strongest Ly$\alpha$  absorption and strongest
low-ionization interstellar absorption lines, which is plotted in red. (right)
The visual difference between the two extreme spectra in the left-hand
panel is confirmed by the mean $E(B-V)$ for each of the four Ly$\alpha$
quartiles, which decreases as a function of increasing Ly$\alpha$
emission.
\label{fig:lyaebv}
}
\end{figure*}

At high redshift, a representative subsample of LBG 
Ly$\alpha$ equivalent widths \citep{steidel2000} show that
only $\sim 20-25 \%$ of LBGs at $z\sim 3$
at a given UV luminosity have Ly$\alpha$ emission lines strong enough
to be flagged as narrow-band excess objects, given typical high-redshift
Ly$\alpha$ emission line search sensitivities \citep{cowie1998,hu1998}.
As discussed below, LBGs with such strong emission have
certain properties which make them distinct from the population of
LBGs as a whole. Narrowband searches
are frequently used to probe redshifts higher than $z\sim3$, so
it is important to understand how the properties of narrow-band-selected objects
relate to those of general population of star-forming galaxies
at similar epochs.

Extending the work of \citet{steidel2000},
we have now measured Ly$\alpha$ equivalent widths for the entire LBG
spectroscopic sample.\footnote{There are actually 17 galaxies
in the spectroscopic sample
for which we did not measure the Ly$\alpha$ equivalent width.
Most of these galaxies are towards the low-redshift end of the
LBG redshift distribution such that Ly$\alpha$ was
not contained in the spectral format of the LRIS detector.
Additionally there are galaxies for which Ly$\alpha$
fell on top of some defect in the two dimensional spectral image,
precluding us from robustly measuring an equivalent width.}
Figure~\ref{fig:lyatothist} shows the rest-frame 
$W_{{\rm Ly}\alpha}$ distribution, which has a median of 
$\sim 0$ \AA\, and ranges from $< -50$ \AA\ in absorption to
$> 100$ \AA\ in emission. The precise measurement of absorption equivalent
widths is difficult, given the broad nature of Ly$\alpha$ absorption,
the presence of strong metal absorption lines at nearby wavelengths
(Si~III~$\lambda 1206$ and Si~II~$\lambda\lambda 1190,1193$), 
and the uncertainty in determining the 
continuum level on the blue side of the line due to Ly$\alpha$
forest absorption. The typical uncertainties in absolute equivalent widths
can be larger than $\sim 50 \%$ for absorption profiles, while
emission equivalent widths are better determined, with $\sim 30 \%$
uncertainties. The Ly$\alpha$ profiles for $38 \%$ of the galaxies 
consist of a combination of both emission and absorption. 
For galaxies characterized by this type of profile, 
two equivalent widths were measured: a total equivalent
width representing the sum of the emission and absorption components,
and also an emission equivalent width representing the emission alone.
The intrinsic rest-frame UV color, and implied dust extinction,
$E(B-V)$, were computed by correcting the observed $G-{\cal R}$ color
for both the total Ly$\alpha$ equivalent width, 
and also the average IGM absorption along the line of sight.

In order to understand the factors which determine the 
emergent Ly$\alpha$ profile in LBGs,
we binned the sample of 794 individual spectra with Ly$\alpha$ measurements
according to $W_{{\rm Ly}\alpha}$ into 4 subsamples of 
equal size. A composite spectrum was constructed
from each subsample. The large number of
galaxies contained in each bin insured that the 
resulting composite spectrum had very high
S/N, yet there were also enough separate bins that 
the Ly$\alpha$ properties of the subsamples at each extreme
were quite distinct: the bin at one extreme was characterized by objects with 
strong absorption, while the other extreme bin was 
dominated by objects with large emission equivalent widths.
We obtain several striking results from binning the sample
according to $W_{{\rm Ly}\alpha}$ and measuring the features
of the resulting composite spectra. These measurements
are summarized in Table~\ref{tab:lya}.

As shown in Figure~\ref{fig:lya},
the average absorption equivalent width of the 6 strongest
low-ionization interstellar lines, $W_{{\rm LIS}}$, decreases in strength
by almost a factor of three as $W_{{\rm Ly}\alpha}$ varies from $-15$~\AA\
in absorption to $50$~\AA\ in emission. Even in the spectrum
with the strongest Ly$\alpha$ emission and weakest absorption lines,
the ratio of Si~II~$\lambda 1260$ and $\lambda 1526$ absorption 
equivalent widths is roughly unity. This indicates that the lines
are still saturated and that the change in low-ionization
equivalent width across the sample is not primarily due 
to a change in metallicity. There is
no significant change in the absorption strength of the high-ionization
interstellar absorption lines, $W_{{\rm SiIV}}$ and $W_{{\rm CIV}}$, as 
$W_{ {\rm Ly}\alpha}$ varies from strong absorption to strong emission,
except in the quartile
of galaxies with $W_{{\rm Ly}\alpha} \geq 20$ \AA, which has line strengths
smaller than the other three quartiles by 50\% 
(still a much less significant change than what is
seen in the low ions). Additionally,
the Si~IV doublet ratio is consistent with the transition
being optically thin in all four subsamples. 

Another important result is that the UV continuum slope becomes
significantly bluer as $W_{{\rm Ly}\alpha}$ varies from strong
absorption to strong emission. This result is most dramatically
illustrated by the left-hand box of Figure~\ref{fig:lyaebv}, which shows
how the strong-absorption and strong-emission composite spectra,
normalized at 1100 \AA, diverge at longer wavelengths. 
To support this visual picture, the right-hand box of Figure~\ref{fig:lyaebv}
shows that the mean $E(B-V)$ value for each
of the four subsamples decreases as a function of 
increasing $W_{{\rm Ly}\alpha}$ emission. 

Furthermore, Figure~\ref{fig:delv_vs_lyaem} shows that
with increasing Ly$\alpha$ emission strength, the kinematic
offset implied by the relative redshifts of Ly$\alpha$ emission
and low-ionization interstellar absorption lines 
decreases monotonically from $\Delta v_{{\rm em-abs}}=800$~\kms\ 
to $\Delta v_{{\rm em-abs}}=480$~\kms. 
The kinematic offsets were measured
directly from the composite spectra. Weak Ly$\alpha$ emission is
detected even in the spectrum composed of the quartile of
galaxies with strong absorption and no emission on an individual
basis. Therefore, a kinematic offset
can be estimated even for this absorption sample.

\begin{inlinefigure}
\centerline{\epsfxsize=9cm\epsffile{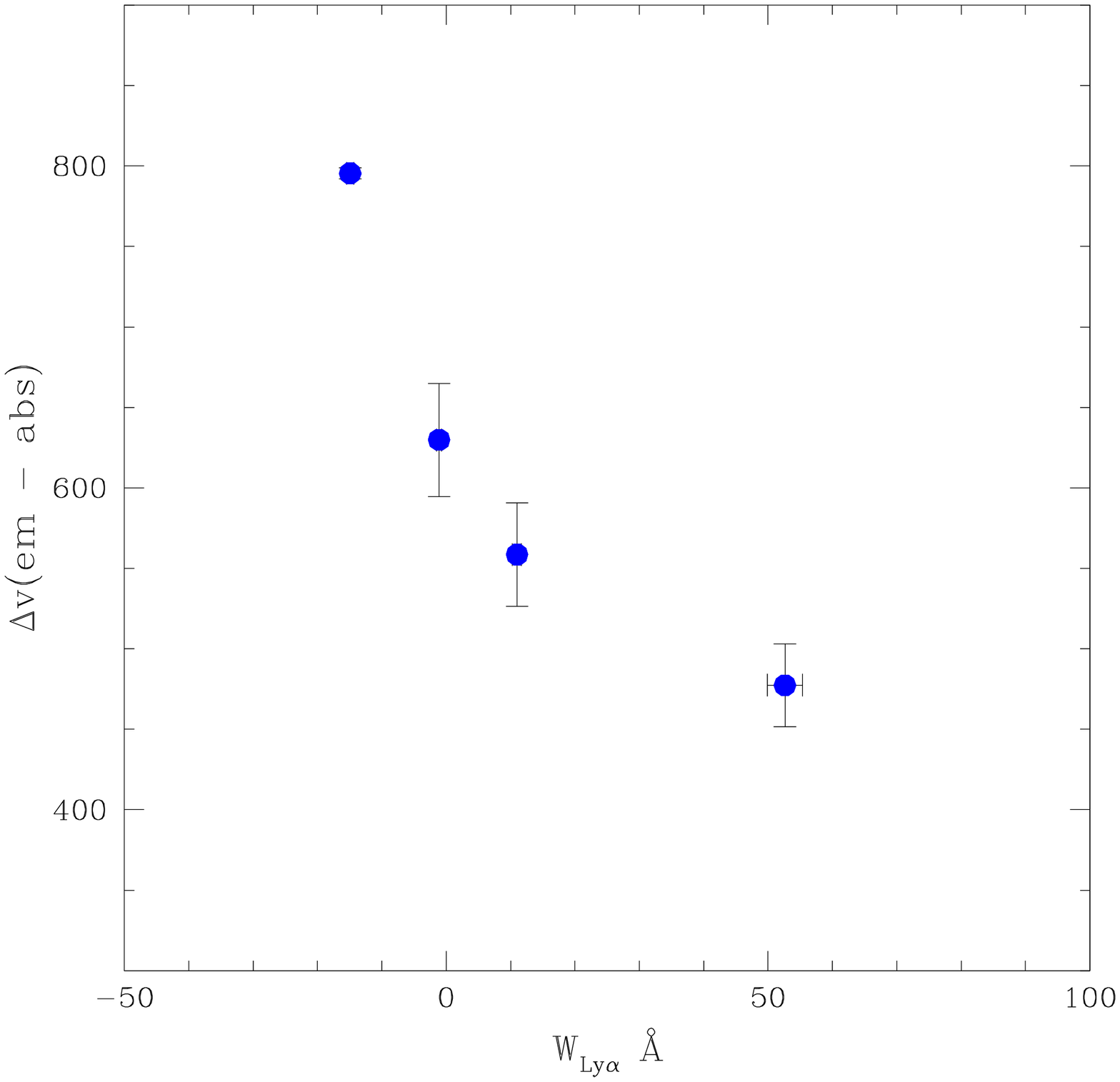}}
\figcaption{The dependence of $\Delta v_{{\rm em-abs}}$ on Ly$\alpha$ equivalent
width. Velocity offsets between Ly$\alpha$ emission and the strongest
low-ionization interstellar absorption lines were measured directly
from the composite spectra of each of the four Ly$\alpha$ subsamples.
As Ly$\alpha$ emission strength increases, the kinematic offset decreases.
The error bars on the velocity offsets represent the scatter among the
velocities of the different individual interstellar absorption features.
\label{fig:delv_vs_lyaem}
}
\end{inlinefigure}

\begin{figure*}
\centerline{\epsfxsize=18cm\epsffile{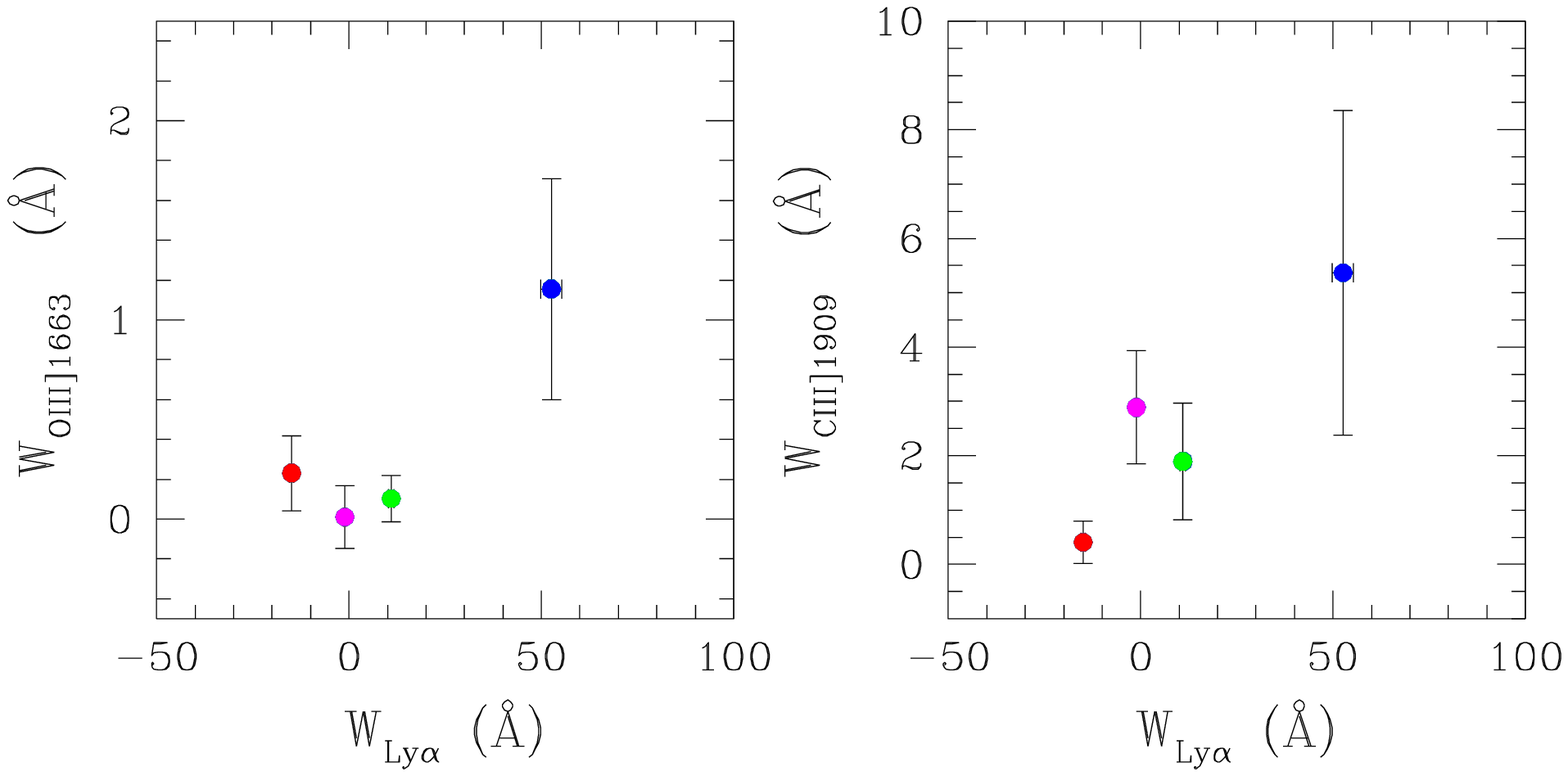}}
\figcaption{The dependence of O~III] and C~III] nebular emission strength on
Ly$\alpha$ equivalent width. The composite spectrum constructed
from the quartile
of galaxies with $W_{{\rm Ly}\alpha}\geq 20$~\AA\ has significantly 
stronger nebular emission than the rest of the sample.
\label{fig:lyaem_co}
}
\end{figure*}

Now we focus on the composite spectrum of the quartile of
galaxies with $W_{{\rm Ly}\alpha} \geq 20$ \AA,
which is strong enough to be selected by current narrow-band emission line
search techniques \citep{hu1998,rhoads2000}.
This composite spectrum has significantly
stronger C~III] and O~III] nebular emission line strengths than the
composite spectra with weaker Ly$\alpha$ emission 
(see Figure~\ref{fig:lyaem_co}). 
In contrast to the strong low-ionization and
high-ionization absorption lines, which probe conditions in the
foreground, outflowing interstellar medium, the nebular lines act as
probes of H~II regions where stars are forming. These strength of these
features should be independent of conditions in the foreground gas and
associated orientation effects, 
and more sensitive to the nebular temperature and metallicity, 
and the nature of the stellar population.

The sample with $W_{{\rm Ly}\alpha} \geq 20$ \AA\ in emission
is much more complete as a function of ${\cal R}$ magnitude
than the spectroscopic sample as a whole, since
we don't fail spectroscopically on objects with observed equivalent widths of
$W_{{\rm Ly}\alpha,{\rm obs}} \geq 80$~\AA\ (the same as rest-frame
$W_{{\rm Ly}\alpha} \geq 20$~\AA\ at $z\sim 3$). 
Therefore, this sample is ideal
to test for the dependence of Ly$\alpha$ emission strength on apparent UV
luminosity. We divide the galaxies with 
$W_{{\rm Ly}\alpha} \geq 20$ \AA\ into three
groups according to ${\cal R}$ magnitude, and compare the
three distributions of Ly$\alpha$ equivalent widths. The mean
$W_{{\rm Ly}\alpha}$ increases towards fainter magnitudes,
and a one-dimensional K-S test shows that there is only
a 0.1\% chance that the brightest distribution of $W_{{\rm Ly}\alpha}$ 
could be drawn from the same parent population as the
faintest sample. Since any individual $W_{{\rm Ly}\alpha}$ 
measurement has a significant uncertainty, especially at faint magnitudes,
we check this statistical result by constructing composite spectra
of the faintest third and brightest third of the subsample 
with $W_{{\rm Ly}\alpha} \geq 20$ \AA. Measured
directly from the bright and faint composite spectra, which are shown 
in Figure~\ref{fig:lyaem4_faintbright}, the Ly$\alpha$ equivalent
widths are $W_{{\rm Ly}\alpha}({\rm bright})=43$~\AA\ and 
$W_{{\rm Ly}\alpha}({\rm faint})=65$~\AA.
Even in the faintest subsample, which has the strongest Ly$\alpha$ emission
of any LBG subsample, the average value of $W_{{\rm Ly}\alpha}({\rm faint})=65$~\AA\ 
is still well below what is expected for Case B recombination,
given a stellar population continuously forming stars with a Salpeter
IMF for $\leq 1$~Gyr \citep{charlot1993}.
Therefore, the escape of Lyman continuum photons from 
LBG star-forming regions is not ruled out by the strength of
Ly$\alpha$ emission, and the escape fraction from LBGs is probably determined
much more by the covering fraction of outflowing neutral clouds at larger radii
\citep{heckman2001a,steidel2001}.

\begin{figure*}
\centerline{\epsfxsize=18cm\epsffile{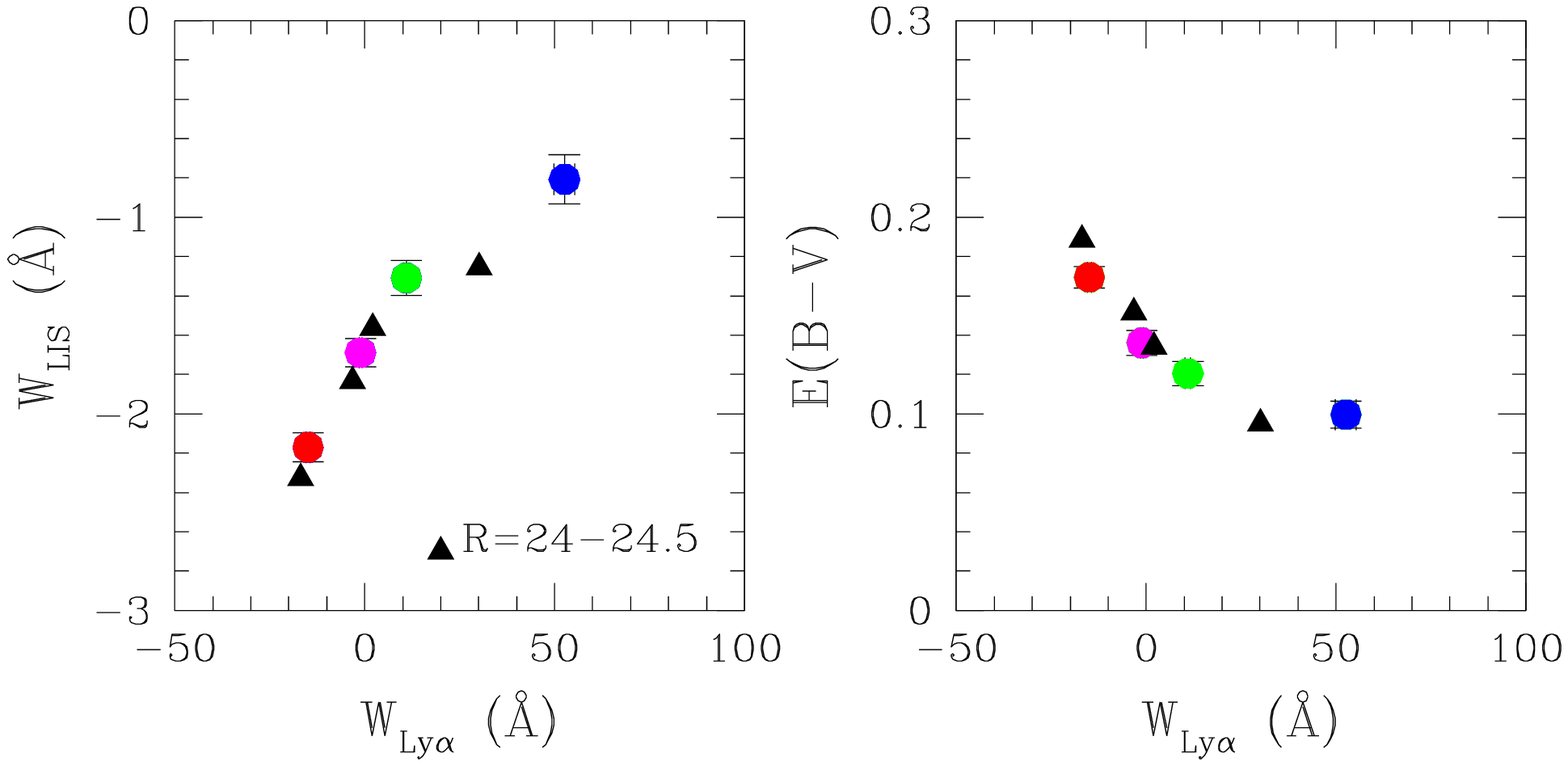}}
\figcaption{Strong Ly$\alpha$-dependent trends at fixed ${\cal R}$ magnitude.
Galaxies with ${\cal R}=24 - 24.5$ are divided into four subsamples
based on Ly$\alpha$ equivalent width, and black triangles indicate
measurements from the corresponding composite spectra.
{\bf Left:} $W_{ {\rm Ly}\alpha}$ vs. $W_{{\rm LIS}}$. Colored symbols
are as in Figure~\ref{fig:lya}.
Galaxies with ${\cal R}=24 - 24.5$ follow the same trend as the
total sample. {\bf Right:} $E(B-V)$ vs. $W_{ {\rm Ly}\alpha}$. Colored
symbols are as in Figure~\ref{fig:lyaebv}, and galaxies with
${\cal R}=24 - 24.5$ obey the same correlation.
\label{fig:comp_allbright}
}
\end{figure*}

At least in part due to selection effects,
the quartile of the sample with the strongest Ly$\alpha$ emission
is significantly fainter on average than the quartile with the 
strongest Ly$\alpha$ absorption. The mean apparent magnitude
of the strong absorption quartile is ${\cal R}=24.44$, 
while the mean apparent magnitude of the strong emission quartile
is ${\cal R}=24.85$. Since grouping the sample by
$W_{{\rm Ly}\alpha}$ also segregates to a certain degree by ${\cal R}$
magnitude, it is important to show that the significant spectroscopic
trends hold at fixed apparent UV luminosity, especially at an apparent magnitude
where all spectroscopic ``types'' are sampled equally. Therefore,
we zero in on the range ${\cal R}=24-24.5$, and bin the 207 galaxies
in this magnitude range into four equal subsamples according to 
$W_{{\rm Ly}\alpha}$. As shown in Figure~\ref{fig:comp_allbright},
the same strong Ly$\alpha$-dependent trends hold
at fixed magnitude: as Ly$\alpha$ changes from strong absorption
to strong emission, the average $W_{{\rm LIS}}$ significantly decreases, and the
UV continuum slope becomes significantly bluer; 
simultaneously, the average $W_{{\rm SiIV}}$ and $W_{{\rm CIV}}$ 
remain roughly independent of $W_{{\rm Ly}\alpha}$.
The fixed-magnitude Ly$\alpha$
sample is roughly 4 times smaller than the total Ly$\alpha$ sample,
and therefore weak nebular emission lines are not detected with much
significance. The same significant Ly$\alpha$-dependent trends
are found to apply as well for
fixed-magnitude bins of ${\cal R}=24.5-25$ and ${\cal R}=25-25.5$.
LBG photometric and spectroscopic
selection effects should not introduce a correlation 
between $W_{{\rm Ly}\alpha}$ and $W_{{\rm LIS}}$, 
or between $W_{{\rm Ly}\alpha}$ and UV continuum slope. 
The consistency of the trends at different ${\cal R}$ 
magnitudes provides direct evidence that
magnitude-dependent selection effects have not 
spuriously introduced correlations; 
we consider the physical significance of these trends
in section~\ref{sec:discussion}.

\subsection{UV Color Dependences}
\label{sec:trends_dust}
At $z\sim 3$, the intrinsic UV color, $(G-{\cal R})_0$, 
is determined by correcting
the observed $G-{\cal R}$ color for the observed Ly$\alpha$ equivalent width, 
and the redshift-dependent average Ly$\alpha$ forest 
opacity \citep{madau1995}. 
A stellar population continuously forming stars
(an appropriate model of the UV colors and spectra of most LBGs)
has an unreddened UV spectral energy distribution shape that remains
fairly constant for ages between 10 Myr and 1 Gyr.
\footnote{If the spectral energy distribution
is parameterized as $F_{\lambda} \propto \lambda^{\beta}$, $\beta$
only ranges between $-2.6$ and $-2.1$ for continuous star-formation
ages between 10 Myr and 1 Gyr \citep{leitherer1999}. } 
Therefore, the range of LBG $(G-{\cal R})_0$ colors can
be parameterized in terms of $E(B-V)$, the amount of dust extinction
reddening the intrinsic UV continuum.
In order to convert $(G-{\cal R})_0$ to $E(B-V)$, we assume
for the unreddened stellar population a 300 Myr old
continuous star-formation model -- 
the median age for a subsample of LBGs with optical-IR colors 
\citep{shapley2001}. We also assume
a relatively grey form for the dust extinction law, which 
accurately describes the dust properties of 
local starburst galaxies \citep{calzetti1997,calzetti2000,meurer1999}.
While the evidence at high redshift is still
preliminary, multi-wavelength
observations show that the starburst extinction law 
predicts the properties of LBGs at X-ray and radio
wavelengths much better than extinction laws
appropriate for the SMC or Ultra-Luminous Infrared Galaxies (ULIRGs)
\citep{seibert2002,meurer1999,nandra2002}.

Based on the range of UV colors of LBGs, and the types of
assumptions described above, \citet{adelberger2000}
find that the UV luminosities of LBGs are attenuated on average
by a factor of $\sim 7$ due to dust extinction, and that this
factor ranges from $0 - 100$. Additionally, LBGs
with more dust extinction have higher intrinsic star-formation
rates and younger stellar populations
\citep{adelberger2000,shapley2001}.
The role played by star-formation induced outflows 
in the evolution of the dust properties of LBGs was
addressed qualitatively by \citet{shapley2001}, but without the benefit
of detailed spectroscopic analysis. Here, we fold in spectroscopic
information to gain a more complete picture of the relationship
between dust extinction and outflows.

\begin{figure*}
\centerline{\epsfxsize=18cm\epsffile{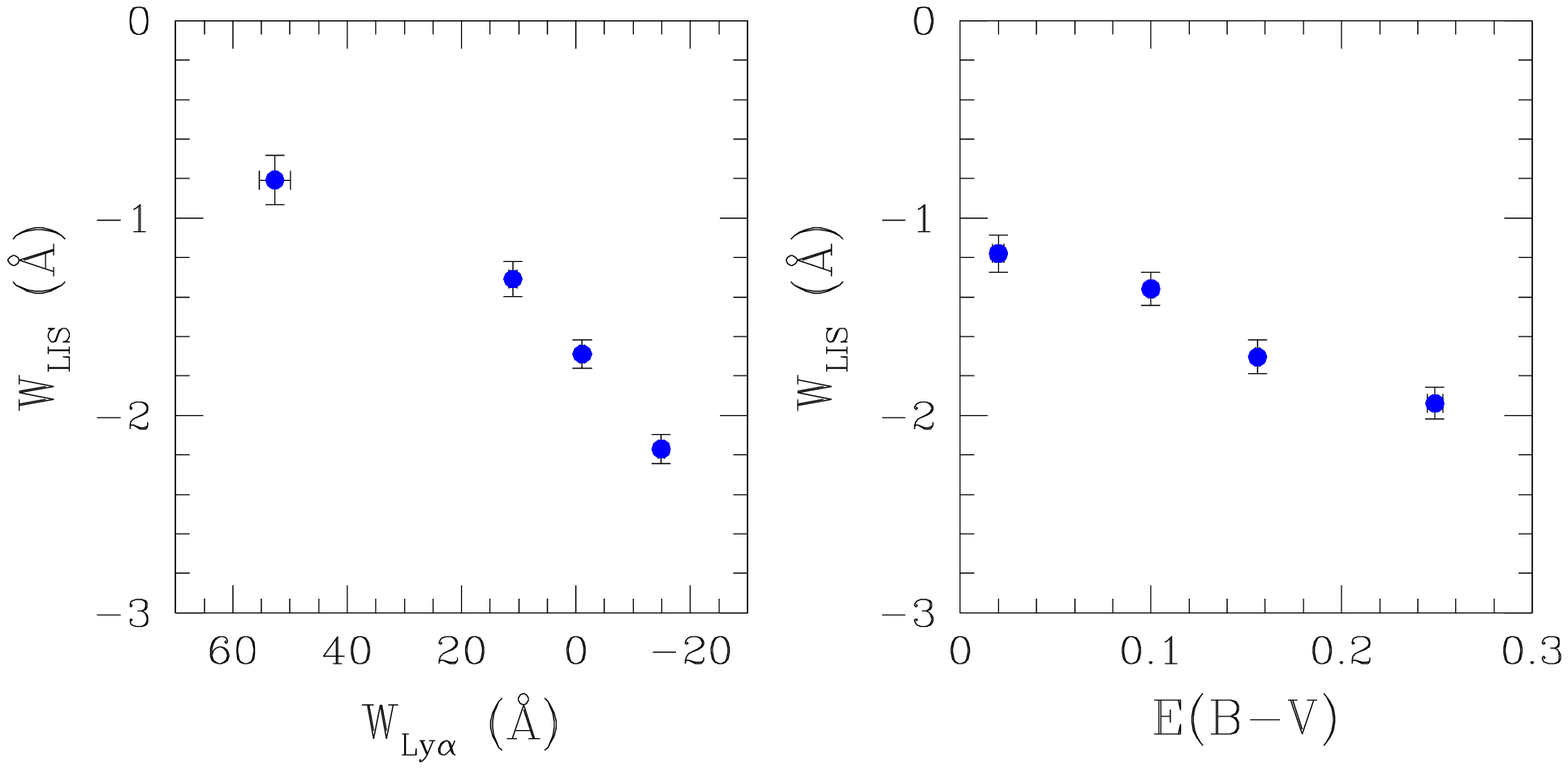}}
\figcaption{The dependence of low-ionization interstellar absorption
strength, $W_{{\rm LIS}}$, on both Ly$\alpha$ equivalent width and
dust reddening. These plots show the results of dividing LBG spectroscopic
sample into quartiles, according to either Ly$\alpha$ equivalent
width (left) or $E(B-V)$ (right). While $W_{{\rm LIS}}$ depends strongly on both
$W_{{\rm Ly}\alpha}$ and $E(B-V)$, there is more variance in $W_{{\rm LIS}}$
when the sample is sorted by $W_{{\rm Ly}\alpha}$. This suggests a stronger
statistical link between $W_{{\rm LIS}}$ and $W_{{\rm Ly}\alpha}$, though
the correlation between $W_{{\rm LIS}}$ and $E(B-V)$ is strong enough
that there may be a direct physical connection among all three
quantities.
\label{fig:wis_wlya_ebv}
}
\end{figure*}

As presented in section~\ref{sec:trends_lya},
when Ly$\alpha$ varies from strong absorption to strong emission:
1) the strength of the low-ionization 
interstellar absorption lines decreases
2) the dust-sensitive UV continuum becomes bluer. 
Given the strong interdependence of $W_{{\rm Ly}\alpha}$, $W_{{\rm LIS}}$,
and $E(B-V)$, it is interesting to consider which 
correlations among the three variables are stronger,
and therefore fundamental, and which arise as a by-product of the
more fundamental correlations.
When the sample is divided into $E(B-V)$
quartiles, we confirm the same interdependence of $W_{{\rm Ly}\alpha}$,
$W_{{\rm LIS}}$, and $E(B-V)$. Figure~\ref{fig:wis_wlya_ebv} also
demonstrates that $W_{{\rm LIS}}$ is more strongly
dependent on $W_{{\rm Ly}\alpha}$ than on $E(B-V)$.
Though $W_{{\rm LIS}}$ becomes significantly weaker (a factor of 1.6)
as $E(B-V)$ decreases from the reddest to the bluest quartile,
there is a more significant change in $W_{{\rm LIS}}$ (a factor of 2.7)
as $W_{{\rm Ly}\alpha}$ varies from strong absorption to strong emission.
The variance in $W_{{\rm LIS}}$ as a function
of $E(B-V)$ cannot be accounted for entirely
by the change in $W_{{\rm Ly}\alpha}$, however; $W_{{\rm Ly}\alpha}$
only changes from $W_{{\rm Ly}\alpha}({\rm red})=5$ to $W_{{\rm Ly}\alpha}({\rm blue})=20$~\AA\ 
as $E(B-V)$ decreases from the
reddest to the bluest quartile (in contrast to the change 
from $W_{{\rm Ly}\alpha}({\rm abs})=-15$~\AA\ 
to $W_{{\rm Ly}\alpha}({\rm em})=50$~\AA\ 
in the four Ly$\alpha$ subsamples). 
According to the relationship
between $W_{{\rm Ly}\alpha}$ and $W_{{\rm LIS}}$ presented in 
section~\ref{sec:trends_lya}, we would predict a smaller
difference between $W_{{\rm LIS}}({\rm red})$ and $W_{{\rm LIS}}({\rm blue})$
to accompany the difference between $W_{{\rm Ly}\alpha}({\rm red})$ and 
$W_{{\rm Ly}\alpha}({\rm blue})$.  We therefore infer a 
direct statistical link between the absorbing gas and the
reddening of the UV continuum, as well as between the
absorbing gas and the Ly$\alpha$ profile.

Due to selection effects, the bluest quartile is fainter 
($\langle {\cal R} \rangle= 24.75$) than
the reddest quartile ($\langle {\cal R}\rangle= 24.51$).
To remove the effects of luminosity, we again
examine whether the trends still hold when we look in fixed magnitude ranges:
${\cal R}=24-24.5, 24.5-25, 25-25.5$. The strength of
correlations with $E(B-V)$ is independent of magnitude,
though there is a systematic offset in the ``zeropoints'',
such that galaxies of a given $E(B-V)$ have stronger Ly$\alpha$ 
and weaker low-ionization interstellar absorption lines at
fainter ${\cal R}$ magnitudes. 
The qualitative similarity of the $E(B-V)$
correlations at all magnitudes confirms their 
connection to actual physical conditions,
rather than to luminosity-dependent selection effects.
For LBGs in general, and especially
for LBGs in a fixed magnitude range,
the dependence of spectroscopic properties on dust extinction is 
roughly equivalent to the dependence on (slightly model-dependent)
dust-corrected star-formation rate. 
Therefore, galaxies with redder UV continua, stronger interstellar absorption
lines, and weaker Ly$\alpha$ emission, also have larger star-formation rates.

\begin{inlinefigure}
\centerline{\epsfxsize=9cm\epsffile{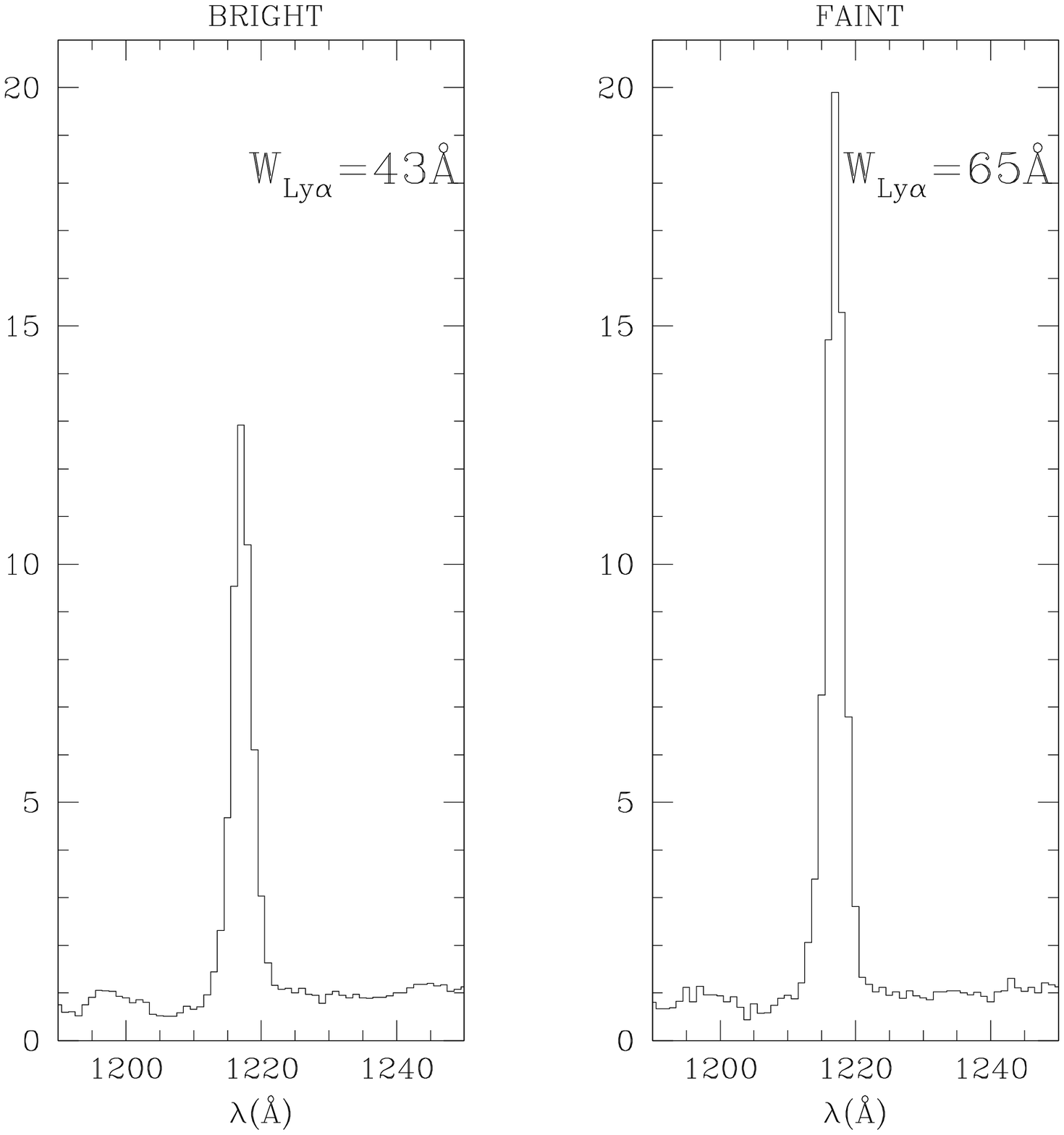}}
\figcaption{The dependence of $W_{{\rm Ly}\alpha}$ on apparent UV luminosity.
Restricting the comparison to galaxies with $W_{{\rm Ly}\alpha} \geq 20$~\AA,
which should not be prone to magnitude-dependent selection effects,
we construct composite spectra for faint (right panel) and
bright (left panel) galaxies. Equivalent widths
are measured directly from the continuum-normalized
composite spectra and indicate that
fainter galaxies have larger Ly$\alpha$
emission equivalent widths than brighter galaxies.
\label{fig:lyaem4_faintbright}
}
\end{inlinefigure}

\subsection{Kinematic $\Delta v_{{\rm em-abs}}$ Dependences}
\label{sec:trends_delv}
The multi-wavelength properties of ``superwinds'' 
in local starburst galaxies have been studied with spatially resolved
imaging and spectroscopy \citep{heckman2000,heckman2001b}.
Also, the hydrodynamics of different phases of the outflowing
ISM have been modeled numerically \citep{strickland2000}.
Due to their small angular sizes and faint magnitudes, 
there is very little two-dimensional
morphological information or spatially resolved spectroscopy of $z\sim3$ 
star-forming galaxies.  However, the one dimensional 
spectroscopic properties of LBGs indicate the presence of 
large velocity fields, consistent with the outflow conditions 
seen in local starbursts. In individual spectra, 
the Ly$\alpha$ and interstellar redshift differ, with an average
offset of $\Delta v_{{\rm em-abs}}  = 650$ \kms, ranging from less than
$0$~\kms to greater than $1000$~\kms
(sections~\ref{sec:z},~\ref{sec:features_outflow}, Figure~\ref{fig:delvhist}). 
The composite spectrum of the total LBG spectroscopic sample,
at rest with respect to stars and H~II regions, has
$\Delta v({\rm Ly}\alpha) = +360$~\kms and $\Delta v({\rm LIS}) = -150$~\kms,
implying $\Delta v_{{\rm em-abs}} = 510$~\kms. This offset is
smaller than the average $\Delta v_{{\rm em-abs}}$ measured from 
individual spectra, but
we actually expect the total sample to have a smaller
average $\Delta v_{{\rm em-abs}}$ than the subsample of objects
with both emission and absorption redshifts.
This is because the total composite spectrum
has stronger Ly$\alpha$ emission than the composite spectrum
constructed only from objects with individual $\Delta v_{{\rm em-abs}}$ measurements,
and section~\ref{sec:trends_lya} demonstrated
a correlation between $W_{{\rm Ly}\alpha}$ and
$\Delta v_{{\rm em-abs}}$.

Since the $\Delta v_{{\rm em-abs}}$ distribution for the 323 galaxies
with both Ly$\alpha$ emission and interstellar absorption redshifts
is fairly broad, it 
affords sufficient dynamic range for us to consider how other
galaxy properties depend on outflow kinematics.
Studies of the radiative transfer of
Ly$\alpha$ in local starburst galaxies \citep{kunth1998}
stress the importance of ISM kinematics on the emergent
Ly$\alpha$ profile. With our sample, we can directly test
the link between Ly$\alpha$ equivalent width and ISM
kinematics. Also, $\Delta v_{{\rm em-abs}}$ should be
related to the velocity FWHM of the outflowing gas.
The FWHM of blue-shifted gas
will determine the range of wavelengths of Ly$\alpha$ 
photons that are scattered. 
Blue-shifted gas with a larger velocity dispersion
will scatter Ly$\alpha$ photons with a larger range of wavelengths, 
pushing the observed centroid of Ly$\alpha$ emission to longer wavelengths,
and causing a larger apparent $\Delta v_{{\rm em-abs}}$  for the same 
outflow velocity (see Figure 3 of Adelberger \et 2002a). 
Ideally, we would like to examine the relationship between
$\Delta v_{{\rm em-abs}}$ and the intrinsic FWHM of the 
strong low-ionization interstellar absorption lines. We are
hampered in this effort by the uncertainty in the effective spectral
resolution of our composite spectra. Even when all of the spectra are 
shifted into the absorption rest frame to minimize the effects of redshift
errors, the FWHM measurements are quite noisy and difficult to interpret.
However, a comparison between the much better determined $W_{{\rm LIS}}$
and $\Delta v_{{\rm em-abs}}$ may help to isolate the kinematic
contribution to the saturated absorption equivalent widths.

\begin{figure*}
\centerline{\epsfxsize=18cm\epsffile{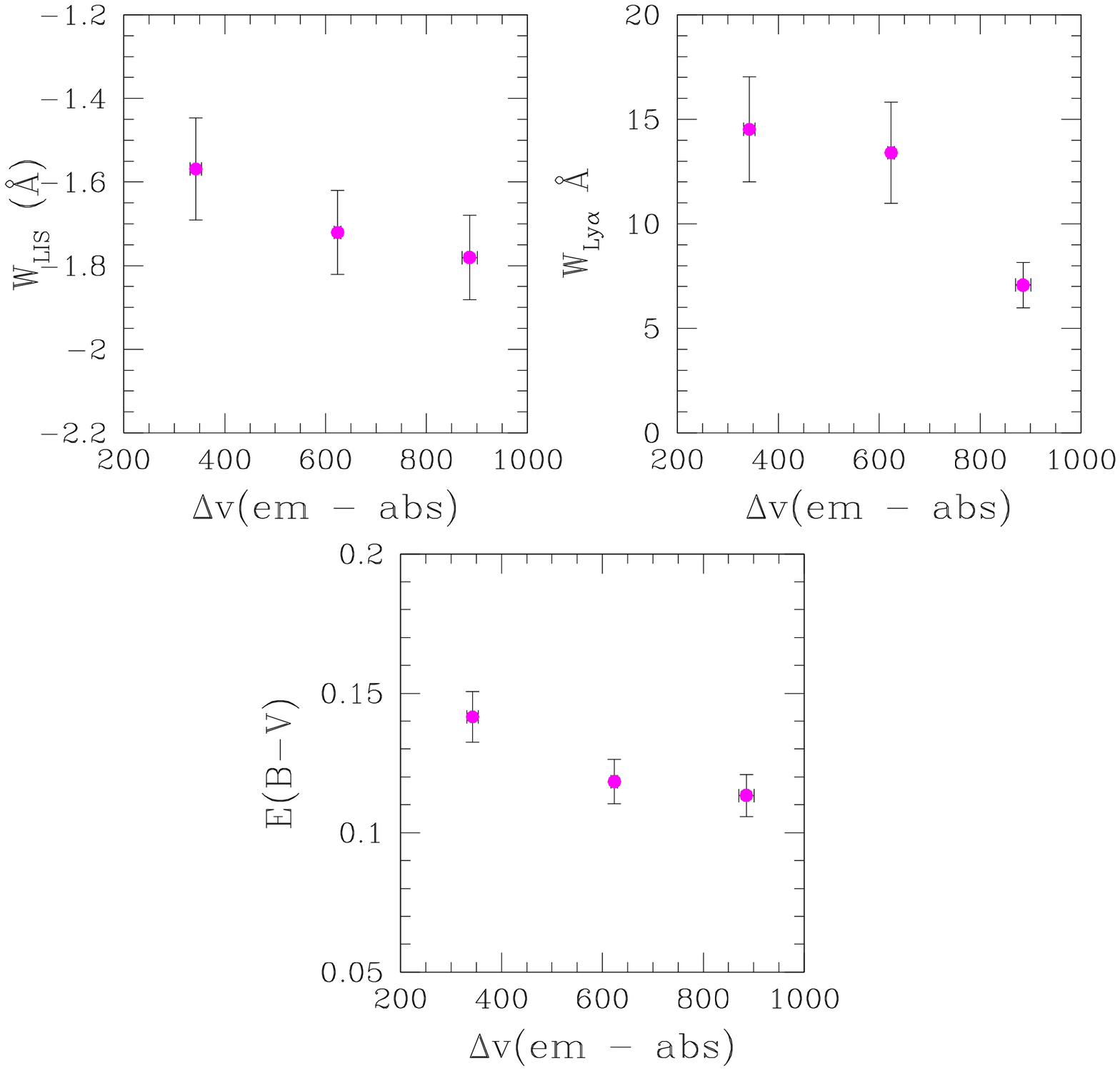}}
\figcaption{Dependences of $W_{{\rm LIS}}$, $W_{{\rm Ly}\alpha}$, and $E(B-V)$ on
$\Delta v_{{\rm em-abs}}$. These plots show the
average $W_{{\rm LIS}}$, $W_{{\rm Ly}\alpha}$, and
$E(B-V)$ values measured from each of the 3 subsets of galaxies
sorted by $\Delta v_{{\rm em-abs}}$, and their associated composite spectra.
$W_{{\rm LIS}}$ and $E(B-V)$ are not significantly dependent on $\Delta v$,
while $W_{{\rm Ly}\alpha}$ is smaller for the subsample with the highest
$\Delta v$ ($\geq 800$~\kms)). This trend is consistent with the
fact that the $\Delta v$ measured in the composite spectrum
with the strongest Ly$\alpha$ absorption is larger than the $\Delta v$
measured in the spectrum with the strongest Ly$\alpha$ emission. However,
the change in $W_{{\rm Ly}\alpha}$ with $\Delta v$ is small compared with the
total variance of Ly$\alpha$ across the whole LBG spectroscopic sample.
\label{fig:delv_sum}
}
\end{figure*}

The galaxies with $\Delta v_{{\rm em-abs}}$ measurements are
divided into three subsamples, with 
$\langle \Delta v_{{\rm em-abs}} \rangle=340,620,890$~\kms,
and a composite spectrum is constructed for each subsample. 
Due to the requirement that both emission and absorption features
are present in order to compute $\Delta v_{{\rm em-abs}}$, the kinematic 
sample is biased towards brighter objects. 
However, there should be no biases which prevent us from including
certain types of objects as a differential function of $\Delta v_{{\rm em-abs}}$.
Consistent with the results of section~\ref{sec:trends_lya},
we find  that $W_{{\rm Ly}\alpha}$  weakly dependent on 
$\Delta v_{{\rm em-abs}}$ (Figure~\ref{fig:delv_sum}). 
While statistically significant, the difference
between $W_{{\rm Ly}\alpha}(\Delta v=340)$ 
and $W_{{\rm Ly}\alpha}(\Delta v=890)$ 
is small compared to the large range of 
$W_{{\rm Ly}\alpha}$ seen in the entire sample.
Also, $W_{{\rm LIS}}$ marginally increases with increasing $\Delta v_{{\rm em-abs}}$,
though the difference is not significant, and may arise as a 
result of the strong correlation between Ly$\alpha$ and
the interstellar absorption lines (Figure~\ref{fig:delv_sum}). 
Somewhat surprisingly, the sample with 
$\langle \Delta v_{{\rm em-abs}} \rangle = 890$ \kms\ has very strong
{\it high}-ionization equivalent widths, though the Si~IV
doublet ratio indicates that Si~IV is still optically thin. Relative to the 
average values for the total LBG sample 
(section~\ref{sec:features_outflow_his}), the high $\Delta v$
sample has $W_{{\rm Si~IV}}$ and $W_{{\rm C~IV}}$ that are
50\% higher. Finally, the UV continuum becomes bluer
with increasing velocity width (Figure~\ref{fig:delv_sum}). 
None of the correlations with $\Delta v_{{\rm em-abs}}$ is as significant
as the trends among $W_{{\rm Ly}\alpha}$, $W_{{\rm LIS}}$, and $E(B-V)$.

\subsection{Comparison with Local Starburst Results}
\label{sec:trends_local}
A systematic analysis of the 
UV spectroscopic properties of local starburst galaxies
was carried out by \citet{heckman1998}, 
intended to guide the interpretation of UV spectroscopic
properties of high-redshift star-forming galaxies.
Several starburst parameters were considered, 
including UV continuum slope, UV and bolometric luminosity, 
UV low-ionization interstellar equivalent width, high-ionization
stellar wind equivalent width, and nebular metallicity. 
One of the parameters carrying a lot of the variance of the 
properties of local starbursts is metallicity, 
measured from the nebular oxygen abundance, O/H. More metal-rich
starbursts have more dust extinction, higher star-formation rates,
and stronger UV absorption lines. Since there are nebular
O/H measurements for only a handful of LBGs, and the C/O
measurements from the composite spectra are very uncertain,
it is not possible to make a direct comparison to the 
metallicity-dependent results for local starbursts. Moreover,
a large part of the analysis in the present work 
addresses the factors which control
the emergent Ly$\alpha$ profile in LBGs, which is, ironically,
much harder to study at low redshift due to contamination
from geocoronal Ly$\alpha$, and therefore not included
in the Heckman \et analysis. 
However, we can directly compare the local and high-redshift
results for the relationship between UV continuum 
reddening and interstellar absorption line strengths.
Both in local starbursts and LBGs, there is a 
strong correlation between $W_{{\rm LIS}}$ and UV continuum reddening.
Since the absorption lines are saturated in both samples, variations in
their equivalent widths reflect changes in the combination
of neutral gas covering fraction and velocity dispersion,
rather than in metal abundance.
For the local starbursts, the correlation is interpreted as 
resulting from the mutual dependence of both $W_{{\rm LIS}}$
and dust extinction on the velocity dispersion
of the absorbing neutral gas. As described more fully in
section~\ref{sec:discussion_physical},
we offer a different interpretation of this trend. 

We can compare the $z\sim3$ results about Ly$\alpha$ emission
with other local studies.
In contrast to the apparent decoupling between Ly$\alpha$ extinction
and continuum reddening found in local starburst galaxies
\citep{giavalisco1996a} we find a significant correlation
between UV continuum extinction and $W_{{\rm Ly}\alpha}$.
Either this difference between the local starbursts and LBGs
points to a significant difference between
the geometries of dusty neutral gas in star-forming
galaxies at low and high-redshift,
or else the low-redshift sample suffered from the vagaries
of small sample statistics.
In the work of \citet{kunth1998}, strong Ly$\alpha$ absorption
is seen in cases where the interstellar absorption lines are static
with respect to the galaxy systemic velocity, whereas
Ly$\alpha$ emission is detected in galaxies with blue-shifted
interstellar absorption lines. 
The blue-shift of interstellar absorption lines is ubiquitous
in the LBG spectroscopic sample, even in 
galaxies with strong Ly$\alpha$ absorption. In contrast to the evidence
at low redshift, the presence of
an outflow in LBGs is not a sufficient criterion for
detecting Ly$\alpha$ emission. In fact, it appears that the
Ly$\alpha$ emission equivalent width {\it increases} as the velocity offset
between Ly$\alpha$ and the interstellar absorption lines
decreases.

In local starbursts, the UV nebular emission lines
such as C~III] $\lambda 1909$ are stronger in starbursts of 
lower metallicity \citep{heckman1998}.
This effect is most likely due to a decrease in the 
nebular electron temperature of higher metallicity gas, 
which causes more of the nebular cooling from 
collisionally excited lines to occur in the infrared
rather than the UV.  Collisionally excited C~III] and O~III]
nebular emission lines are stronger than average in
the composite spectrum of LBGs with $W_{{\rm Ly}\alpha} \geq 20$ \AA.
By analogy with the local results, 
this subsample may be composed of objects with 
lower than average LBG metallicites. 
The C/O ratio in the strong emission composite spectrum
implies an O/H abundance which is lower than the average O/H
observed in the small sample of LBGs with rest-frame optical
spectroscopic measurements \citep{pettini2001}, though the uncertainties in
both measurements are large. Important future observations include
nebular metallicity measurements of statistical
samples of LBGs, when multi-object rest-frame optical spectroscopy 
becomes possible. Also important are
outflow metallicity measurements from
unsaturated interstellar absorption
lines, which will be possible with
higher S/N and resolution rest-frame UV spectra.

\section{Summary and Discussion}
\label{sec:discussion}
We have presented a summary of the spectroscopic
features contained in LBG rest-frame UV spectra, and 
the important trends among UV
continuum reddening; outflow kinematics; and the equivalent widths of
Ly$\alpha$, low- and high-ionization interstellar absorption, 
and nebular emission.
The most important results are:

1. LBGs with 
stronger Ly$\alpha$ emission have bluer UV continua,
weaker low-ionization interstellar absorption lines,
smaller kinematic offsets between Ly$\alpha$ emission and interstellar
absorption lines, and lower star-formation rates 
(the last property may be due in part to selection effects).

2. Low- and high-ionization absorption equivalent widths exhibit different
behaviors as functions of other spectral properties.

3. Galaxies with rest-frame $W_{{\rm Ly}\alpha} \geq 20$ \AA\ 
in emission have
weaker than average high-ionization lines, and nebular
emission lines which are significantly stronger than in the 
rest of the sample. In the subsample with strong emission lines,
which is itself fainter than average for the
spectroscopic sample, there is evidence that Ly$\alpha$
emission strength increases towards fainter magnitudes (and lower
star-formation rates). 
To study the dependence of the full distribution of Ly$\alpha$ equivalent
widths on luminosity, a more careful treatment of 
photometric and spectroscopic incompleteness is required.

\subsection{A Physical Picture}
\label{sec:discussion_physical}
Ultimately, these spectroscopic trends are not very interesting
unless we consider what they imply about the physical
conditions in LBGs and their outflows.
Here we offer a picture which
is consistent with all of the observations, and highlight
empirical results which still require further explanation.
In high-resolution optical and near-IR images,
LBGs have typical half-light radii of $1.6 h^{-1}$ kpc 
\citep{giavalisco1996b}. This typical half-light
radius represents the scale of the H~II regions where massive
star formation takes place. 
The UV stellar continuum, photospheric and wind features, 
and UV and optical nebular emission lines 
are produced here. Negligible systemic velocities of nebular 
emission lines with respect to stellar photospheric absorption features
indicate that the massive stars and H~II regions are at rest with respect to
each other (sections~\ref{sec:z} 
and~\ref{sec:features_neb_co}). 
Also produced in the H~II regions through recombinations
of ionized hydrogen are the original
Ly$\alpha$ photons which subsequently diffuse through 
frequency space and through dusty 
neutral gas at larger radii. 

A generic feature of galaxies selected with Lyman Break color criteria
is a star-formation surface density
high enough that the mechanical energy input from the 
resulting large number of 
Type II supernova explosions drives a large
bubble of hot gas out of the galaxy \citep{heckman2002}. 
This superbubble expands at several hundred \kms,
and may eventually escape the galactic gravitational
potential, possibly without radiating away most of the mechanical energy
input from supernovae \citep{adelberger2002a}. 
The location, relative to the central
star-formation regions, of the neutral and ionized
gas giving rise to strong blueshifted absorption
is an important component of any physical picture
describing the spectra of LBGs.

Neutral gas
displays a range of blue-shifted line-of-sight velocities with an average of
$\Delta v \sim -200$ \kms, but extending to $\Delta v \sim -600$~\kms
\citep{pettini2002}.
If cold neutral interstellar clouds are entrained
and accelerated by the flow of hot gas, they may reside at smaller
distances, $r$, than the expanding shock front of the superbubble.
A plausible lower limit to $r$ is the typical LBG 
half-light radius, $1.6 h^{-1}$ kpc, given
that the majority of LBG rest-frame UV spectra
exhibit such strong absorption lines. If the absorbing gas were
at much smaller radii, more mixed in with the stars, we would
expect a much smaller covering factor, and weaker absorption on average.
An upper limit on {\it r} may be obtained from consideration
of close pairs of LBGs, by searching for absorption from
the lower redshift member of a pair in the spectrum of
the higher redshift one.
There are 17 LBG pairs with projected separations of 
$r_{\theta} < 160h^{-1}$ comoving kpc. When the spectra of the 
higher-redshift pair members are shifted into the rest-frame
of the lower-redshift galaxies and averaged, little Ly$\alpha$ or metal
absorption is detected. The mean impact parameter in these pairs
is $110h^{-1}$~kpc, which puts an upper limit of $r \simlt 25h^{-1}$
proper kpc on the physical dimensions of the
gas giving rise to strong blue-shifted
absorption in LBG spectra \citep{adelbergeret2002a}.

Along with neutral material, there is also ionized 
gas with a similar mean blueshift and range
of velocities over which absorption takes place.
Blue-shifted Si~IV and C~IV absorption features are produced
in this gas, which has been ionized by a mixture of radiation
and collisional processes. In contrast to the low-ionization features,
the Si~IV doublet appears to be optically thin. While the neutral
and ionized phases have comparable kinematic properties, this similarity does
not necessarily constrain their relative physical distributions.
In alternate scenarios consistent with the kinematics of the low and high ions,
1) the Si~IV and C~IV absorption may be produced in the
outer ionized regions of the outflowing clouds giving rise 
to low-ionization absorption,
or 2) it may originate in more diffuse, 
but still outflowing, ionized gas in which
the outflowing neutral clouds are embedded.
There may also be evidence for a third, 
hotter phase of gas at $T \sim 3\times 10^5$~K,
if the O~VI producing absorption
is collisionally ionized. In the future, it will be interesting to compare
the kinematics of the O~VI absorption with that of the
low-ions and other high-ions, using higher spectral
resolution data.

\citet{heckman2001b} and \citet{strickland2000} have considered the outflow
scenario in detail for local starbursts, and \citet{adelberger2002a}
and \citet{pettini2002} have done the same for LBGs at $z\sim 3$. 
The composite LBG spectra provide some new results
about outflows at high redshift.
The low-ionization interstellar
absorption lines are the most direct probe of the outflowing
neutral gas, and there is clearly a direct link between $W_{{\rm LIS}}$ and
the emergent Ly$\alpha$ profile. Quite strikingly, the interstellar
absorption strength monotonically decreases as the Ly$\alpha$
emission strength increases. Also important is
the decoupling of the behavior of neutral and ionized absorbing
gas. When the sample is divided according to $W_{{\rm Ly}\alpha}$, 
$W_{{\rm LIS}}$ varies by almost 
a factor of three, whereas $W_{{\rm SiIV}}$
stays roughly constant except for the sample with strong Ly$\alpha$
emission, in which the high-ions are 50\% weaker.
The difference between the behavior in low and high ion line strengths
is especially intriguing since the two sets of lines have 
similar mean blue-shifts and velocity FWHMs.
This result may be evidence for a variable neutral
gas covering fraction in the outflow, whereas
the ionized gas maintains a more constant covering fraction.
It also favors the scenario in which patchy neutral clouds are embedded in
an ionized gas phase with unity covering fraction, rather than the one in which
Si~IV and C~IV absorption are produced in the outer regions of the patchy
neutral clouds.

Dust has been observed in the outflows from local star-forming
galaxies \citep{heckman2000}. In this work, 
we also find strong evidence for dust in the outflowing neutral clouds
at $z\sim 3$. 
The correlation between the $W_{{\rm LIS}}$ and $E(B-V)$ is most
naturally explained if some fraction of the
reddening of the stellar continuum takes place in the outflowing
gas. The strong correlation between $W_{{\rm LIS}}$ and $E(B-V)$
also argues for an outflow geometry which
is at least comparable in size
to the galaxy half-light radii --
i.e. the absorbing neutral clouds must be distributed in 
front of the entire face of the galaxy, affecting the total region of
UV continuum surface brightness, and not only a small central region.
LBGs differ from local galaxies hosting starbursts in terms
of two important geometrical properties, which may also distinguish
the nature and evolution of LBG outflows from local ones. 
First, local starbursts often
occur in a central nuclear region with  $r_* \leq 1$ kpc, which
is small compared to the galaxy half-light radius, whereas LBGs
indicate high star-formation surface-densities over a larger size-scale. 
Second, local star-forming galaxies often have large gas disks. 
Superwinds expand perpendicular to these disks, in the
direction of the steepest pressure gradient \citep{heckman2000,heckman2002}.
Based on the morphological information about LBGs
and models of galaxy formation, such disks
were probably not in place at $z\sim 3$.

\subsection{The Absorbing Gas}
\label{sec:discussion_gas}
The properties of the blue-shifted neutral gas are crucial in
determining the appearance of LBG rest-frame UV spectra. 
The equivalent widths of these low-ionization lines 
correlate strongly with two key spectroscopic
properties: $W_{{\rm Ly}\alpha}$ and $E(B-V)$.
An important question is: what physical
parameter of the low-ionization gas governs the equivalent widths?
As stated before, 
the four strongest low-ionization lines must be saturated
based on the ratio of equivalent widths of different
Si~II transitions. Therefore, the change in average equivalent width
by a factor of $\sim 3$ reflects a change in the neutral clouds'
velocity dispersion, covering fraction, or both.
Using all of the available information, we attempt to infer
which of these properties is reflected by
the change in $W_{{\rm LIS}}$.

Since $W_{{\rm LIS}}$ is so strongly correlated with $W_{{\rm Ly}\alpha}$,
we focus on the properties of the
composite spectra described in section~\ref{sec:trends_lya},
constructed from subsamples sorted by $W_{{\rm Ly}\alpha}$.
This is, for practical purposes, 
equivalent to sorting by $W_{{\rm LIS}}$, which
we would like to do, but cannot, given the quality of the individual spectra.
The relevant measurements are the observed residual intensities
and FWHMs of the low-ionization interstellar absorption 
lines. For saturated lines, the deconvolved residual intensity
is an empirical estimate of the gas covering fraction, (i.e. $I=1-C_f$,
where $I$ is the residual intensity, and $C_f$ is the covering fraction).
The observed residual intensities smoothly increase as $W_{{\rm Ly}\alpha}$ 
increases from strong absorption to strong emission
(Figure~\ref{fig:lya}). Also, the observed
FWHMs of the composite spectrum with the strongest Ly$\alpha$
emission are narrower than those in the strongest absorption
composite, though there is not a smooth trend with $W_{{\rm Ly}\alpha}$.

Ideally, we would like to determine simultaneously
the deconvolved FWHMs and residual intensities of the low-ionization
interstellar absorption lines. The deconvolved
quantities depend very sensitively on the effective spectral
resolutions of the four Ly$\alpha$ composite spectra, however, and these
four spectral resolutions may not even be equivalent. 
For example, spectra with Ly$\alpha$ emission 
could have been spectroscopically identified under conditions of
worse atmospheric seeing than absorption-only spectra. Another effect is
that the systemic redshift formulae from \citet{adelbergeret2002a}
have different levels of precision for different types of spectra. The
effective spectral resolution is only constrained to be 
between $\Delta \lambda \sim 2.0-3.2$~\AA\ for the four spectra,
which is of the same order as the intrinsic FWHMs.
When the spectral resolution is so uncertain, yet comparable
to the intrinsic FWHMs, resolution effects are degenerate with
both intrinsic FWHM and residual intensity. 
Thus, we can only say that
there is a significant reduction in
the {\it product} of covering fraction and velocity spread
as the absorption lines decrease in equivalent width, 

\begin{inlinefigure}
\centerline{\epsfxsize=9cm\epsffile{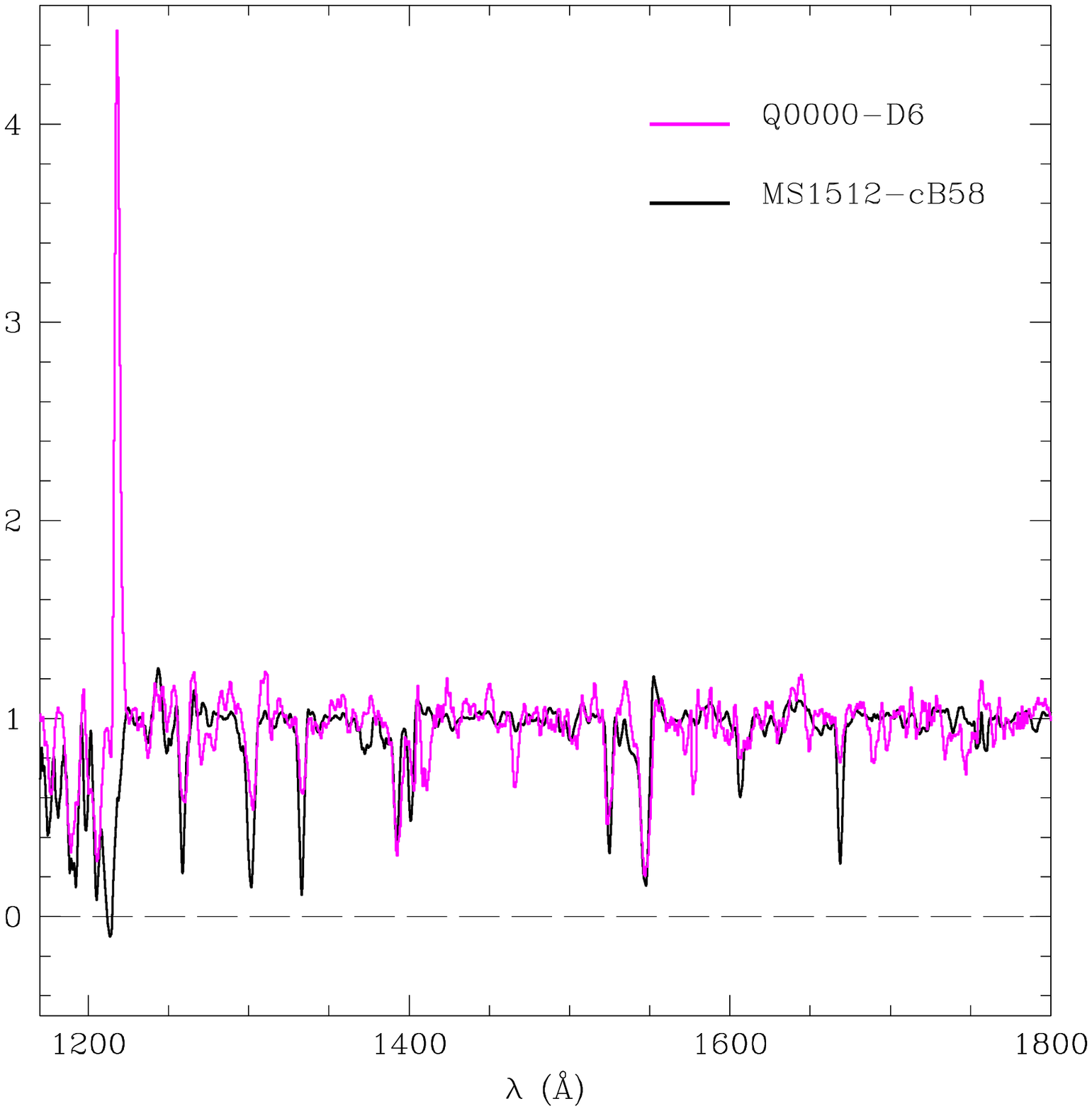}}
\figcaption{Continuum-normalized spectra of the two highest S/N LBGs in the
spectroscopic sample: MS1512-cB58 and Q0000-D6. cB58 (plotted in black)
belongs in the quartile of LBGs with the strongest Ly$\alpha$ absorption,
while Q0000-D6 (plotted in magenta)
is in the 30\% of the sample with the strongest Ly$\alpha$ emission.
The contrast between Ly$\alpha$ profiles is clear, as is the
contrast between $W_{{\rm LIS}}$ absorption
strengths.
On the other hand, the
strength of the high-ionization absorption lines is roughly
comparable. In this case, where we know the spectral resolution
accurately for both spectra, we can show that the intrinsic velocity
widths of the absorption lines
are comparable, and that the difference in $W_{{\rm LIS}}$ between cB58
and Q0000-D6 is due to a difference in the covering fraction
of neutral gas.
\label{fig:d6_cb58}
}
\end{inlinefigure}

Without a quantitative determination of the relative
significance of velocity FWHM and covering fraction in moderating
$W_{{\rm LIS}}$, we offer two pieces of circumstantial evidence that
covering fraction may be the more important effect of the two.
First, we compare the spectroscopic properties of the two
brightest spectroscopically confirmed LBGs that are not AGN: 
MS1512-cB58 $({\cal R}=20.6)$
and Q0000-D6 $({\cal R}=22.88)$. 
cB58 is in the
quartile of LBGs with the strongest Ly$\alpha$ and low-ionization
interstellar absorption. In contrast, Q0000-D6 has 
$W_{{\rm Ly}\alpha}$ and $W_{{\rm LIS}}$ which place it in the second highest
quartile of $W_{{\rm Ly}\alpha}$ emission. As shown in 
Figure~\ref{fig:d6_cb58}, the difference
in Ly$\alpha$ properties of the two objects is reflected in the difference in
low-ionization equivalent widths, which are $\sim 1.6$ times stronger
in cB58 than in Q0000-D6 (though, just as in the composite spectra,
the high-ionization line-strengths are roughly the same strength).
In this case, we explicitly measure the velocity widths of the 
low-ionization lines in cB58 and Q0000-D6. The C~II~$\lambda 1334$
FWHM is measured directly from the high-resolution spectrum of cB58
as FWHM(cB58)=655~\kms. Based on a careful analysis of the spatial extent
of the object along the slit in
the 2D spectrogram of the low-resolution spectrum of Q0000-D6, we estimate
a rest-frame resolution of $\Delta \lambda = 2.5$~\AA\, 
and the deconvolved C~II~$\lambda 1334$ velocity width is
FWHM(D6)=665~\kms. In this case, cB58 and Q0000-D6 have the same
low-ionization velocity dispersions, and the difference in equivalent width
can only be explained as a difference in the covering fraction of 
blueshifted neutral clouds. While the cB58 C~II line
has a residual intensity of $\sim 0$, implying a unity covering fraction,
the intrinsic residual intensity in Q0000-D6 is 0.4,
implying a covering fraction of only 60\%.
The second piece of circumstantial evidence is the strong correlation
between $W_{{\rm LIS}}$ and $E(B-V)$
(sections ~\ref{sec:trends_lya} and~\ref{sec:trends_dust}).
\citet{heckman1998} attribute this correlation in local starbursts to the 
fact that the velocity dispersion in the absorbing gas must be
larger in galaxies with more dust extinction. This explanation is plausible
in that galaxies with higher star-formation rates can
drive winds with larger velocity spreads \citep{heckman2000},
and they are also dustier \citep{adelberger2000,shapley2001}.
However, a more direct explanation for the correlation results
if the blue-shifted neutral gas is dusty. Accordingly, galaxies with
a larger covering fraction of dusty clouds suffer
more extinction of the UV stellar continuum, as well as 
exhibiting larger saturated equivalent widths, and weaker Ly$\alpha$
emission. 

\subsection{Lyman Continuum Leakage}
\label{sec:discussion_lyc}
One of the reasons we have devoted so much of the discussion
to the covering fraction of neutral gas is that this property
may determine how optically thick LBGs are to their own H~I-ionizing
radiation. The optical depth of LBGs to Lyman continuum photons is a
cosmologically interesting question, given the 
controversy surrounding the contribution of galaxies
to the ionizing background at $z\sim 3$. 
Escaping Lyman continuum flux was apparently detected
by \citet{steidel2001}
in a composite spectrum of 29 LBGs drawn from the high-redshift 
tail of the current spectroscopic
sample. This spectrum
is very similar to the composite spectrum of the total LBG
spectroscopic sample (Figure~\ref{fig:plotall}) 
in terms of $W_{{\rm Ly}\alpha}$, $W_{{\rm LIS}}$ and $W_{{\rm HIS}}$,
though it indicates less reddening by dust 
($\langle E(B-V)  = 0.07$ for the 
galaxies in the Lyman continuum sample, and $\langle E(B-V) \rangle = 0.13$
for galaxies in the total LBG sample). As discussed before,
the composite spectrum of the total LBG sample suffers from
its own selection biases with respect to the LBG UV luminosity function.
Therefore, the similarity of the Lyman
continuum composite spectrum to the total LBG composite spectrum
is not proof that the Lyman continuum spectrum represents a
true ``average''  $z\sim 3.4$ galaxy spectrum. 

Furthermore, in contradiction to the apparent detection of 
Lyman continuum emission, \citet{giallongo2002} derive an upper limit for
the escape fraction in two bright LBGs which is four
times lower than Steidel et al (2001) detection. One of these two
galaxies is Q0000-D6, (section~\ref{sec:discussion_gas}), which
apparently has only 60\% covering fraction of gas capable
of producing significant low-ionization metal absorption lines!
If there are indeed no unresolved saturated components
in the low-ionization features which are missed due to finite
spectral resolution, the case of Q0000-D6 may imply that
low-ionization lines which do not reach zero intensity
are a necessary but not sufficient condition for the escape
of Lyman continuum emission. For example, we note that the cores of
the high-ionization
lines in the spectrum of Q0000-D6 are black (when viewed at 
sufficient spectral resolution using the Keck Echelle Spectrograph
and Imager). If these transitions arise in gas with sufficient
neutral hydrogen column density to be optically thick to
Lyman continuum photons, very little ionizing radiation will escape
from Q0000-D6.
It is clearly of great interest to push the
Lyman continuum observations of this galaxy
to more sensitive limits than those
achieved so far. 

In order to arrive at the global contribution of LBGs to 
the ionizing background, 
high-resolution spectra are necessary to measure the
neutral gas velocity dispersions and covering fractions
without degeneracy as a function of $W_{{\rm Ly}\alpha}$, $W_{{\rm LIS}}$,
$W_{{\rm HIS}}$, $E(B-V)$, and $\Delta v_{{\rm em-abs}}$. Once it is understood how
the gas covering fraction depends on other spectroscopic
parameters, we need to construct an unbiased estimate
of the average spectrum of the light associated with LBGs. 
Proper weighting of each individual spectrum as a function of
its ${\cal R}$ magnitude and Ly$\alpha$ equivalent 
width should compensate for all of
the photometric and spectroscopic biases presented in 
section~\ref{sec:trends_seleffects} yielding the true average spectrum.
Based on the spectroscopic properties of the average spectrum,
the average LBG neutral gas covering fraction can be inferred.
The final step is to use direct observations of the Lyman continuum
region to calibrate the relationship between gas covering fraction
and Lyman continuum leakage. If it is possible to 
calibrate this relationship, 
the average LBG covering fraction can be converted
into a true average Lyman continuum escape fraction.

\subsection{Future Observations}
While this work has illuminated the range of rest-frame
UV spectroscopic properties of LBGs,
there are many limitations in the data which prevented us from
drawing quantitative conclusions, and which highlight the
need for several specific future observations. 
Data of higher spectral resolution are necessary
to measure the velocity distribution and covering
fraction of the absorbing neutral gas without
degeneracy. Properties such as stellar population age
should be folded into the analysis to search for temporal
evolution in the UV spectroscopic properties.
Comparing the {\it HST} morphologies of galaxies
of different spectroscopic types will also aid
in understanding the effects of geometry and orientation. 
With such observations we will truly be able to characterize the detailed
effects of star-formation at $z\sim 3$, both on galaxies and the
intergalactic medium.

\bigskip
\bigskip
We wish to extend special thanks to those of Hawaiian ancestry on
whose sacred mountain we are privileged to be guests. Without their generous
hospitality, most of the observations presented herein would not
have been possible. We also thank 
Mark Dickinson, Mauro Giavalisco, Mindy Kellogg,
Matthew Hunt, and Dawn Erb, who assisted in 
the observations and reductions, and the referee, Claus
Leitherer for careful and helpful suggestions which 
improved the quality of the paper.
CCS and AES have been supported by grant
AST-0070773 from the U.S. National Science 
Foundation and by the David and Lucile
Packard Foundation. KLA acknowledges support 
from the Harvard Society of Fellows.

\clearpage

\begin{deluxetable}{cllrrc}
\tablewidth{0pc}
\tablewidth{0pt}
\tablecaption{\textsc{Strong LBG Interstellar Absorption Features}\label{tab:is}}
\tablehead{
  \multicolumn{1}{c}{Ion}
& \multicolumn{1}{c}{$\lambda_{\rm lab}$\tablenotemark{a}}
& \multicolumn{1}{c}{$f$\tablenotemark{b}}
& \multicolumn{1}{c}{$W_0$\tablenotemark{c}}
& \multicolumn{1}{c}{$\sigma$\tablenotemark{c}}
& \multicolumn{1}{c}{$\Delta v$\tablenotemark{d}} \\
  \colhead{ }
& \multicolumn{1}{c}{(\AA)}
& \colhead{ }
& \multicolumn{1}{c}{(\AA)}
& \multicolumn{1}{c}{(\AA)}
& \colhead{ (\kms)}
 }
\startdata
Si II       &  1260.4221 & 1.007    & $-1.63$ & 0.10 &  $-110$ \\
O   I       &  1302.1685 & 0.04887  & $-2.20$\tablenotemark{e} & 0.12\tablenotemark{e} &  [$-270$]\tablenotemark{e} \\
Si II       &  1304.3702 & 0.094    & $-2.20$\tablenotemark{e} & 0.12\tablenotemark{e} &  [$-270$]\tablenotemark{e} \\
C  II       &  1334.5323 &  0.1278  & $-1.72$ & 0.11 &  $-150$ \\
Si IV       &  1393.76018  & 0.5140 & $-1.83$ & 0.16 & $-180$ \\ 
Si IV       &  1402.77291  & 0.2553 & $-0.81$ & 0.10 &  $-180$ \\ 
Si II       &  1526.70698 & 0.130   & $-1.72$ & 0.18 &  $-110$ \\
C IV       &  1548.204   & 0.1908   & $-3.03$\tablenotemark{f} & 0.21\tablenotemark{f} & [$-390$]\tablenotemark{f} \\ 
C IV       &  1550.781   & 0.09522  & $-3.03$\tablenotemark{f} & 0.21\tablenotemark{f} &  [$-390$]\tablenotemark{f} \\ 
Fe II       &  1608.45085 & 0.058   & $-0.91$ & 0.15 &  ~$-60$ \\
Al II       &  1670.7886 & 1.833    & $-1.04$ & 0.15 &  $-100$ \\
\enddata
\tablenotetext{a}{Vacuum wavelengths.}
\tablenotetext{b}{Transition oscillator strengths as in \citet{pettini2002}.}
\tablenotetext{c}{Rest frame equivalent width and $1 \sigma$ error. The error takes into account both sample variance and the S/N of the composite spectrum (see section~\ref{sec:trends_uncertainties}).}
\tablenotetext{d}{Relative velocity measured in the systemic rest frame of the composite spectrum, equivalent to the rest frame of the stars.} 
\tablenotetext{e}{$W_0$, $\sigma$, and $\Delta v$ values listed for 
O~I~$\lambda 1302$ and Si~II~$\lambda 1304$ refer to the total measured
for the blend of these two features. The value of $\Delta v$ assumes
that the rest wavelength of the blend is $\lambda=1303.2694$~\AA.}
\tablenotetext{f}{$W_0$, $\sigma$, and $\Delta v$ values listed for 
both members of the C~IV $\lambda 1548, 1550$ doublet 
refer to the blend of these two features. There may be
a contribution to $W_0$ from stellar wind absorption 
which has not been subtracted out. The value of $\Delta v$ assumes
that the rest wavelength of the blend is $\lambda=1549.479$~\AA.}
\end{deluxetable}

\begin{deluxetable}{rllrrc}
\tablewidth{0pc}
\tablewidth{0pt}
\tablecaption{\textsc{Weak LBG Emission Features}\label{tab:nebem}}
\tablehead{
  \multicolumn{1}{c}{Ion}
& \multicolumn{1}{c}{$\lambda_{\rm lab}$\tablenotemark{a}}
& \multicolumn{1}{c}{$A_{ul}$\tablenotemark{b}}
& \multicolumn{1}{c}{$W_0$\tablenotemark{c}}
& \multicolumn{1}{c}{$\sigma$\tablenotemark{c}}
& \multicolumn{1}{c}{$\Delta v$\tablenotemark{d}} \\
  \colhead{ }
& \multicolumn{1}{c}{(\AA)}
& \multicolumn{1}{c}{($10^8$s$^{-1}$)}
& \multicolumn{1}{c}{(\AA)}
& \multicolumn{1}{c}{(\AA)}
& \multicolumn{1}{c}{ (\kms)}
 }
\startdata
Si II*       &  1264.738 & 2.30e$+$01 & 0.34 & 0.09 &  $+130$ \\
Si II*       &  1309.276 & 7.00e$+$00 & 0.35 & 0.10 &  ~$+50$ \\
Si II*       &  1533.431 & 7.40e$+$00 & 0.21 & 0.09 &  $+110$ \\
O III]       &  1660.809  & 2.20e$-$06 & 0.23\tablenotemark{e} & 0.13\tablenotemark{e} &  ~$+40$ \\
O III]       &  1666.150  & 5.48e$-$06 & 0.23\tablenotemark{e} & 0.13\tablenotemark{e} &  ~$-50$ \\ 
CIII]       &  1908.734  & 1.14e$-$06  & 1.67 & 0.59 &  ~$+40$ \\ 
\enddata
\tablenotetext{a}{Vacuum wavelengths.}
\tablenotetext{b}{Einstein $A$-coefficients from the NIST Atomic Spectra
Database (http://physics.nist.gov/cgi-bin/AtData/main\_asd).}
\tablenotetext{c}{Rest frame equivalent width and $1 \sigma$ error. The error takes into account both sample variance and the S/N of the composite spectrum 
(see section~\ref{sec:trends_uncertainties}).}
\tablenotetext{d}{Relative velocity measured in the systemic rest frame of the composite spectrum, equivalent to the rest frame of the stars.} 
\tablenotetext{e}{$W_0$ and $\sigma$ values are given for the flux integrated over the whole O~III] doublet.}
\end{deluxetable}

\begin{deluxetable}{crrrr}
\tablewidth{0pc}
\tablewidth{0pt}
\tablecaption{\textsc{Spectroscopic Properties of Ly$\alpha$ Subsamples}\label{tab:lya}}
\tablehead{
  \multicolumn{1}{c}{}
& \multicolumn{1}{r}{Group 1\tablenotemark{a}}
& \multicolumn{1}{r}{Group 2\tablenotemark{a}}
& \multicolumn{1}{r}{Group 3\tablenotemark{a}}
& \multicolumn{1}{r}{Group 4\tablenotemark{a}}
 }
\startdata
$N_{gal}$                                & 199                & 198              & 199             & 198              \\
$W_{ {\rm Ly}\alpha}$\tablenotemark{b}   &  $-14.92 \pm 0.56$ & $-1.10 \pm 0.38$ & $11.00 \pm 0.71$ & $52.63 \pm 2.74$  \\
$W_{ {\rm SiII}, 1260}$\tablenotemark{b}  &  $-1.85 \pm 0.16$ & $-1.59 \pm 0.16$  & $-1.36 \pm 0.19$& $-1.05 \pm 0.22$  \\
$W_{ {\rm OI+SiII}, 1303}$\tablenotemark{b} &  $-3.24 \pm 0.16$ & $-2.71 \pm 0.16$  & $-1.98 \pm 0.19$ & $-1.21 \pm 0.21$  \\
$W_{ {\rm CII}, 1334}$\tablenotemark{b}     &  $-2.34 \pm 0.16$ &  $-1.91 \pm 0.15$ & $-1.43 \pm 0.19$ & $-0.83 \pm 0.23$  \\
$W_{ {\rm SiII}, 1526}$\tablenotemark{b}    &  $-2.38 \pm 0.19$ & $-1.82 \pm 0.19$ & $-1.33 \pm 0.23$ & $-1.21 \pm 0.33$  \\
$W_{ {\rm FeII}, 1608}$\tablenotemark{b}      &  $-1.57 \pm 0.21$ & $-1.03 \pm 0.18$   & $-0.69 \pm 0.25$ & $-0.59 \pm 0.34$  \\
$W_{ {\rm AlII}, 1670}$\tablenotemark{b}      &  $-1.64 \pm 0.22$ & $-1.08 \pm 0.20$ & $-1.07 \pm 0.25$ & $0.04 \pm 0.44$  \\
$W_{ {\rm SiIV}, 1393}$\tablenotemark{b}    &  $-1.83 \pm 0.23$  & $-1.87 \pm 0.21$ & $-1.93 \pm 0.27$ & $-1.22 \pm 0.30$  \\ 
$W_{ {\rm SiIV}, 1402}$\tablenotemark{b}    &  $-1.01 \pm 0.17$  & $-0.98 \pm 0.16$ & $-0.75 \pm 0.21$ & $-0.54 \pm 0.23$  \\ 
$W_{ {\rm CIV},  1549}$\tablenotemark{b}      &  $-3.56 \pm 0.30$   & $-2.97 \pm 0.32$   & $-3.22 \pm 0.39$ & $-2.43 \pm 0.47$  \\ 
$W_{ {\rm OIII]}, 1663}$\tablenotemark{b}     &  $0.23 \pm 0.19$ & $0.01 \pm 0.16$ & $0.10 \pm 0.12$ & $1.16 \pm 0.56$  \\
$W_{ {\rm CIII]}, 1909}$\tablenotemark{b}     &  $0.41 \pm 0.39$ & $2.89 \pm 1.04$ & $1.90 \pm 1.07$ & $5.37 \pm 2.99$  \\
$E(B-V)$\tablenotemark{c}                     &  $0.169\pm 0.006$ & $0.136\pm 0.006$ & $0.120\pm0.006$ & $0.099\pm0.007$  \\
$\beta$\tablenotemark{c}                      &  $-0.73\pm0.03$   & $-0.88\pm0.04$   & $-0.98\pm0.03$  & $-1.09\pm0.05$  \\
$\Delta v_{em-abs}$~(\kms)   &  $795\pm3$      & $630\pm35$ & $560\pm30$         &   $475\pm25$ \\
${\cal R}_{AB}$              &  $24.44\pm0.04$ & $24.49\pm0.04$   & $24.64\pm0.04$  & $24.85\pm0.04$  \\
SFR$_0$~($h^{-2}M_{\odot}$~yr$^{-1}$)\tablenotemark{d}   &  $52\pm 5$     & $38\pm2$         & $29\pm3$        & $25\pm3$  \\
\enddata
\tablenotetext{a}{Groups 1 -- 4 are the four quartiles of the LBG spectroscopic sample, divided according to
Ly$\alpha$ equivalent width.}
\tablenotetext{b}{Rest-frame equivalent width in \AA, measured from the composite spectra. 
Positive values indicate emission, while negative
values indicate absorption. Uncertainties are calculated as described in section~\ref{sec:trends_uncertainties}.}
\tablenotetext{c}{Estimates of reddening, based on the intrinsic UV color, $(G-{\cal R})_0$. $E(B-V)$ assumes
the Calzetti law for dust extinction, and an underlying stellar population with 300 Myr of constant star-formation. 
$\beta$ is derived directly from the $(G-{\cal R})_0$ color, assuming that the UV spectrum can be described by the
form, $F_{\lambda} \propto \lambda^{\beta}$.}
\tablenotetext{d}{Dust-corrected star-formation rate, derived from the apparent ${\cal R}$ magnitude, the redshift,
and the amount of UV extinction inferred from the $(G-{\cal R})_0$ color.}
\end{deluxetable}

\end{document}